\documentclass[12pt]{book}
\usepackage{amssymb}

\begin{document}

\thispagestyle{empty}

\vspace*{-1.5truecm}

\hspace*{-0.6cm}%
\makebox[\textwidth]{
		\raisebox{0.68cm}{
	\begin{minipage}[h]{13truecm}
        \hspace*{0.2cm}{\LARGE UNIVERSIT\`A DEGLI STUDI DI LECCE 
                }
    \end{minipage}}}

\vspace{0.8cm}

\begin{center}
FACOLT\`A DI SCIENZE \\[0pt]
Dipartimento di Fisica \\[0pt]
\end{center}

\vspace{2cm}

\begin{center}
{\LARGE \textbf{Ring Division Algebras, Self Duality and Supersymmetry }}
\end{center}

\vspace{1.5cm}

\begin{center}
{\large Tesi di dottorato di ricerca in Fisica} \\[0pt]
presentata da \\[0pt]
{\large \emph{Khaled Safwat Abdel-Khalek Mostafa}}
\end{center}

\vspace{1.5cm} \noindent Relatori \newline
{\large Professor \emph{Pietro Rotelli}} \newline
The Exam Commission \newline
{\large Professor \emph{Federico Cesaroni}} \newline
{\large Professor \emph{Boris Konopelchenko}} \newline
{\large Professor \emph{Emilia D'Anna}} \newline
External Member \newline
{\large Professor \emph{Antoine Van Proeyen}} \newline

\vspace{1cm}

\begin{center}
\underline{Ciclo XII} \\[0pt]
\vspace{0.5cm} Anno Accademico 1999-2000
\end{center}

\newpage
\thispagestyle{empty}
\vspace*{2cm}

\begin{center}
{\LARGE \textbf{\emph{Acknowledgements}}}
\end{center}

\vspace{1cm}

{\em
First and formost my gratitude goes to Allah my god for his continuous help, for 
helping me in 
choosing the topic of the thesis,
the supervisors and the place. Prof. M. Shalaby (Ain Shames University/Cairo) was responsible for 
introducing me to theoretical physics and for this I owe him much. I feel really grateful 
to Prof. P. Rotelli for
giving me the opportunity to work with and learn from him many beautiful ideas and I 
would like to thank him for many enjoyable discussions during the last four 
years. I feel indebted to Prof. C. Imbimbo for many constructive criticisims of this thesis
 and 
for helping me to have a better understanding of my work.
Over the last four years, I have had several fruitful discussions with Prof. G. 
Thompson  to whom I am thankful.
I would like to acknowledge Prof. A. Zichichi and the third world laboratory
for financial support at a critical period of my thesis.
Many thanks also to Prof. S. Marchiafava and F. Englert for encouragements.
My gratitude  to the members of  the Physics department at 
the
{\rm Universit\`a di Lecce} for the very stimulating
environment I have found here. Last but not least, this thesis is dedicated to my family.
}

\tableofcontents

\hyphenation{Jor-dan} \hyphenation{al-ge-bras} \hyphenation{quan-tum}

\chapter{INTRODUCTION}

\pagenumbering{arabic} \setcounter{page}{1}

\section{Mathematics}

Imaginary numbers appeared in mathematics a long time ago. For example,
Nicolas Chuquet (1445--1500) wrote ``Triparty en la science des nombres''
where he introduced an exponential notation, allowing positive, negative and
zero powers. He showed that some equations lead to imaginary solutions but
rejected them ``tel nombre est ineperible''. Geronimo Cardano (1501--1576)
wrote ``Ars magna'', found solutions to polynomials which lead to square
roots of negative quantities but also rejected them ``as subtle as it is
useless''. The first to consider imaginary numbers was Rafael Bombelli
(1530--1590) who published ``Algebra'' and proposes the ``wild idea'' that
one can use these square roots of negative numbers to get to the real
solutions by using conjugation. Albert Girard (1595--1632) publishes
``Invention nouvelle en l'algebra'' retaining all imaginary roots because
they show the general principles in the formation of an equation from its
roots. Rene Descartes (1596--1650) coins the term ``imaginary'' for terms
involving square roots of negative numbers but takes their existence as a
sign that the problem is insoluble. Reviving some speculation, Gottfried
Leibniz (1646--1716) says that imaginary numbers are halfway between
existence and nonexistence. Sustaining algebra by geometry, John Wallis
(1616--1703) was the first to represent complex numbers geometrically in his
book ``Algebra'' published in 1673. Roger Cotes (1682--1716) deduces that $%
\exp \left( \sqrt{-1}\;a\right) =\cos \left( a\right) +\sqrt{-1}\sin \left(
a\right) $ but his result was largely ignored. But then a new era begins, it
was Leonhard Euler who brought complex numbers from the shadow to the
daylight, he invents the symbol $i$ for $\sqrt{-1}$ and works extensively
with imaginary numbers, for example, he shows that a complex number to the
power of a complex number is also a complex number. Jean d'Alembert's
(1717--1783) constructs functions of complex variables, obtaining what later
is called the Cauchy--Riemann equation. Caspar Wessel (1745--1818) discovers
that complex numbers can be represented graphically on a two dimensional
plane, what we now call the ``Argand'' or ``Guassian'' representation of
complex numbers. Modern complex analysis may be dated to the book of
Augustin Cauchy (1789--1857) ``Memoire sur les integrales definies, prises
entre des limites imaginaires'' which contains his integral theorems on
residues. Then the work of Augustus de Morgan (1806--1871) and Carl Gauss
(1777--1855) opens the way to what later becomes complex numbers analysis.
So finally, what was rejected as useless quantities become the heart of
mathematics.

While the discovery and acceptance of complex numbers took a long time, the
histroy of quaternions and octonions is much shorter. Quaternions were
discovered by a single man \cite{ham}, William Hamilton (1805--1865). Trying
to generalize his ``Theory of Algebraic Couples'', where he constructs a
rigorous algebra of complex numbers as number pairs for the first time, he
identifies $x+iy$ with its $\Bbb{R}^{2}$ coordinates $\left( x,y\right) $.
After many years of trial and error, Hamilton discovers quaternions on
Monday 16 October 1843 and defines a vector subspace $ai+bj+ck$ by elements
which may be interpreted as an $R^{3}$ coordinate system $\left(
a,b,c\right) $ but $i,j,k$ are not commutative. As early as 1845, Octonions
were \ introduced by Arthur Cayley and John Graves independently \cite{gra}%
\cite{cay}.

Quaternions \ and octonions may be presented as a linear algebra over the
field of real numbers $\Bbb{R}$ with a general element of the form 
\begin{equation}
Y=y_{0}e_{0}+y_{i}e_{i},\;\;\;\;\;\;\;y_{0},y_{i}\in \Bbb{R}  \label{gen}
\end{equation}
where $i=1,2,3$ for quaternions $\Bbb{H}$ and $i=1..7$ for octonions $\Bbb{O}
$. \ We always use Einstein's summation convention. The $e_{i}$ are
imaginary units, for quaternions 
\begin{eqnarray}
e_{i}e_{j} &=&-\delta _{ij}+\epsilon _{ijk}e_{k}, \\
e_{i}e_{0} &=&e_{0}e_{i}=e_{i}, \\
e_{0}e_{0} &=&e_{0},
\end{eqnarray}
where $\delta _{ij}$ is the Kronecker delta and $\epsilon _{ijk}$ is the
three dimensional Levi--Cevita tensor, as $e_{0}=1$ when there is no
confusion we omit it. Octonions have the same structure, only we must
replace $\epsilon _{ijk}$ by the octonionic structure constant $f_{ijk}$
which is completely antisymmetric and equal to one for any of the following
three cycles 
\begin{equation}
123,\;145,\;176,\;246,\;257,\;347,\;\;365.
\end{equation}
The important feature of real, complex, quaternions and octonions is the
existence of an inverse for any non-zero element. For the generic
quaternionic or octonionic element given in (\ref{gen}), we define the
conjugate $Y^{*}$ as an involution $\left( Y^{*}\right) ^{*}=Y$, such that 
\begin{equation}
Y^{*}=y_{0}e_{0}-y_{i}e_{i},
\end{equation}
introducing the norm as $N\left( Y\right) \equiv \left\| Y\right\|
=YY^{*}=Y^{*}Y$ \ then the inverse is 
\begin{equation}
Y^{-1}=\frac{Y^{*}}{\left\| Y\right\| }.
\end{equation}
The Norm is nondegenerate and positively definite. We have the decomposition
property 
\begin{equation}
\left\| XY\right\| =\left\| X\right\| \;\;\left\| Y\right\|  \label{dec}
\end{equation}
$N\left( xy\right) $ being nondegenerate \ and positive definite obeys the
axioms of the scalar product and our algebra is called a normed algebra. The
uniqueness and beauty of real, complex, quaternionic and octonionic\ numbers
stem from Hurwitz' theorem \cite{hur}:

\begin{itemize}
\item[ ]  Each normed composition algebra with a unit element is isomorphic
to one of the following algebras: to the algebra of real numbers, to the
algebra of complex numbers, to the quaternion algebra or to the octonion
algebra.
\end{itemize}

Another important mathematical property of the ring algebra is the
following: For the set $\mathcal{S\;\;\;}$defined by 
\begin{equation}
\mathcal{S\;\;}=\left\{ X\;\;|\;\;\left\| X\right\| =1\right\}
\end{equation}
where $X$ is a ring division element then from the decomposition property (%
\ref{dec}), we have a closure structure 
\begin{equation}
X,Y\in \mathcal{S\;\;\;\;}\rightarrow Z=XY\in \mathcal{S}
\end{equation}
even for octonions which do not admit a group structure (group is defined
for associative algebra). This beautiful closure can be extended to a
generic ring division element by scaling. For any two generic ring division
elements $W,V$ we construct 
\begin{equation}
\widetilde{W}=\frac{W}{\left\| W\right\| },\;\;\;\;\;\widetilde{V}=\frac{V}{%
\left\| V\right\| }
\end{equation}
hence 
\begin{equation}
\left\| \widetilde{W}\right\| =\left\| \widetilde{V}\right\| =\left\| 
\widetilde{W}\widetilde{V}\right\| =1.
\end{equation}
A geometric meaning of this closure is the parallelizability of ring
division spheres. For $\left\| X\right\| =1$%
\begin{equation}
\begin{array}{lllllll}
\Bbb{C} &  & x_{0}^{2}+x_{1}^{2} & =1 & defines & a\;unit & S^{1} \\ 
\Bbb{H} &  & x_{0}^{1}+x_{1}^{2}+x_{2}^{2}+x_{3}^{2} & =1 & defines & a\;unit
& S^{3} \\ 
\Bbb{O} &  & 
x_{0}^{1}+x_{1}^{2}+x_{2}^{2}+x_{3}^{2}+x_{4}^{2}+x_{5}^{2}+x_{6}^{2}+x_{7}^{2}
& =1 & defines & a\;unit & S^{7}
\end{array}
\end{equation}
the parallelizability means that there is such an $X$ that defines \emph{%
globally} $1,3,7$ vector fields for $S^{1},S^{3},S^{7}$ respectively \cite
{adams}\cite{husm}.

Another importance of ring division algebras is their relevance to the
classification of real Clifford algebras $Cliff\left( m,n\right) .$ A task
that had been achieved by Atiyah, Bott and Shapiro \cite{abs}. Their
appearance is clear through the Bott periodicity. We also wish to mention
the work of Milnor\footnote{%
According to legend, Milnor presented his first acheivements as an
assignment. On one occasion, he was late for \ the class of Fox (the Father
of the american Knot theory). During that lesson, Fox explained his way of
doing research. Usually he writes the most difficult 10 questions and tries
to solve any one of them. As an example, he wrote for his students the 10
questions that kept him busy at that time. Fortunatley, John Milnor came
late that day, he saw the quetions, he thought they were homework. After the
class he worked hard untill the next morning. Then before the class of the
next day, he approached Fox expressing his desire to change his field of
research because he only managed to solve one of the 10 questions. Fox was
totally surprised. But then later Milnor continued to surprise the world
especially by his Field Medal's diffeomorphic non-homeomorphic structure.},
where for the first time in history a diffeomorphic non homeomorphic
structure was found. There are 28 of such structures over $S^{7}$, the fake $%
S^{7}$ \ \cite{milnor}. As a matter of fact, these fake $S^{7}$ are the
higher bundles of the four dimensional $SO\left( 4\right) $ instanton
solutions\cite{yoniama}. Donaldson received in 1986 the Field Medal for his
work about the infinite diffeomorphic non-homeomorphic $\Bbb{R}^{4}$. An
idea that he got by carefully studying the space of solutions of the four
dimensional $n=1$ ``quaternionic'' instanton. Finally, ring division
algebras have a connection with homotopies, Hopf fibrarion and many other
interesting topics.

It is clear that the history of these other elements of the ring division
algebra, quaternions and octonions, are much shorter than the complex one.
Maybe, it has not been yet fully written. With the hindsight of complex
numbers, these new hypercomplex numbers were immediately accepted, perhaps
only the question of their utility in physics is still to be discovered and
may yet involve much discussion and take a long time.

\section{Physics}

Non associative algebra appeared for the first time in physics when Jordan,
van Neuman and Wigner introduced commutative but non-associative operators
-- Jordan algebras -- for the construction of \ a new quantum mechanics\cite
{c1}. More recently ,after the proposed eightfold way by Gell-Mann and
Ne'eman, there were some octonionic rivals for $SU\left( 3\right) $ such as $%
G_{2},\;SO\left( 7\right) ,\;SO\left( 8\right) $ and others\cite{c2}.
Possible octonionic internal symmetries were considered by different people
such as Souriau and Kastler \cite{c3}, Pais \cite{c4}, Tiomno \cite{c5},
Gamba \cite{c6}, and Penny \cite{c7}. The inclusion $SU\left( 3\right)
\subset G_{2}\subset SO\left( 7\right) \subset SO\left( 8\right) $ lead many
authors to consider the relationship between the nuclear strong field and
octonions. There are a lot of papers dealing with the $SO\left( 8\right) $
symmetry \cite{c8}. The potential significance of non-associative algebras
for the generalization of classical dynamics was pointed out by Nambu \cite
{c9}. Generalizing the Liouville theorem, about the conservation of the
phase space volume, Nambu introduced a new generalization of Poisson
brackets which may be interpreted as an associator. Nambu in his paper made
some remarks about non-associative algebras, octonions and Jordan algebras.
As another application of octonions, allowing parastatistics and the
paraquark model, Freund showed in \cite{c9a} that color gauging is only
possible in the octonionic realization of the paraquark model. Octonions
have even been applied to gravitation. Vollendorf \cite{vollendorf}
constructed a bilocal field theory with the group $SO\left( 4,4\right) $
instead of the Lorentz group and with a system of 24 coupled differential
equations to explain the five conservation laws of charge, hypercharge,
baryon number and the two lepton number then known.

From the early seventies and up to the present time, octonions have been
applied with some success to different important problems such as quark
confinement, grand unified models (GUT). A serious step was taken by
G\"{u}naydin and G\"{u}rsey \cite{c10}\cite{c10a} when they present in their
work a systematic study of the octonionic algebraic structure. If we follow
the theory of observable states developed by Birkhoff and van Neumann \cite
{c11}, we can only have observable states in Hilbert spaces over associative
normed algebras. The standard quantum mechanics explores the Hilbert space
over complex numbers. The quaternion case with a quaternionic scalar product
was developed by Finkelstein, Jauch and Emch \cite{c12}. The general hope
was to introduce isospin degrees of freedom by enlarging the quantum Hilbert
space. But Jauch proved that quaternionic representations of the
poincar\'{e} group did not generate any new states. The reader is referred
to the book of Adler for a modern formulation of quaternionic field theory 
\cite{adler}. Another interesting idea is the use of hypercomplex Hilbert
space \ but with a complex scalar product, Goldstein, Horwitz and Biedenharn 
\cite{c13}, considered octonionic Hilbert space with a complex scalar
product as early as 1962. The abandon of associativity means the existence
of nonobservable states. G\"{u}naydin and G\"{u}rsey used these
nonobservable states to explain the quark confinement phenomena.

Later on octonions entered the Grand Unified Theories era with different
applications. In \cite{c14}, G\"{u}rsey suggested that the exceptional group 
$F_{4}$ might be used to describe the internal charge space of particles.
Even the other exceptional groups $E_{6},E_{7},E_{8}$ have been utilized to
provide larger GUT models \cite{e6e7} 
\begin{equation}
SU(3)_{c}\times SU(2)_{L}\times U(1)_{Y}\subset SU(5)\subset SO(10)\subset
E_{6}\subset E_{8}\,.
\end{equation}
As a matter of fact, nowadays, $E_{6}\times U\left( 1\right) $ is the most
promising scheme from the superstrings phenomenological point of view.
G\"{u}naydin constructed an exceptional realization of the Lorentz group in 
\cite{c15}. The exceptional $SL\left( 2,\Bbb{C}\right) $ multiplets generate
a non-associative algebra. Exceptional supergroups were also introduced and
investigated in \cite{c15}.

Starting from the eighties, new applications of ring division algebras in
physics were found. The instanton problem, supersymmetry, supergravity,
superstrings and recently branes technology. We give references in the
appropriate chapters of this thesis to the history of the first two topics.
Application of quaternions and octonions to supergravity spontaneous
compactification was a very important and active field of research during
the mid eighties. Especially compactification of $d=11$ supergravity over $%
S^{7}$ to 4 dimensions. It is an impossible task to list all the relevant
papers, so we direct the interested reader to the physics report \cite
{duffpope} written by Duff, Nilsson and Pope where a lot of references are
given. We just mention that the first indication of the octonionic nature of
this problem appeared in the Englert solution of $d=11$ supergravity
compactification over $S^{7}$\cite{englert} and a systematic study along
this line has been carried out in \cite{rom}\cite{grutze}\cite{tukersudbery}%
\cite{dewittnicol}. The relations between superstrings (p-branes) and
octonions had been considered from many different points of view, the reader
may consult the references given in \cite{duff} for details.\newpage

\section{Outline of The Thesis}

Superstrings promise the possibility of unifying all the fundamental forces
of nature. Abandoning the idea of point particle seems necessary to
incorporate general relativity with renormalizable field theory. One of the
puzzles of this string program is its double facet \cite{gsw}\cite{polsh}.
On the one hand, we can work over the two dimensional string sheet where we
can use the powerful methods of conformal field theory. On the other hand
working with the ten dimensional space-time Green--Shwarz (GS)
supersymmetric action, non--perturbative effects may be seen. Understanding
the relations between these two different formalisms of string theory is
important but many features of the GS formalism still have to be elucidated.
For example, the the algebra is given on--shell and the action exists only
in certain space--time dimensions, $d=3,4,6,10$. Only for ten dimensions,
the quantum anomalies of the model cancel i.e. $d=10$ is a very special case
of an already a very restrictive class. The relationship between gamma
matrices needed for the existence of the GS picture is the same as that used
to prove the existence of simple supersymmetric Yang-Mills models
(containing only a gauge field and a spinor) in the same dimensions $%
d=3,4,6,10$ \cite{r0}. A complete comprehension of this fact is important.

At the quantum level, we know how to proceed in a perturbative fashion using
Feynman diagrams. Starting from quantum chromodynamics QCD, and its quark
confinement problem, theoretician searched for non--perturbative phenomena.
Self--dual Yang--Mills solutions, \emph{instantons} \cite{bpst}\cite{hooft1}%
, can never be evaluated using perturbation theory. It is widely believed
that in superstrings, non--perturbative solutions are of great significance
and they may give interesting phenomenological applications that can be
tested experimentally.

In short, off--shell formalism of SSYM and self--duality are very critical
topics which may improve our theoretical knowledge and can be used as toy
models for testing new approaches.

The object of this thesis is to investigate in a systematic way the
relations between ring division algebras, off--shell SSYM and higher
dimensional self--duality. The starting point is to understand how ring
division algebras are specific representations of Clifford algebras. We
present this analysis in chapter II. For complex numbers the discussion is
simple. Even for quaternionic numbers, once the non-commutativity is taken
into account, the formulation can be fully analyzed and understood easily.
As octonions are non-associative numbers, we need to work harder to clarify
many different subtleties. In chapter II we concentrate upon the Clifford
structure that can be extracted from octonions. To make the picture clearer,
we show what will happen if we go beyond the ring division algebras limit.
The Clifford structure is no longer faithful.

In chapter III, we continue our study of octonions, we show that they are
endowed with additional useful characteristics. Fixing the direction of
action, octonions exhibit soft Lie algebra properties which we call a soft
seven sphere \cite{estps}\cite{soh}. Soft Lie algebras are elements that
close under the action of the commutator with structure functions of
coordinates system that parameterize a hidden space (the gauge manifold). We
study this scheme in full details, we compute the structure functions
explicitly with different degrees of complication.

In Chapter IV, we start to investigate the physical applicability of the
soft seven sphere. The \ self duality conditions play a fundamental role for
any non--perturbative effects in point particle or string field theory \cite
{duality}. The higher dimensional self dual constraints have properties
distinct from the four dimensional one. As a first exercise in the use of
the soft seven sphere, we show how to reformulate a quartic eight
dimensional self duality condition into a quadratic form. Thus, we put the
Grossman--Kephart--Stasheff condition (GKS) \cite{gks} into a form much
similar to the four dimensional equation.

In Chapter V, we will discuss supersymmetry. In particular, the off-shell
simple supersymmetric Yang--Mills models SSYM. We shall show that important
characteristics can be seen clearly only by using the ring division
approach. The ten dimensional case will be very special. For example, we
recover Berkovits formulation for the $d=10$ off--shell SSYM in a very
transparent way.

\chapter{Hypercomplex Structures}

The starting point of this chapter is to know how to translate some real $%
n\times n$ matrices $\Bbb{R}(n)$ to their corresponding complex,
quaternionic and octonionic representations ( shaeffer bimodule
representation \ for octonions \cite{shaeffer}). It is well known from a
mathematical point of view that any $\Bbb{R}^{2n}$ is trivially a $\Bbb{C}%
^{n}$ complex manifold and any $\Bbb{R}^{4n}$ is also a trivial quaternionic
manifold $\Bbb{H}^{n}$. Furthermore, any $\Bbb{R}^{8n}$ is a trivial $\Bbb{O}%
^{n}$ octonionic manifold, in the sense that the seven sphere can always be
embedded in $\Bbb{R}^{8}$. As any $\Bbb{R}^{n}\times \Bbb{R}^{n}$ is
isomorphic as a vector space to the space of $n\times n$ matrices $\Bbb{R}%
(n) $ \cite{gilmore}, we would expect 
\begin{eqnarray}
\Bbb{C}(n)\times \Bbb{C}(n) &\rightarrow &\Bbb{R}(2n);  \label{cl0} \\
\Bbb{H}(n)\times \Bbb{H}(n) &\rightarrow &\Bbb{R}(4n);  \label{cl1} \\
\Bbb{O}(n)\times \Bbb{O}(n) &\rightarrow &\Bbb{R}(8n).  \label{cl2}
\end{eqnarray}
Even if in this thesis we only work with matrices, there is a hidden
geometric and topological underlying structure behind this algebraic
construction.

Any hypercomplex manifold has a well defined $\emph{local}$ hypercomplex
structure that can be put into the matrix form that we shall develop in this
chapter. Lifting this local hypercomplex structure to a global one is not
always possible. It amounts to dividing the manifold into local patches
where the almost structure is well defined and gluing together these
different patches insuring the existence of a (differentiable) structure
function that transfers the local hypercomplex structure from one patch to
another. If this can be achieved over all the manifold then our space admits
a global hypercomplex structure. From the geometric point of view, one
should prove the vanishing of the ``Nehijinus tensors'' \cite{yano}\cite
{marcha}. From the topological point of view, one should overcome global
obstructions. The story is very similar to the existence of spinorial
manifolds. Actually, our almost hypercomplex structures, when represented as
matrices, close as Clifford algebra over certain Euclidean spaces.

\section{Complex Structure}

For complex variables, one can represent any complex number z as an element
of $\Bbb{R}^{2}$

\[
z=z_{0}e_{0}+z_{1}e_{1}\equiv Z=\left( 
\begin{array}{c}
z_{0} \\ 
z_{1}
\end{array}
\right) . 
\]
The action of $e_{0}=1$ and $e_{1}$ induce the following matrix
transformations on Z , 
\[
e_{0}z=ze_{0}=z\equiv \Bbb{E}_{0}Z=\mathbf{1}_{2}Z=Z, 
\]
where $\mathbf{1}_{n}$ will always mean the $n\times n$ identity matrix,
while 
\begin{eqnarray}
e_{1}z &=&ze_{1}=z_{0}e_{1}-z_{1} \\
&\equiv &\Bbb{E}_{1}Z \\
&=&\left( 
\begin{array}{cc}
0 & -1 \\ 
1 & 0
\end{array}
\right) \left( 
\begin{array}{c}
z_{0} \\ 
z_{1}
\end{array}
\right) =\left( 
\begin{array}{c}
-z_{1} \\ 
z_{0}
\end{array}
\right) .
\end{eqnarray}
Of course, we have 
\begin{equation}
\left( \Bbb{E}_{0}\right) ^{2}=\mathbf{1}_{2},\quad \quad \left( \Bbb{E}%
_{1}\right) ^{2}=-\mathbf{1}_{2}.
\end{equation}
Now, there is a problem, these two matrices $\Bbb{E}_{0}$ and $\Bbb{E}_{1}$
alone are not sufficient to form a basis for $R(2)$. The solution of our
dilemma is straightforward. If we also take into account that 
\[
z^{*}=z_{0}-z_{1}e_{1}\equiv Z^{*}=\left( 
\begin{array}{c}
z_{0} \\ 
-z_{1}
\end{array}
\right) 
\]
we find 
\begin{eqnarray}
Z^{*} &=&{\tilde{\Bbb{E}}}_{0}Z \\
&=&\left( 
\begin{array}{cc}
1 & 0 \\ 
0 & -1
\end{array}
\right) \left( 
\begin{array}{c}
z_{0} \\ 
z_{1}
\end{array}
\right) =\left( 
\begin{array}{c}
z_{0} \\ 
-z_{1}
\end{array}
\right) ,
\end{eqnarray}
and 
\begin{eqnarray}
e_{1}z^{*} &=&z^{*}e_{1}=z_{0}e_{1}+z_{1}=e_{1}^{*}z \\
&\equiv &\Bbb{E}_{1}Z^{*}=\tilde{\Bbb{E}}_{1}Z \\
&=&\left( 
\begin{array}{cc}
0 & -1 \\ 
1 & 0
\end{array}
\right) \left( 
\begin{array}{c}
z_{0} \\ 
-z_{1}
\end{array}
\right) =\left( 
\begin{array}{cc}
0 & 1 \\ 
1 & 0
\end{array}
\right) \left( 
\begin{array}{c}
z_{0} \\ 
z_{1}
\end{array}
\right) =\left( 
\begin{array}{c}
z_{1} \\ 
z_{0}
\end{array}
\right) .
\end{eqnarray}
Obviously with these four matrices $\{\Bbb{E}_{0},\tilde{\Bbb{E}}_{0},\Bbb{E}%
_{1},\tilde{\Bbb{E}}_{1}\}$, $\Bbb{C}(n)\rightarrow \Bbb{R}(2n)$ is proved.

\section{Sp(1) Structure}

The quaternionic algebra is given by $e_{i}.e_{j}=-\delta _{ij}+\epsilon
_{ijk}e_{k}\Longleftrightarrow \left[ e_{i},e_{j}\right] =2\epsilon
_{ijk}e_{k},$ where$\ \epsilon _{ijk}$ is the three dimensional Levi-Civita$%
\ $tensor $\left( \epsilon _{123}=1\right) $ and $i,j,k=1,2,3$. Being non
commutative, one must distinguish between right and left multiplication. In
1989, writing a quaternionic Dirac equation~\cite{rot}, Rotelli introduced a
``barred'' momentum operator with right action of $e_{1}$ 
\begin{equation}
-\partial _{\mu }\mid e_{1}\quad
\end{equation}
such that 
\begin{equation}
\lbrack ~(-\partial _{\mu }\mid e_{1})\psi \equiv -\partial _{\mu }\psi
e_{1}~]\quad .
\end{equation}
In recent papers~\cite{deleos}, partially barred quaternions 
\begin{equation}
q+p\mid e_{1}\quad \quad [~q,\;p\in \Bbb{H}~]\quad ,
\end{equation}
have been used to formulate a quaternionic quantum mechanics and
quaternionic field theory. From the viewpoint of group structure, these
barred operators are very similar to complexified quaternions~\cite{morita} 
\begin{equation}
q+\mathcal{I}p
\end{equation}
where the imaginary unit $\mathcal{I}$ commutes with the quaternionic
imaginary units ($e_{1},$ $e_{2},$ $e_{3}$), but in physical problems, like
eigenvalue calculations, tensor products, relativistic equations solutions,
they give different results. A complete generalization for quaternionic
multiplication is represented by the following barred operators 
\begin{equation}
\Bbb{H}|\Bbb{H}=q_{1}+q_{2}\mid e_{1}+q_{3}\mid e_{2}+q_{4}\mid e_{3}\quad
\quad [~q_{1,...,4}\in \Bbb{H}~]\quad ,  \label{gen1}
\end{equation}
and was developed a long time ago. As early as 1912, they had been used by
Conway and Silberstein \cite{conway}\cite{silberstein} to reformulate
special relativity and electromagnetism in a pure quaternionic language.
Look to Synge \cite{synge} for a review. The set of $\Bbb{H}|\Bbb{H}$ \
numbers with its 16 linearly independent elements, form a basis of $GL(4,%
\Bbb{R})$. They have been revived recently to write down a one-component
Dirac equation~\cite{deleos2}.

Let us now show how to represent these quaternionic operators as real $%
4\times 4$ matrices. Like in the complex case, we represent \emph{any
quaternionic number (as distinct from an ``operator'') } as a column vector 
\[
q=q_{0}+q_{1}e_{1}+q_{2}e_{2}+q_{3}e_{3}\equiv Q=\left( 
\begin{array}{c}
q_{0} \\ 
q_{1} \\ 
q_{2} \\ 
q_{3}
\end{array}
\right) \quad , 
\]
then 
\begin{equation}
e_{1}.q=q_{0}e_{1}-q_{1}+q_{2}e_{3}-q_{3}e_{2}\equiv \Bbb{E}_{1}Q
\end{equation}

and so on forth. The canonical left quaternionic structures \cite{yano}\cite
{kh1} over $\Bbb{R}^{4}$ are 
\[
\Bbb{E}_{1}~=~\left( 
\begin{array}{cccc}
0 & -1 & 0 & 0 \\ 
1 & 0 & 0 & 0 \\ 
0 & 0 & 0 & -1 \\ 
0 & 0 & 1 & 0
\end{array}
\right) ;\Bbb{E}_{2}~=~\left( 
\begin{array}{cccc}
0 & 0 & -1 & 0 \\ 
0 & 0 & 0 & 1 \\ 
1 & 0 & 0 & 0 \\ 
0 & -1 & 0 & 0
\end{array}
\right) ;\quad 
\]
\begin{equation}
\Bbb{E}_{3}~=~\left( 
\begin{array}{cccc}
0 & 0 & 0 & -1 \\ 
0 & 0 & -1 & 0 \\ 
0 & 1 & 0 & 0 \\ 
1 & 0 & 0 & 0
\end{array}
\right) ;  \label{q1}
\end{equation}
such that 
\begin{equation}
\Bbb{E}_{i}\Bbb{E}_{j}=(-\delta _{ij}+\epsilon _{ijk}\Bbb{E}_{k})\qquad
,\qquad (\Bbb{E}_{i})^{2}=-\mathbf{1}_{4},  \label{q2}
\end{equation}
Using Rotelli's notation, right action is given by 
\begin{eqnarray}
(1|e_{1})q &=&qe_{1}=q_{0}e_{1}-q_{1}-q_{2}e_{3}+q_{3}e_{2} \\
&\equiv &1|\Bbb{E}_{1}~Q
\end{eqnarray}
and so on for $1|e_{2},1|e_{3}$. Our canonical right quaternionic structures
are 
\[
{1|}\Bbb{E}_{1}~=~\left( 
\begin{array}{cccc}
0 & -1 & 0 & 0 \\ 
1 & 0 & 0 & 0 \\ 
0 & 0 & 0 & 1 \\ 
0 & 0 & -1 & 0
\end{array}
\right) ;{1|}\Bbb{E}_{2}~=~\left( 
\begin{array}{cccc}
0 & 0 & -1 & 0 \\ 
0 & 0 & 0 & -1 \\ 
1 & 0 & 0 & 0 \\ 
0 & 1 & 0 & 0
\end{array}
\right) ; 
\]
\begin{equation}
{1|}\Bbb{E}_{3}~=~\left( 
\begin{array}{cccc}
0 & 0 & 0 & -1 \\ 
0 & 0 & 1 & 0 \\ 
0 & -1 & 0 & 0 \\ 
1 & 0 & 0 & 0
\end{array}
\right) ,  \label{q3}
\end{equation}
and 
\begin{equation}
1|\Bbb{E}_{i}~1|\Bbb{E}_{j}=(-\delta _{ij}-\epsilon _{ijk}1|\Bbb{E}%
_{k})\qquad ,\qquad (1|\Bbb{E}_{i})^{2}=-\mathbf{1}_{4}\qquad .  \label{q4}
\end{equation}
We can write these left/right quaternionic structures compactly as 
\begin{eqnarray}
(\Bbb{E}_{i})_{\mu \nu } &=&(\delta _{0\mu }\delta _{i\nu }-\delta _{0\nu
}\delta _{i\mu }-\epsilon _{i\mu \nu })  \label{cqs1} \\
({1|}\Bbb{E}_{i})_{\mu \nu } &=&(\delta _{0\mu }\delta _{i\nu }-\delta
_{0\nu }\delta _{i\mu }+\epsilon _{i\mu \nu })
\end{eqnarray}
where $\mu ,\nu $ run from 0 to 3 or explicitly 
\begin{eqnarray}
\Bbb{E}_{i\mu \nu } &=&-\epsilon _{i\mu \nu }\mbox{\qquad if }\mu ,\nu
=1,2,3.  \label{cqs2} \\
\Bbb{E}_{i0\nu } &=&-\delta _{i\nu }\qquad ,\qquad \Bbb{E}_{i\mu 0}=\delta
_{i\mu }\qquad ,\qquad \Bbb{E}_{i00}=0.
\end{eqnarray}
and 
\begin{eqnarray}
1|\Bbb{E}_{i\mu \nu } &=&\epsilon _{i\mu \nu }\mbox{\qquad if }\mu ,\nu
=1,2,3.  \label{cqs3} \\
1|\Bbb{E}_{i0\nu } &=&-\delta _{i\nu }\qquad ,\qquad 1|\Bbb{E}_{i\mu
0}=\delta _{i\mu }\qquad ,\qquad 1|\Bbb{E}_{i00}=0.
\end{eqnarray}
These mathematical quaternionic structures are the 't~Hooft eta symbols \cite
{hooft1} well known in physics and we can check that 
\begin{equation}
1|\Bbb{E}_{i\mu \nu }=\frac{1}{2}\epsilon _{\mu \nu \alpha \beta }~1|\Bbb{E}%
_{i\alpha \beta }\qquad and\qquad \Bbb{E}_{i\mu \nu }=-\frac{1}{2}\epsilon
_{\mu \nu \alpha \beta }~\Bbb{E}_{i\alpha \beta }.
\end{equation}
Having $\Bbb{E}_{i},1|\Bbb{E}_{i}$ enables us to find any generic operator $%
\Bbb{E}_{i}|\Bbb{E}_{j}$ corresponding to $e_{i}|e_{j}$ 
\begin{equation}
\left( e_{i}|e_{j}\right) \;\;q=e_{i}\;\;\left( 1|e_{j}\right)
\;\;\;q=e_{i}qe_{j}\equiv \left( \Bbb{E}_{i}|\Bbb{E}_{j}\right) \;\;\;Q=\Bbb{%
E}_{i}\;\left( 1|\Bbb{E}_{j}\right) \;\;\;\;\ Q,
\end{equation}
then we have the 16 base elements of the operator in $\Bbb{H}|\Bbb{H}$ 
\begin{equation}
\left\{ 
\begin{array}{c}
1,e_{1},e_{2},e_{3}, \\ 
1|e_{1},e_{1}|e_{1},e_{2}|e_{1},e_{3}|e_{1}, \\ 
1|e_{2},e_{1}|e_{2},e_{2}|e_{2},e_{3}|e_{2}, \\ 
1|e_{3},e_{1}|e_{3},e_{2}|e_{3},e_{3}|e_{3}
\end{array}
\right\} .
\end{equation}
And their corresponding matrix representations 
\begin{equation}
\left\{ 
\begin{array}{c}
\mathbf{1}_{4},\Bbb{E}_{1},\Bbb{E}_{2},\Bbb{E}_{3}, \\ 
1|\Bbb{E}_{1},\Bbb{E}_{1}|\Bbb{E}_{1},\Bbb{E}_{2}|\Bbb{E}_{1},\Bbb{E}_{3}|%
\Bbb{E}_{1}, \\ 
1|\Bbb{E}_{2},\Bbb{E}_{1}|\Bbb{E}_{2},\Bbb{E}_{2}|\Bbb{E}_{2},\Bbb{E}_{3}|%
\Bbb{E}_{2}, \\ 
1|\Bbb{E}_{3},\Bbb{E}_{1}|\Bbb{E}_{3},\Bbb{E}_{2}|\Bbb{E}_{3},\Bbb{E}_{3}|%
\Bbb{E}_{3}
\end{array}
\right\} .  \label{ml2}
\end{equation}

We can thus deduce the following group structure for our quaternionic
operators

\begin{itemize}
\item  $su(2)_{Left}$ 
\begin{eqnarray}
e_{i}e_{j} &=&-\delta _{ij}+\epsilon _{ijk}e_{k},  \label{p1} \\
su(2)_{Left} &\sim &~\{e_{1},e_{2},e_{3}\}\sim \left\{ \Bbb{E}_{1},\Bbb{E}%
_{2},\Bbb{E}_{3}\right\} .
\end{eqnarray}

\item  $su(2)_{Right}$ 
\begin{eqnarray}
1|e_{j}1|e_{i} &=&1|(e_{i}e_{j})=-\delta _{ij}-\epsilon _{ijk}1|e_{k},
\label{p2} \\
su(2)_{Right} &\sim &~\{a_{1}=1|e_{2},a_{2}=1|e_{1},a_{3}=1|e_{3}\} 
\nonumber \\
&\sim &\left\{ a_{1}=1|\Bbb{E}_{2},a_{2}=1|\Bbb{E}_{1},a_{3}=1|\Bbb{E}%
_{3}\right\}
\end{eqnarray}
such that \footnote{%
We rename $a_{1}=1|e_{2}$ and $a_{2}=1|e_{1}$ so that equation (\ref{right})
comes with the standard sign.} 
\begin{equation}
a_{i}a_{j}=-\delta _{ij}+\epsilon _{ijk}a_{k}.  \label{right}
\end{equation}

\item  $so(4)\sim su(2)_{Left}\times su(2)_{Right}$, as 
\begin{equation}
e_{i}.1|e_{j}=1|e_{j}.e_{i}=e_{i}|e_{j},\quad i.e.\quad [e_{i},~1|e_{j}]=0,
\end{equation}
thus 
\begin{equation}
so(4)\sim ~\{e_{1},e_{2},e_{3},1|e_{1},1|e_{2},1|e_{3}\}.
\end{equation}

\item  $spin(2,3)$ - and its subgroups - can be realized by a Clifford
algebra construction, e.g. 
\begin{eqnarray}
~\gamma _{1} &=&e_{3},~~\gamma _{2}=e_{2},~~\gamma _{3}=e_{1}|e_{1},~~\gamma
_{4}=e_{1}|e_{2},~~\gamma _{5}=e_{1}|e_{3},  \nonumber \\
&&  \label{gamma5}
\end{eqnarray}
\begin{equation}
\{\gamma _{\alpha },\gamma _{\beta }\}=2diag(-,-,+,+,+).  \label{clif1}
\end{equation}
By explicit calculation, one finds (in the basis given above) 
\begin{eqnarray}
spin(2,3) &\sim &~the\;set\;of\;[\gamma _{\alpha },\gamma _{\beta }]\quad
\quad {\ \alpha ,\beta =1..5}\quad ,  \nonumber \\
&\sim
&~%
\{e_{1},1|e_{1},1|e_{2},1|e_{3},e_{2}|e_{1},e_{3}|e_{1},e_{2}|e_{2},e_{3}|e_{2},e_{2}|e_{3},e_{3}|e_{3}\}.
\nonumber \\
&&
\end{eqnarray}
The reason that eqn.(\ref{gamma5}) can lead to eqn.(\ref{clif1}) is that 
\[
e_{i}e_{j}~\left( 1|e_{k}\right) +e_{j}e_{i}~\left( 1|e_{k}\right) =0. 
\]
This construction was first introduced by Synge \cite{synge} to give a
quaternionic formulation of special relativity ($so(1,3)$).

\item  Also at the matrix level the full set $\Bbb{H}|\Bbb{H}$ closes as an
algebra, indeed using the above equations we find 
\begin{eqnarray}
1|e_{i}\;\;\;e_{j}|e_{k} &=&\epsilon _{kil}\;\;e_{j}|e_{l}, \\
e_{i}\;\;\;e_{j}|e_{k} &=&\epsilon _{ijl}\;\;e_{l}|e_{k}, \\
e_{i}|e_{j}\;\;\;e_{m}|e_{n} &=&\epsilon _{iml}\;\;\epsilon
_{njp}\;\;e_{l}|e_{p}.  \label{il5}
\end{eqnarray}
We have used Maple \cite{math} to prove that the 16 matrices of $\left\{ 
\Bbb{H}|\Bbb{H}\right\} $ are linearly independent so that they can form a
basis for any $\Bbb{R}(4)$ as we claimed in (\ref{cl1}). Omitting the
identity, the set of 15 elements $\left\{ \Bbb{H}|\Bbb{H}\right\} \backslash
1$ closes an $sl(4,\Bbb{R})$ algebra.
\end{itemize}

\newpage

\section{Octonionic Structure}

We now summarize our notation for the octonionic algebra. There is a number
of equivalent ways to represent the octonions multiplication table.
Fortunately, it is always possible to choose an orthonormal basis $%
(e_{0},\ldots ,e_{7})$ such that 
\begin{equation}
\varphi =\varphi _{0}+\varphi _{m}e_{m}\quad \quad \quad (~\varphi
_{0,...,7}~~\in \Bbb{R}~),
\end{equation}
where $e_{m}$ are elements obeying the noncommutative and nonassociative
algebra 
\begin{equation}
e_{m}e_{n}=-\delta _{mn}+f_{mnp}e_{p}(~{\ m,n,p=1..7}~),
\end{equation}
with $f_{mnp}$ totally antisymmetric and equal to unity for the seven
different three cycles 
\[
123,\;145,\;176,\;246,\;257,\;347,\;\;365 
\]
(each cycle represents a quaternionic subalgebra). We can define an
associator as follows, for any three octonionic numbers $a,b$ and $c$, 
\begin{equation}
\{a,\;b,\;c\}\equiv (ab)c-a(bc),
\end{equation}
where in each term on the right-hand we must, first of all, perform the
multiplication in brackets. Note that for real, complex and quaternionic
numbers the associator is trivially null. For octonionic imaginary units
however we have 
\begin{equation}
\{e_{m},\;e_{n},\;e_{p}\}\equiv
(e_{m}e_{n})e_{p}-e_{m}(e_{n}e_{p})=2C_{mnps}e_{s},  \label{assoc}
\end{equation}
with $C_{mnps}$ totally antisymmetric and equal to unity for the seven
combinations 
\[
1247,\;1265,\;2345,\;2376,\;3146,\;3157,\;\;4567. 
\]
Working with octonionic numbers the associator~(\ref{assoc}) is
non-vanishing for any three elements which are not in the same three cycles,
however, the ``alternative condition'' is always fulfilled 
\begin{equation}
\{a,\;b,\;c\}+\{c,\;b,\;a\}=0.  \label{rul}
\end{equation}

Due to the non-associativity, representing any form of octonions by matrices
seems impossible. Nevertheless, we overcome these problems by introducing
left/right-octonionic operators and fixing the direction of action. We
discuss in the next subsection their relation to $GL(8,\Bbb{R})$ where we
present our translation idea and give some explicit examples which allow us
to establish the isomorphism between our left/right octonionic operators and 
$GL(8,\Bbb{R})$.

Let us summarize the main points of the translation idea leaving the details
to the next subsection. Exactly as in the quaternionic case, it seems
natural to define and investigate the existence of barred operators 
\begin{equation}
\Bbb{O}_{0}\;\;+\;\;\Bbb{O}_{m}|e_{m}\quad \quad [~\Bbb{O}_{0,...,7}~~%
\mbox{octonions}~]\quad .
\end{equation}
We first observe that an octonionic barred operator, $a|b$, which acts on
octonionic functions, $\varphi $, 
\[
a|b\;\;\varphi \equiv a\varphi b\quad , 
\]
is not a well defined object. For $a\neq b$ the triple product $a\varphi b$
could be either $(a\varphi )b$ or $a(\varphi b)$. So, in order to avoid this
ambiguity (due to the nonassociativity of the octonionic numbers) we need to
introduce left/right-barred operators. We will define left-barred operators
by $a)b$, with $a$ and $b$ which represent octonionic numbers \cite{shaeffer}%
\cite{kh2}. They act on octonionic functions $\varphi $ as follows

\begin{equation}
a\mathbf{)}b\;\;\varphi =(a\varphi )b\quad .
\end{equation}
In a similar way we can introduce right-barred operators $a\mathbf{(}b$,
defined by 
\begin{equation}
a\mathbf{(}b\;\;\varphi =a(\varphi b)\quad .
\end{equation}
Obviously, there are barred-operators which are associative like 
\[
1\mathbf{)}a=1\mathbf{(}a\equiv 1|a\quad . 
\]
Furthermore, because of the alternativity condition (\ref{rul}) , 
\[
a\mathbf{)}a=a\mathbf{(}a\equiv a|a\quad . 
\]
At first glance it seems that we must consider the following 106
barred-operators:

\begin{center}
\begin{tabular}{lr}
$1,~e_{m},~1|e_{m}$ & {\ ~~~~~(15 elements)} , \\ 
$e_{m}|e_{m}$ & {\ (7)} , \\ 
$e_{m}\mathbf{)}e_{n}$~~~~{\ $(m\neq n)$} & {\ (42)} , \\ 
$e_{m}\mathbf{(}e_{n}$~~~~{\ $(m\neq n)$} & {\ (42)} , \\ 
{\ $(m,\;n=1,...,7)\quad .$} & 
\end{tabular}
\end{center}

Nevertheless, \emph{it is possible to prove that each right-barred operator
can be expressed by a suitable combination of left-barred operators.} For
example, from eq.~(\ref{rul}), by posing $a=e_{m}$ and $c=e_{n}$, we quickly
obtain 
\begin{equation}
e_{m}\mathbf{(}e_{n}+e_{n}\mathbf{(}e_{m}~~\equiv ~~e_{m}\mathbf{)}%
e_{n}+e_{n}\mathbf{)}e_{m}\quad .  \label{f1}
\end{equation}
So we can represent the most general octonionic operator by only left-barred
objects 
\begin{equation}
\Bbb{O}_{0}+\sum_{m=1}^{7}\Bbb{O}_{m}\mathbf{)}e_{m}\quad \quad [~\Bbb{O}%
_{0,...,7}~~\mbox{octonions}~]\quad ,  \label{go}
\end{equation}
reducing to 64 independent elements the previous 106. This number of 64
suggests a correspondence between our barred octonions~(\ref{go}) and $GL(8,%
\Bbb{R})$.

In subsection (2.3.2), we focus our attention on the group $GL(4,\Bbb{C}%
)\subset GL(8,\Bbb{R})$. In doing so, we will find that only particular
combinations of octonionic barred operators give us suitable candidates for
the $GL(4,\Bbb{C})$ \ translation.

\subsection{Octonionic Operators and $8\times 8$ Real Matrices}

In order to explain the idea of \emph{translation}, let us look explicitly
at the action of the operators $1\mid e_{1}$ and $e_{2}$, on a generic
octonionic function $\varphi $ 
\begin{equation}
\varphi =\varphi _{0}+e_{1}\varphi _{1}+e_{2}\varphi _{2}+e_{3}\varphi
_{3}+e_{4}\varphi _{4}+e_{5}\varphi _{5}+e_{6}\varphi _{6}+e_{7}\varphi
_{7}\quad [~\varphi _{0,\dots ,7}\in \Bbb{R}~].
\end{equation}
We have

\begin{eqnarray}
1|e_{1}\;\;\;\varphi &\equiv &\varphi e_{1}=e_{1}\varphi _{0}-\varphi
_{1}-e_{3}\varphi _{2}+e_{2}\varphi _{3}-e_{5}\varphi _{4}+e_{4}\varphi
_{5}+e_{7}\varphi _{6}-e_{6}\varphi _{7},  \nonumber \\
&&  \label{opa} \\
e_{2}\varphi &=&e_{2}\varphi _{0}-e_{3}\varphi _{1}-\varphi
_{2}+e_{1}\varphi _{3}+e_{6}\varphi _{4}+e_{7}\varphi _{5}-e_{4}\varphi
_{6}-e_{5}\varphi _{7}.  \nonumber \\
&&  \label{opa2}
\end{eqnarray}

If we represent our octonionic (``state'') function $\varphi $ by the
following real column vector 
\begin{equation}
\Phi ~\leftrightarrow ~\left( 
\begin{array}{c}
\varphi _{0} \\ 
\varphi _{1} \\ 
\varphi _{2} \\ 
\varphi _{3} \\ 
\varphi _{4} \\ 
\varphi _{5} \\ 
\varphi _{6} \\ 
\varphi _{7}
\end{array}
\right) \quad ,
\end{equation}
we can rewrite eqs.~(\ref{opa}--\ref{opa2}) in matrix form,

\begin{eqnarray}
1|\Bbb{E}_{1}\Phi &=&\left( 
\begin{array}{l}
-\varphi _{1} \\ 
\varphi _{0} \\ 
\varphi _{3} \\ 
-\varphi _{2} \\ 
\varphi _{5} \\ 
-\varphi _{4} \\ 
-\varphi _{7} \\ 
\varphi _{6}
\end{array}
\right) =\left( 
\begin{array}{llllllll}
0 & -1 & 0 & 0 & 0 & 0 & 0 & 0 \\ 
1 & 0 & 0 & 0 & 0 & 0 & 0 & 0 \\ 
0 & 0 & 0 & 1 & 0 & 0 & 0 & 0 \\ 
0 & 0 & -1 & 0 & 0 & 0 & 0 & 0 \\ 
0 & 0 & 0 & 0 & 0 & 1 & 0 & 0 \\ 
0 & 0 & 0 & 0 & -1 & 0 & 0 & 0 \\ 
0 & 0 & 0 & 0 & 0 & 0 & 0 & -1 \\ 
0 & 0 & 0 & 0 & 0 & 0 & 1 & 0
\end{array}
\right) \left( 
\begin{array}{l}
\varphi _{0} \\ 
\varphi _{1} \\ 
\varphi _{2} \\ 
\varphi _{3} \\ 
\varphi _{4} \\ 
\varphi _{5} \\ 
\varphi _{6} \\ 
\varphi _{7}
\end{array}
\right)  \nonumber \\
&& \\
\Bbb{E}_{2}\Phi &=&\left( 
\begin{array}{l}
-\varphi _{2} \\ 
\varphi _{3} \\ 
\varphi _{0} \\ 
-\varphi _{1} \\ 
-\varphi _{6} \\ 
-\varphi _{7} \\ 
\varphi _{4} \\ 
\varphi _{5}
\end{array}
\right) =\left( 
\begin{array}{llllllll}
0 & 0 & -1 & 0 & 0 & 0 & 0 & 0 \\ 
0 & 0 & 0 & 1 & 0 & 0 & 0 & 0 \\ 
1 & 0 & 0 & 0 & 0 & 0 & 0 & 0 \\ 
0 & -1 & 0 & 0 & 0 & 0 & 0 & 0 \\ 
0 & 0 & 0 & 0 & 0 & 0 & -1 & 0 \\ 
0 & 0 & 0 & 0 & 0 & 0 & 0 & -1 \\ 
0 & 0 & 0 & 0 & 1 & 0 & 0 & 0 \\ 
0 & 0 & 0 & 0 & 0 & 1 & 0 & 0
\end{array}
\right) \left( 
\begin{array}{l}
\varphi _{0} \\ 
\varphi _{1} \\ 
\varphi _{2} \\ 
\varphi _{3} \\ 
\varphi _{4} \\ 
\varphi _{5} \\ 
\varphi _{6} \\ 
\varphi _{7}
\end{array}
\right)  \nonumber \\
&&
\end{eqnarray}

In this way we can immediately obtain a real matrix representation for the
octonionic barred operators $1\mid e_{1}$ and $e_{2}$. Following this
procedure we can construct the complete set of translation rules for the
imaginary unit operators $e_{m}$ and the barred operators $1\mid e_{m}$
(appendix A).

At first glance it seems that our translation doesn't work. If we extract
the matrices corresponding to $e_{1}$, $e_{2}$ and $e_{3}$, namely, 
\begin{eqnarray*}
\Bbb{E}_{1} &=&\left( 
\begin{array}{llllllll}
0 & -1 & 0 & 0 & 0 & 0 & 0 & 0 \\ 
1 & 0 & 0 & 0 & 0 & 0 & 0 & 0 \\ 
0 & 0 & 0 & -1 & 0 & 0 & 0 & 0 \\ 
0 & 0 & 1 & 0 & 0 & 0 & 0 & 0 \\ 
0 & 0 & 0 & 0 & 0 & -1 & 0 & 0 \\ 
0 & 0 & 0 & 0 & 1 & 0 & 0 & 0 \\ 
0 & 0 & 0 & 0 & 0 & 0 & 0 & 1 \\ 
0 & 0 & 0 & 0 & 0 & 0 & -1 & 0
\end{array}
\right) , \\
\Bbb{E}_{2} &=&\left( 
\begin{array}{llllllll}
0 & 0 & -1 & 0 & 0 & 0 & 0 & 0 \\ 
0 & 0 & 0 & 1 & 0 & 0 & 0 & 0 \\ 
1 & 0 & 0 & 0 & 0 & 0 & 0 & 0 \\ 
0 & -1 & 0 & 0 & 0 & 0 & 0 & 0 \\ 
0 & 0 & 0 & 0 & 0 & 0 & -1 & 0 \\ 
0 & 0 & 0 & 0 & 0 & 0 & 0 & -1 \\ 
0 & 0 & 0 & 0 & 1 & 0 & 0 & 0 \\ 
0 & 0 & 0 & 0 & 0 & 1 & 0 & 0
\end{array}
\right) , \\
\Bbb{E}_{3} &=&\left( 
\begin{array}{llllllll}
0 & 0 & 0 & -1 & 0 & 0 & 0 & 0 \\ 
0 & 0 & -1 & 0 & 0 & 0 & 0 & 0 \\ 
0 & 1 & 0 & 0 & 0 & 0 & 0 & 0 \\ 
1 & 0 & 0 & 0 & 0 & 0 & 0 & 0 \\ 
0 & 0 & 0 & 0 & 0 & 0 & 0 & -1 \\ 
0 & 0 & 0 & 0 & 0 & 0 & 1 & 0 \\ 
0 & 0 & 0 & 0 & 0 & -1 & 0 & 0 \\ 
0 & 0 & 0 & 0 & 1 & 0 & 0 & 0
\end{array}
\right) ,
\end{eqnarray*}
we find 
\begin{equation}
\Bbb{E}_{1}\Bbb{E}_{2}\neq \Bbb{E}_{3}\quad .
\end{equation}
In obvious contrast with the octonionic relation 
\begin{equation}
e_{1}e_{2}=e_{3}\quad .
\end{equation}
This paradox is easily understood. In deducing our translation rules, we
understand octonions as operators, and so they must be applied to a certain
octonionic function, or state, $\varphi $, and \texttt{not} upon another
``operator''. So the octonionic relation

\begin{equation}
e_{3}\varphi ~~=(e_{1}e_{2})\varphi ~
\end{equation}
is indeed translated by 
\begin{equation}
\Bbb{E}_{3}\varphi \quad ,
\end{equation}
whereas, 
\begin{equation}
e_{1}(e_{2}\varphi )~~[~\neq e_{3}\varphi ~]
\end{equation}
becomes 
\begin{equation}
\Bbb{E}_{1}\Bbb{E}_{2}\varphi ~~[~\neq \Bbb{E}_{3}\varphi ~]\quad .
\end{equation}
For $e_{m}$ and $1|e_{n}$, we have simple multiplication rules. In fact,
utilizing the associator properties we find

\begin{eqnarray}
e_{m}\;\;\left[ e_{n}\varphi \right] &~=~&(e_{m}e_{n})\;\;\varphi
+(e_{m}\varphi )\;\;e_{n}-e_{m}\;\;(\varphi e_{n})\;,  \label{pp} \\
\left[ \varphi e_{m}\right] \;\;e_{n} &~=~&\varphi
\;\;(e_{m}e_{n})-(e_{m}\varphi )\;\;e_{n}+e_{m}\;\;(\varphi e_{n})\;.
\end{eqnarray}
Thus\footnote{%
We have used square brackets on the L.H.S. in the previous two equations in
order to avoid confusion between the L.H.S. and the last term on the R.H.S.
in the following two equations which might occur if we had employed \ $%
e_{m}\left( e_{n}\varphi \right) $ for the L.H.S. of \ (\ref{pp}) etc.}, 
\begin{eqnarray}
e_{m}~.~e_{n} &~\equiv ~&-\delta _{mn}+f_{mnp}e_{p}+e_{m}\mathbf{)}%
e_{n}-e_{m}\mathbf{(}e_{n}\;, \\
1|e_{n}\;\;.\;1|e_{m} &~\equiv ~&-\delta _{mn}+f_{mnp}e_{p}-e_{m}\mathbf{)}%
e_{n}+e_{m}\mathbf{(}e_{n}\;.
\end{eqnarray}

The previous relation can be immediately rewritten in matrix form as follows 
\cite{shaeffer}

\begin{eqnarray}
\Bbb{E}_{m}\Bbb{E}_{n} &~\equiv ~&-\delta _{mn}+f_{mnp}\Bbb{E}_{p}+[1|\Bbb{E}%
_{n},\;\Bbb{E}_{m}]\quad , \\
1|\Bbb{E}_{n}1|\Bbb{E}_{m} &~\equiv ~&-\delta _{mn}+f_{mnp}1|\Bbb{E}_{p}+[%
\Bbb{E}_{m},\;1|\Bbb{E}_{n}]\quad .
\end{eqnarray}

Introducing a new matrix multiplication, ``~$\circ $~'', related to the
standard matrix multiplication (row by column) by 
\begin{equation}
\Bbb{E}_{m}\circ \Bbb{E}_{n}\equiv \Bbb{E}_{m}\Bbb{E}_{n}-[1|\Bbb{E}_{n},\;%
\Bbb{E}_{m}]\quad ,  \label{new}
\end{equation}
we can quickly reformulate the nonassociative octonionic algebra (in terms
of matrices this time) by 
\begin{equation}
\Bbb{E}_{m}\circ \Bbb{E}_{n}=-\delta _{mn}+f_{mnp}\Bbb{E}_{p}\quad .
\end{equation}

Working with left/right barred operators we now show how the
nonassociativity is realized with our matrix translation. Such operators
enable us to reproduce the octonions nonassociativity by the matrix algebra.
Consider for example 
\begin{equation}
e_{3}\mathbf{)}e_{1}\;\;\;\varphi ~\equiv ~(e_{3}\varphi
)\;\;e_{1}~=~e_{2}\varphi _{0}-e_{3}\varphi _{1}+\varphi _{2}-e_{1}\varphi
_{3}-e_{6}\varphi _{4}-e_{7}\varphi _{5}+e_{4}\varphi _{6}+e_{5}\varphi
_{7}\quad .
\end{equation}
This equation will be translated into 
\begin{equation}
\left( 
\begin{array}{llllllll}
0 & 0 & 1 & 0 & 0 & 0 & 0 & 0 \\ 
0 & 0 & 0 & -1 & 0 & 0 & 0 & 0 \\ 
1 & 0 & 0 & 0 & 0 & 0 & 0 & 0 \\ 
0 & -1 & 0 & 0 & 0 & 0 & 0 & 0 \\ 
0 & 0 & 0 & 0 & 0 & 0 & 1 & 0 \\ 
0 & 0 & 0 & 0 & 0 & 0 & 0 & 1 \\ 
0 & 0 & 0 & 0 & -1 & 0 & 0 & 0 \\ 
0 & 0 & 0 & 0 & 0 & -1 & 0 & 0
\end{array}
\right) \left( 
\begin{array}{l}
\varphi _{0} \\ 
\varphi _{1} \\ 
\varphi _{2} \\ 
\varphi _{3} \\ 
\varphi _{4} \\ 
\varphi _{5} \\ 
\varphi _{6} \\ 
\varphi _{7}
\end{array}
\right) =\left( 
\begin{array}{l}
\varphi _{2} \\ 
-\varphi _{3} \\ 
\varphi _{0} \\ 
-\varphi _{1} \\ 
\varphi _{6} \\ 
\varphi _{7} \\ 
-\varphi _{4} \\ 
-\varphi _{5}
\end{array}
\right)
\end{equation}
Whereas, 
\begin{equation}
e_{3}\mathbf{(}e_{1}\;\;\;\varphi ~\equiv ~e_{3}(\varphi
e_{1})~=~e_{2}\varphi _{0}-e_{3}\varphi _{1}+\varphi _{2}-e_{1}\varphi
_{3}+e_{6}\varphi _{4}+e_{7}\varphi _{5}-e_{4}\varphi _{6}-e_{5}\varphi
_{7}\quad ,
\end{equation}
will become 
\begin{equation}
\left( 
\begin{array}{llllllll}
0 & 0 & 1 & 0 & 0 & 0 & 0 & 0 \\ 
0 & 0 & 0 & -1 & 0 & 0 & 0 & 0 \\ 
1 & 0 & 0 & 0 & 0 & 0 & 0 & 0 \\ 
0 & -1 & 0 & 0 & 0 & 0 & 0 & 0 \\ 
0 & 0 & 0 & 0 & 0 & 0 & -1 & 0 \\ 
0 & 0 & 0 & 0 & 0 & 0 & 0 & -1 \\ 
0 & 0 & 0 & 0 & 1 & 0 & 0 & 0 \\ 
0 & 0 & 0 & 0 & 0 & 1 & 0 & 0
\end{array}
\right) \left( 
\begin{array}{l}
\varphi _{0} \\ 
\varphi _{1} \\ 
\varphi _{2} \\ 
\varphi _{3} \\ 
\varphi _{4} \\ 
\varphi _{5} \\ 
\varphi _{6} \\ 
\varphi _{7}
\end{array}
\right) =\left( 
\begin{array}{l}
\varphi _{2} \\ 
-\varphi _{3} \\ 
\varphi _{0} \\ 
-\varphi _{1} \\ 
-\varphi _{6} \\ 
-\varphi _{7} \\ 
\varphi _{4} \\ 
\varphi _{5}
\end{array}
\right)
\end{equation}
The nonassociativity is then reproduced since left and right barred
operators, like 
\[
e_{3}\mathbf{)}e_{1}\quad \mbox{and}\quad e_{3}\mathbf{(}e_{1} 
\]
are represented by different matrices. The complete set of translation rules
for left/right-barred operators is given in appendix A.

The full matrix representation for left/right barred operators can be
quickly obtained by suitable multiplications of the matrices $\Bbb{E}_{m}$
and $1|\Bbb{E}_{n}$. By direct calculations we can extract the matrices
which correspond to the operators 
\[
e_{m}\mathbf{)}e_{n}\quad \quad and\quad \quad e_{m}\mathbf{(}e_{n}\quad
\quad , 
\]
which we call, respectively, 
\[
\Bbb{E}_{m}\mathbf{)}\Bbb{E}_{n}\quad \quad and\quad \quad \Bbb{E}_{m}%
\mathbf{(}\Bbb{E}_{n}\quad . 
\]
Since our left/right barred operators can be represented by an ordered
action of the operators $e_{m}$ and $1\mid e_{n}$, we can relate the
matrices $\Bbb{E}_{m}\mathbf{)}\Bbb{E}_{n}$ and $\Bbb{E}_{m}\mathbf{(}\Bbb{E}%
_{n}$ to the matrices $\Bbb{E}_{m}$ and $1|\Bbb{E}_{m}$:

\begin{eqnarray}
\Bbb{E}_{m}\mathbf{)}\Bbb{E}_{n} &~\equiv ~&1|\Bbb{E}_{n}\;\;\;\Bbb{E}%
_{m}\quad , \\
\Bbb{E}_{m}\mathbf{(}\Bbb{E}_{n} &~\equiv ~&\Bbb{E}_{m}\;\;\;1|\Bbb{E}%
_{n}\quad .
\end{eqnarray}

The previous discussions concerning the octonions nonassociativity and the
isomorphism between $GL(8,\Bbb{R})$ and barred octonions, can now be
elegantly summarized as follows.\newline
\texttt{1 - Matrix representation for octonions nonassociativity acting }%
\newline
\texttt{on certain octonionic number.} 
\[
\Bbb{E}_{m}\mathbf{)}\Bbb{E}_{n}~\neq ~\Bbb{E}_{m}\mathbf{(}\Bbb{E}_{n}\quad
\quad [~1|\Bbb{E}_{n}\Bbb{E}_{m}\neq \Bbb{E}_{m}1|\Bbb{E}_{n}~~%
\mbox{{ for
$m\neq n$}}]\quad . 
\]
\texttt{2 - Isomorphism between} \mbox{$GL(8, \Bbb{R})$} \texttt{and barred
octonions.}\newline
If we rewrite our 106 barred operators by real matrices:

\begin{center}
\begin{tabular}{lr}
$1,\;\Bbb{E}_{m},\;1|\Bbb{E}_{m}$ & {\ ~~~~~(15 matrices)} , \\ 
$\Bbb{E}_{m}|\Bbb{E}_{m}\equiv \Bbb{E}_{m}\;\;\;1|\Bbb{E}_{m}=1|\Bbb{E}%
_{m}\;\;\;\Bbb{E}_{m}$ & {\ (7)} , \\ 
$\Bbb{E}_{m}\mathbf{)}\Bbb{E}_{n}\equiv 1|\Bbb{E}_{n}\;\;\;\Bbb{E}_{m}$ ~~~~{%
\ $(m\neq n)$} & {\ (42)} , \\ 
$\Bbb{E}_{m}\mathbf{(}\Bbb{E}_{n}\equiv \Bbb{E}_{n}\;\;\;1|\Bbb{E}_{m}$ ~~~~{%
\ $(m\neq n)$} & {\ (42)} , \\ 
{\ $(m,\;n=1,...,7)\quad ;$} & 
\end{tabular}
\end{center}

we have two different basis for $GL(8, \Bbb{R})$:

\begin{center}
\begin{tabular}{cl}
{\ (1)} & ~~~~~$1\;,~\Bbb{E}_{m}\;,~1|\Bbb{E}_{m}\;,~1|\Bbb{E}_{n}\;\;\Bbb{E}%
_{m},\;\;1|\Bbb{E}_{m}\;\;\Bbb{E}_{m}\quad ,$ \\ 
&  \\ 
{\ (2)} & ~~~~~$1\;,~\Bbb{E}_{m}\;,~1|\Bbb{E}_{m}\;,~\Bbb{E}_{m}\;\;1|\Bbb{E}%
_{n},\quad 1|\Bbb{E}_{m}\;\;\Bbb{E}_{m}.$%
\end{tabular}
\end{center}

We now note some difficulties due to the nonassociativity of octonions. When
we translate from barred octonions to $8\times 8$ real matrices there is no
problem. For example, in the octonionic equation 
\begin{equation}
e_{4}\{[(e_{6}\varphi )e_{1}]e_{5}\}\quad ,  \label{ww}
\end{equation}
we quickly recognize the following left/right octonionic operators, 
\[
e_{6}\mathbf{)}e_{1}\;\;\;\;followed\;by\;\;\;e_{4}\mathbf{(}e_{5}\quad
\quad . 
\]
Hence we can translate eq.~(\ref{ww}) into 
\begin{equation}
\left[ \Bbb{E}_{4}\mathbf{(}\Bbb{E}_{5}\right] \quad \left[ \Bbb{E}_{6}%
\mathbf{)}\Bbb{E}_{1}\right] ~~\varphi \quad .
\end{equation}
But in going from $8\times 8$ real matrices to octonions we must be careful
of the ordering. For example, with A, B matrices 
\begin{equation}
AB~\varphi ~~~  \label{mm}
\end{equation}
can be understood for translation purposes as

\begin{equation}  \label{mm1}
(AB) \varphi \quad ,
\end{equation}
or 
\begin{equation}  \label{mm2}
A (B \varphi) \quad .
\end{equation}

In order to avoid confusion we translate eq.~(\ref{mm}) by eq.~(\ref{mm2}).
In general when brackets are absent we shall choose the convention that 
\begin{equation}
ABC\ldots Z\varphi \equiv A(B(C\ldots (Z\varphi )\ldots ))\quad .
\end{equation}

\subsection{Octonionic Operators and $4\times 4$ Complex Matrices}

Some complex groups play a critical role in physics. No one can deny the
importance of $U(1,\Bbb{C})$ and $SU(2,\Bbb{C})$. In relativistic quantum
mechanics, $GL(4,\Bbb{C})$ is implicit in writing the Dirac equation.
Starting from our $GL(8,\Bbb{R})$, we should be able to extract its subgroup 
$GL(4,\Bbb{C})$. Whence we should be able to translate the famous
Dirac-gamma matrices and write down a four dimensional one-component
octonionic Dirac equation~\cite{odir}.

Let us show how we can extract our 32 basis of $GL(4,\Bbb{C})$: Working with
the symplectic decomposition of octonionic ``states'' 
\begin{equation}
\psi =\left( 
\begin{array}{c}
{\psi _{1}} \\ 
{\psi _{2}} \\ 
{\psi _{3}} \\ 
{\psi _{4}}
\end{array}
\right) ~\leftrightarrow ~\psi _{1}+e_{2}\psi _{2}+e_{4}\psi _{3}+e_{6}\psi
_{4}\quad \quad [~\psi _{1,...,4}\in \Bbb{C}(1,\;e_{1})~]\quad .
\end{equation}
we analyze the action of left-barred operators on our octonionic wave
functions $\psi $. For example, we find 
\begin{eqnarray}
1|e_{1}\;\;\;\psi ~\equiv ~ &\psi e_{1}&~=~e_{1}\psi _{1}+e_{2}(e_{1}\psi
_{2})+e_{4}(e_{1}\psi _{3})+e_{6}(e_{1}\psi _{4})\quad , \\
&e_{2}\psi &~=~-\psi _{2}+e_{2}\psi _{1}-e_{4}\psi _{4}^{*}+e_{6}\psi
_{3}^{*}\quad , \\
e_{3}\mathbf{)}e_{1}\;\;\;\psi ~\equiv ~ &(e_{3}\psi )e_{1}&~=~\psi
_{2}+e_{2}\psi _{1}+e_{4}\psi _{4}^{*}-e_{6}\psi _{3}^{*}\quad .
\end{eqnarray}

Following the same methodology of the previous section, we can immediately
note a correspondence between the complex matrix $i\mathbf{1}_{4\times 4}$%
and the octonionic operator $1\mid e_{1}$%
\begin{equation}
\left( 
\begin{array}{cccc}
{i} & {0} & {0} & {0} \\ 
0 & i & 0 & 0 \\ 
0 & 0 & i & 0 \\ 
0 & 0 & 0 & i
\end{array}
\right) ~\leftrightarrow ~1|e_{1}\quad .
\end{equation}

This translation does not work for all barred operators. Let us show this,
explicitly. For example, we cannot find a $4\times 4$ complex matrix which,
acting upon 
\[
\left( 
\begin{array}{c}
{\psi _{1}} \\ 
{\psi _{2}} \\ 
{\psi _{3}} \\ 
{\psi _{4}}
\end{array}
\right) \quad , 
\]
gives the column vector 
\[
e_{2}\psi =\left( 
\begin{array}{c}
-{\psi _{2}} \\ 
{\psi _{1}} \\ 
-{\psi _{4}^{*}} \\ 
{\psi _{3}^{*}}
\end{array}
\right) \quad or\quad e_{3}\mathbf{)}e_{1}\psi =\left( 
\begin{array}{c}
{\psi _{2}} \\ 
{\psi _{1}} \\ 
{\psi _{4}^{*}} \\ 
-{\psi _{3}^{*}}
\end{array}
\right) 
\]
and so we have not the possibility to relate 
\[
e_{2}\quad \quad or\quad \quad e_{3}\mathbf{)}e_{1} 
\]
with a complex matrix. Nevertheless, a combined action of these two
operators gives us 
\[
e_{2}\psi +(e_{3}\psi )\;\;e_{1}=2e_{2}\psi _{1}\quad , 
\]
and allows us to represent the octonionic barred sum

\begin{equation}
e_{2}\;+\;e_{3}\mathbf{)}e_{1}\quad ,
\end{equation}
by the $4\times 4$ complex matrix 
\begin{equation}
\left( 
\begin{array}{cccc}
{0} & {0} & {0} & {0} \\ 
2 & 0 & 0 & 0 \\ 
0 & 0 & 0 & 0 \\ 
0 & 0 & 0 & 0
\end{array}
\right) \quad .  \label{ct}
\end{equation}

Following this procedure we can represent our generic $4\times 4$ complex
matrix by octonionic barred operators (but not necessarily the contrary).
The explicit correspondence rules are given in appendix B.

We conclude our discussion upon the relationship between barred operators
and $4\times 4$ complex matrices, by noting that the 32 basis elements of $%
GL(4,\Bbb{C})$ can be extracted in a different way from the 64 generators of 
$GL(8,\Bbb{R})$. It is well known that any complex matrix can be rewritten
as a real matrix by the following isomorphism ($\sigma $ are the standard
Pauli matrices), 
\[
1~\leftrightarrow ~\mathbf{1}_{2\times 2}\quad \quad \mbox{and}\quad \quad
i~\leftrightarrow ~-i\sigma _{2}\quad . 
\]
The situation at the lowest order is 
\begin{eqnarray}
GL(2,\Bbb{R}) &~~~~~\mbox{generators :}~~~~~&\mathbf{1}_{2\times 2}~,~\sigma
_{1}~,~-i\sigma _{2}~,~\sigma _{3}\quad ; \\
GL(1,\Bbb{C})~ &~~~~~\mbox{isomorphic :}~~~~~&\mathbf{1}_{2\times
2}~,~-i\sigma _{2}\quad .
\end{eqnarray}
In a similar way (choosing appropriate combinations of left-barred
octonionic operators, in which only $\pm \mathbf{1}_{2\times 2}$and $\pm
i\sigma _{2}$appear) we can extract from $GL(8,\Bbb{R})$ the 32 basis
elements of $GL(4,\Bbb{C})$. For further details see appendix B.

\section{Beyond Octonions and Clifford Algebras}

Going to higher dimensions, we define ``hexagonions'' ($\Bbb{X}$) by
introducing a new element $e_{8}$ such that 
\begin{equation}
\begin{array}{llll}
\Bbb{X} & = & \Bbb{O}_{1}+\Bbb{O}_{2}e_{8} &  \\ 
& = & x_{0}e_{0}+\ldots +x_{16}e_{16}. & x_{\mu }\in \Bbb{R}
\end{array}
\label{ffff}
\end{equation}
and 
\begin{equation}
e_{i}e_{j}=-\delta _{ij}+C_{ijk}e_{k}.
\end{equation}
Now, we have to find a suitable form of $C_{ijk}$. Recalling how the
structure constant is written for octonions 
\begin{eqnarray}
\Bbb{O} &=&\Bbb{Q}_{1}+\Bbb{Q}_{2}e_{4}  \nonumber \\
&=&x_{0}e_{0}+\ldots +x_{7}e_{7},  \label{a4}
\end{eqnarray}
where $\Bbb{Q}$ are quaternions, we have already chosen the convention $%
e_{1}e_{2}=e_{3}$ which is extendable to (\ref{a4}), we set $%
e_{1}e_{4}=e_{5} $, $e_{2}e_{4}=e_{6}$and $e_{3}e_{4}=e_{7}$, but we still
lack the relationships between the remaining possible triplets, $%
\{e_{1},e_{6},e_{7}\};$ $\{e_{2},e_{5},e_{7}\};$ $\{e_{3},e_{5},e_{6}\}$
which can be fixed by using 
\[
\begin{array}{c}
e_{1}e_{6}=e_{1}(e_{2}e_{4})=-(e_{1}e_{2})e_{4}=-e_{3}e_{4}=-e_{7}, \\ 
e_{2}e_{5}=e_{2}(e_{1}e_{4})=-(e_{2}e_{1})e_{4}=+e_{3}e_{4}=+e_{7}, \\ 
e_{3}e_{5}=e_{3}(e_{1}e_{4})=-(e_{3}e_{1})e_{4}=-e_{2}e_{4}=-e_{6}.
\end{array}
\]
These define all the structure constants for octonions. Returning to $\Bbb{X}
$, we have the seven octonionic conditions, and the decomposition (\ref{ffff}%
). We set $e_{1}e_{8}=e_{9},\;e_{2}e_{8}=e_{A},\;e_{3}e_{8}=e_{B},%
\;e_{4}e_{8}=e_{C},\;e_{5}e_{8}=e_{D},\;e_{6}e_{8}=e_{E},\;e_{7}e_{8}=e_{F}$
where $A=10,\;B=11,\;C=12,\;D=13,\;E=14$ and $F=15$. The other elements of
the multiplication table may be chosen in analogy with (\ref{a4}).
Explicitly, the 35 hexagonionic triplets are 
\[
\begin{array}{ccccccc}
(123), & (145), & (246), & (347), & (257), & (176), & (365), \\ 
(189), & (28A), & (38B), & (48C), & (58D), & (68E), & (78F), \\ 
(1BA), & (1DC), & (1EF), & (29B), & (2EC), & (2FD), & (3A9), \\ 
(49D), & (4AE), & (4BF), & (3FC), & (3DE), & (5C9), & (5AF), \\ 
(5EB), & (6FD), & (6CA), & (6BD), & (79E), & (7DA), & (7CB).
\end{array}
\]
This can be extended for any generic higher dimensional field $\Bbb{F}^{n}$.

It can be shown by using some combinatorics that the number of such triplets 
$N$ for a general $\Bbb{F}^{n}$field is ($n>1$) 
\begin{equation}
N={\frac{~~~\left( 2^{n}-1\right) !~~~}{~\left( 2^{n}-3\right) !~~~3!~}},
\end{equation}
giving 
\[
\begin{array}{cccc}
\Bbb{F}^{n} & n & ~~~~dim~~~~ & N \\ 
\Bbb{Q} & 2 & 4 & 1 \\ 
\Bbb{O} & 3 & 8 & 7 \\ 
\Bbb{X} & 4 & 16 & 35 \\ 
&  & and\ so\ on. & 
\end{array}
\]
One may notice that for any non-ring division algebra $\left( \Bbb{F},\
n>3\right) $,\ $N>dim(\Bbb{F}^{n})$ except when dim = $\infty ,$ i.e. a
functional Hilbert space with a Cliff(0,$\infty $) structure. Does this
inequality have any connection with the ring division structure of the ($%
S^{1},S^{3},S^{7}$) spheres~? Yes, that is what we are going to show now.

Following, the same translation idea projecting our algebra $\Bbb{X}$ over $%
\Bbb{R}^{16}$, any $\Bbb{E}_{i}$ is given by a relation similar to that for $%
\Bbb{Q}$%
\begin{equation}
(\Bbb{E}_{i})_{\alpha \beta }=\delta _{i\alpha }\delta _{\beta 0}-\delta
_{i\beta }\delta _{\alpha 0}+C_{i\alpha \beta }.
\end{equation}
But contrary to the quaternions and octonions, the Clifford algebra closes
only for a subset of these $E_{i}$'s, namely 
\begin{equation}
\{\Bbb{E}_{i},\Bbb{E}_{j}\}=-2\delta _{ij}\quad \mbox{for}\quad
i,j,k=1\ldots 8\;\;not\;1...15.  \label{nrd}
\end{equation}
Because we have lost the ring division structure. By careful investigation,
we find that another ninth $\Bbb{E}_{i}$ \ can be constructed, in agreement
with the Clifford algebra classification \cite{abs}. There is no standard$%
\footnote{%
Look to \cite{kh1} for a non standard representation.}$ 16 dimensional
representation for $Cliff\left( 15\right) $. Following this procedure, we
can give a simple way to write real Clifford algebras over any arbitrary
dimensions.

Sometimes, a specific multiplication table may be favored. For example in
soliton theory, the existence of a symplectic structure related to the
bihamiltonian formulation of integrable models is welcome. It is known from
the Darboux theorem, that locally a symplectic structure is given up to a
minus sign by 
\begin{equation}
\mathcal{J}_{dim\times dim}=\left( 
\begin{array}{cc}
0 & -\mathbf{1}_{\frac{dim}{2}} \\ 
\mathbf{1}_{\frac{dim}{2}} & 0
\end{array}
\right) ,
\end{equation}
this fixes the following structure constants 
\begin{eqnarray}
&&C_{\left( {\frac{dim}{2}}\right) 1\left( {\frac{dim}{2}}+1\right) }=-1, \\
&&C_{\left( {\frac{dim}{2}}\right) 2\left( {\frac{dim}{2}}+2\right) }=-1, \\
&&~~~~~~~\vdots \\
&&C_{\left( {\frac{dim}{2}}\right) \left( {\frac{dim}{2}}-1\right) \left(
dim-1\right) }=-1,
\end{eqnarray}
which is the decomposition that we have chosen in (\ref{a4}) for octonions 
\begin{equation}
C_{415}=C_{426}=C_{437}=-1.
\end{equation}
Generally our symplectic structure is 
\begin{equation}
\left( 1|\Bbb{E}_{\left( {\frac{dim}{2}}\right) }\right) _{\alpha \beta
}=\delta _{0\alpha }\delta _{\beta \left( {\frac{dim}{2}}\right) }-\delta
_{0\beta }\delta _{\alpha \left( {\frac{dim}{2}}\right) }-\epsilon _{\alpha
\beta \left( {\frac{dim}{2}}\right) }.
\end{equation}
Moreover some other choices may exhibit a relation with number theory and
Galois fields \cite{dix}. It is highly non-trivial how Clifford algebraic
language can be used to unify many distinct mathematical notions such as
Grassmanian \cite{kh3}, complex, quaternionic and symplectic structures.

The main result of this section, the non-existence of 16 dimensional
representation of $Cliff\left( 0,15\right) $ is in agreement with the
Atiyah--Bott--Shapiro classification of real Clifford algebras \cite{abs}.
In this context, the importance of ring division algebras can also be
deduced from the Bott periodicity \cite{martucci}. Another interesting
observation, if we interpret the complex, quaternions, octonions
eigenfunctions as real spinors, we find that for 
\begin{eqnarray}
complex\;\;\;\;\;\;\;\;\;Z^{t}E_{1}Z &=&0, \\
quaternions\;\;\;\;\;Q^{t}\Bbb{E}_{i}Q &=&0\;\;\;i=1..3, \\
octonions\;\;\;\;\;\;\;\Phi ^{t}\Bbb{E}_{i}\Phi &=&0\;\;\;i=1..7.
\end{eqnarray}
These states ($Z$, $Q$, $\Phi $) are called pure spinors as first coined by
Cartan. These pure spinors play an important role for minimal surfaces,
integrable models, twistor calculus, and string theory \cite{budinich}.

\chapter{The Soft Seven Sphere}

Ring division algebras play fundamental roles in mathematics from algebra to
geometry and topology with many different applications. The \ applicability
of real and complex numbers in physics is not in question. Quaternions which
may be represented as Pauli matrices are also important. the use of
octonions in physics is the problem. In the first section of this chapter,
we introduce what we mean by the word ``soft'' algebra, generally we follow
closely the presentation of Sohnius\cite{soh}. Sohnius used soft algebras
with structure functions that vary over space-time as well as \ over an
internal gauge manifold. In this thesis we use only soft algebras with
structure functions that vary over the internal gauge manifold (the fiber)
not the base space-time manifold. In the second section, we introduce the
seven sphere as a soft algebra \cite{estps}, we calculate its structure
functions explicitly and also discuss some relevant points such as the
validity of the Jacobi identity. Furthermore, we emphasis some important
features such as the pointwise reduction, closure, and some other
consistency checks. In the last section, we reformulate some standard Lie
group results putting them in a form suitable to subsequent applications.

\section{Sohnuis' Idea}

Trying to find a suitable framework for supersymmetric theories, Sohnius
introduced the notion of soft gauge algebras i.e. algebras where the
structure constants become structure functions of space-time and
gauge-dependent fields. He proceeded as follows: For any Lie group, we know
that a generic element can be written as $exp(i\varepsilon _{i}L_{i})$ where 
$\varepsilon $ are finite numbers of parameters and $L_{i}$ are our Lie
algebras elements. A field $A$, that lives in a certain representation of
our algebras, transforms infinitesimally as 
\begin{equation}
A\longrightarrow A+\delta _{\varepsilon _{1}}A\quad \mbox{with}\quad \delta
_{\varepsilon _{1}}A=-i[A,\varepsilon _{1}^{i}L_{i}]\qquad .
\end{equation}
Sohnuis considered the special case when the commutator of two successive
transformations leads to 
\begin{equation}
\lbrack \delta _{\varepsilon _{1}},\delta _{\varepsilon _{2}}]\Phi =-i[\Phi
,\varepsilon _{1}^{i}\varepsilon _{2}^{j}f_{ij}^{\ \ k}(\varphi
)L_{k}]=-i\left[ \Phi ,\varepsilon _{3}^{k}L_{k}\right] \qquad ,
\end{equation}
where $\varphi $ is a coordinate system for the internal gauge manifold and $%
f_{ij}^{\ \ k}(\varphi )$ are ``the structure functions'' of our soft
algebra defined by 
\begin{equation}
\lbrack \delta _{\varepsilon _{1}^{i}},\delta _{\varepsilon
_{2}^{j}}]=\delta _{f_{\ \ k}^{ij}(\varphi )\varepsilon _{3}^{k}}.
\end{equation}
In standard local gauge theory, we have $\varepsilon (x)$ and we need a
gauge field that transforms inhomogeneously: 
\begin{equation}
\delta _{\varepsilon }A_{\mu }^{i}=\partial _{\mu }\varepsilon (x)+A_{\mu
}^{j}\varepsilon (x)^{i}f_{ij}^{\ \ k}(\varphi )\qquad .
\end{equation}
In this thesis, we will always assume 
\begin{equation}
\partial _{\mu }f_{ijk}(\varphi )=0.
\end{equation}
By the $\varphi $ dependence of $f_{ijk}(\varphi )$, we mean a dependence on
the internal gauged space which is another manifold distinct from our
space-time $x$. To develop a representation, we use 
\begin{equation}
\delta _{\varepsilon }\Phi =\varepsilon (x)\delta (L_{i})\Phi ,
\end{equation}
and 
\begin{equation}
\lbrack \delta (L_{i}),\delta (L_{j})]=af_{ij}^{\ \ k}(\varphi )\delta
(L_{k})  \label{sla}
\end{equation}
where $a\in \Bbb{R}$. In this thesis we set $a=2$. Let's see how octonions
may be treated as a soft algebra exactly in the sense of eq. (\ref{sla}).

\section{Octonions and the Soft Seven Sphere}

We start by recalling the non-associative octonion algebra. A generic
octonion number is 
\begin{equation}
\varphi =\varphi _{0}e_{0}+\varphi _{i}e_{i}=\varphi _{\mu }e_{\mu
}.\;\;\;\;\;\left[ i=1..7,\mu =0..7,\varphi _{\mu }\in \Bbb{R}\right]
\end{equation}
and its associator 
\begin{equation}
\lbrack e_{i},e_{j},e_{k}]=(e_{i}e_{j})e_{k}-e_{i}(e_{j}e_{k})\qquad ,
\end{equation}
is non-zero for any three elements that are not in the same three cycles and
is completely antisymmetric The following formula or any of its
generalization is thus ambiguous 
\begin{equation}
e_{1}e_{5}e_{7}=\left\{ 
\begin{array}{c}
(e_{1}e_{5})e_{7}=-e_{3} \\ 
\mbox{or} \\ 
e_{1}(e_{5}e_{7})=e_{3}
\end{array}
\right.  \label{hj1}
\end{equation}
so the best way is to define the action of the imaginary units in a certain
direction. In the spirit of Englert, Sevrin, Troost, Van Proeyen and Spindel%
\cite{estps} (also look at \cite{shaeffer}) we define the left action of
octonionic operators $\delta _{i}$ by 
\begin{equation}
\delta _{i}\varphi =(e_{i}\varphi )
\end{equation}
implying that the following equation is well defined 
\begin{equation}
\delta _{i}\delta _{j}\varphi =\delta _{i}(\delta _{j}\varphi )=\delta
_{i}(e_{j}\varphi )=e_{i}(e_{j}\varphi ),
\end{equation}
then eq. (\ref{hj1}) reads unambiguously as 
\begin{equation}
\delta _{1}\delta _{5}e_{7}=(e_{1}(e_{5}e_{7}))=e_{3}.
\end{equation}
Since octonions are also non-commutative, we must also differentiate between
left and right action. Using the barred notation \cite{rot}, we introduce
right action as 
\begin{equation}
1|\delta _{i}~\varphi =(\varphi e_{i})\qquad ,
\end{equation}
for example 
\begin{equation}
(1|\delta _{i})(1|\delta _{j})\varphi =(1|\delta _{i})(1|\delta _{j}\varphi
)=1|\delta _{i}(\varphi e_{j})=((\varphi e_{j})e_{i})\qquad .
\end{equation}
As we shall see shortly, we can express the associator in terms of left and
right operators. The imaginary octonionic units generate the seven sphere $%
S^{7}$which has many properties similar to Lie algebras and/or Lie groups. $%
S^{7}$and Lie groups are the only non-flat compact parallelizable manifolds 
\cite{cart1}\cite{cart2}\cite{wolf}.

The important point for evaluating any Lie algebra is the commutator, so
let's examine 
\begin{eqnarray}
\lbrack \delta _{i},\delta _{j}]\varphi &=&e_{i}(e_{j}\varphi
)-e_{j}(e_{i}\varphi ) \\
&=&2f_{ijk}e_{k}\varphi -2[e_{i},e_{j},\varphi ] \\
&=&2f_{ijk}e_{k}\varphi +2[e_{i},\varphi ,e_{j}].
\end{eqnarray}
now the last term can be written as 
\begin{eqnarray}
\lbrack e_{i},\varphi ,e_{j}] &=&(e_{i}\varphi )e_{j}-e_{i}(\varphi e_{j}) \\
&=&-\delta _{i}(1|\delta _{j})\varphi +(1|\delta _{j})\delta _{i}\varphi \\
&=&-[\delta _{i},1|\delta _{j}]\varphi ,
\end{eqnarray}
thus our commutator can be rewritten as 
\begin{eqnarray}
\lbrack \delta _{i},\delta _{j}]\varphi &=&2f_{ijk}e_{k}\varphi
+2[e_{i},\varphi ,e_{j}]  \label{jk1} \\
&=&2f_{ijk}\delta _{k}\varphi -2[\delta _{i},1|\delta _{j}]\varphi .
\label{jk2}
\end{eqnarray}
Note that right operators are necessary because the last term the associator
can never be written in terms of left operators alone.

After simple calculations, one concludes that the octonionic imaginary units
are determined completely by (\ref{jk2}) and the following equations 
\begin{eqnarray}
\left[ 1|\delta _{i},1|\delta _{j}\right] \varphi &=&-2f_{ijk}1|\delta
_{k}\varphi -2[\delta _{i},1|\delta _{j}]\varphi  \label{ll2} \\
\{\delta _{i},\delta _{j}\}\varphi &=&-2\delta _{ij}\varphi  \label{ll3} \\
\{1|\delta _{i},1|\delta _{j}\}\varphi &=&-2\delta _{ij}\varphi \qquad ,
\label{ll4}
\end{eqnarray}
where the $\delta _{ij}$ in (\ref{ll3}) and (\ref{ll4}) are the standard
Kronecker delta tensor.

It has been proved in \cite{estps}\cite{ced1}, using three different ways,
that the $\delta _{i}$ algebra is associative. Thus a representation theory
in terms of matrices should be, in principle, possible. Indeed, in the last
chapter, we have derived an algebra completely isomorphic to (\ref{jk2},\ref
{ll2},\ref{ll3},\ref{ll4}) by exploiting the idea that octonions can be used
as a basis for any $8\times 8$ real matrix. we have two sets of matrices,
essentially, 
\begin{equation}
\begin{array}{ccccc}
\delta _{i} & \Longleftrightarrow & (\Bbb{E}_{i})_{\mu \nu } & = & \delta
_{0\mu }\delta _{i\nu }-\delta _{0\nu }\delta _{i\mu }-f_{i\mu \nu }, \\ 
1|\delta _{i} & \Longleftrightarrow & (1|\Bbb{E}_{i})_{\mu \nu } & = & 
\delta _{0\mu }\delta _{i\nu }-\delta _{0\nu }\delta _{i\mu }+f_{i\mu \nu }.
\end{array}
\label{ee1}
\end{equation}
The set of matrices $\Bbb{E}_{i}$ and $1|\Bbb{E}_{i}\,\ $have appeared in
different octonionic works e.g. \cite{c10}\cite{rom}\cite{dewittnicol}\cite
{luk}\cite{gunn1}\cite{gunket}. Furthermore, they correspond, as we have
already said, to the 't~Hooft eta symbols (\ref{cqs1}-\ref{cqs3}). We
suggest that their most appropriate names should be \emph{the canonical left
and right octonionic structure at the north/south pole of the seven sphere}.
By explicit calculation, one finds that 
\begin{eqnarray}
\lbrack \Bbb{E}_{i},\Bbb{E}_{j}]\varphi &=&2f_{ijk}\Bbb{E}_{k}\varphi -2[%
\Bbb{E}_{i},1|\Bbb{E}_{j}]~\varphi  \label{lll2} \\
\left[ 1|\Bbb{E}_{i},1|\Bbb{E}_{j}\right] \varphi &=&-2f_{ijk}1|\Bbb{E}%
_{k}~\varphi -2[\Bbb{E}_{i},1|\Bbb{E}_{j}]~\varphi \\
\{\Bbb{E}_{i},\Bbb{E}_{j}\}\varphi &=&-2\delta _{ij}\varphi  \label{lll3} \\
\{1|\Bbb{E}_{i},1|\Bbb{E}_{j}\}\varphi &=&-2\delta _{ij}\varphi  \label{lll4}
\end{eqnarray}
where $\varphi $ is represented by a column matrix 
\[
\varphi ^{t}=\left( 
\begin{array}{llllllll}
\varphi _{0} & \varphi _{1} & \varphi _{2} & \varphi _{3} & \varphi _{4} & 
\varphi _{5} & \varphi _{6} & \varphi _{7}
\end{array}
\right) \quad 
\]
The word ``isomorphic'' above is justified since \{$\delta _{i}$\} is
associative \cite{estps}\cite{ced1} and the same holds obviously for our \{$%
\Bbb{E}_{i}$ \} as they are written in terms of matrices. Our Jacobian
identities are 
\begin{equation}
\lbrack \delta _{i},[\delta _{j},\delta _{k}]]\varphi +[\delta _{j},[\delta
_{k},\delta _{i}]]\varphi +[\delta _{k},[\delta _{i},\delta _{j}]]\varphi
=0\qquad ,  \label{jac1}
\end{equation}
or 
\begin{equation}
\lbrack \Bbb{E}_{i},[\Bbb{E}_{j},\Bbb{E}_{k}]]\varphi +[\Bbb{E}_{j},[\Bbb{E}%
_{k},\Bbb{E}_{i}]]\varphi +[\Bbb{E}_{k},[\Bbb{E}_{i},\Bbb{E}_{j}]]\varphi
=0\qquad .  \label{jac2}
\end{equation}
We shall return to these identities again at the end of this section.

Fixing the direction of the application for any imaginary octonionic units
extracts a part of the algebra that respects associativity, but a certain
price has to be paid. The presence of the $\varphi $ is essential and either
of \{$\Bbb{E}_{i}$\} or \{$1|\Bbb{E}_{i}$\} is an open algebra, they don't
close upon the action of the commutator. Here comes the second step, the
soft Lie algebra idea. It is clear that the right hand side of (\ref{jk1} or 
\ref{lll2}) has a complicated $\varphi $ dependence. Knowing that the seven
sphere has a torsion that varies from one point to another \cite{rom}\cite
{cart1}\cite{cart2} and mimicking the Lie group case where the structure
constants are proportional to the fixed group torsion, it is natural to
propose that (\ref{jk1}) may be redefined \cite{estps} as 
\begin{equation}
\lbrack \delta _{i},\delta _{j}]\varphi =2f_{ijk}^{\left( +\right) }(\varphi
)\delta _{k}\varphi \qquad ,
\end{equation}
where $f_{ijk}^{(+)}(\varphi )$ are structure functions that vary over the
whole $S^{7}$manifold. It is clear that our $\delta _{i}$ play the same role
of the $\delta (L_{i})$ defined in the previous subsection eq.(\ref{sla}).
These structure functions $f_{ijk}^{(+)}(\varphi )$ were computed previously
using different properties of the associator and some other octonionic
identities in \cite{rom}\cite{estps}\cite{cart1}\cite{cart2}\cite{ced1}.
Here we use our matrix representation to give another alternative way to
calculate $f_{ijk}^{(+)}(\varphi )$%
\begin{equation}
\lbrack \Bbb{E}_{i},\Bbb{E}_{j}]{\ \varphi }=2f_{ijk}^{(+)}(\varphi )\Bbb{E}%
_{k}{\ \varphi }\qquad .  \label{tor1}
\end{equation}
Let's do it for the following example 
\begin{equation}
\lbrack \Bbb{E}_{1},\Bbb{E}_{2}]{\ \varphi }=2f_{12k}^{(+)}(\varphi )\Bbb{E}%
_{k}{\ \varphi }\qquad ,
\end{equation}
which is equivalent to the following eight equations , 
\begin{eqnarray}
\varphi _{3} &=&\varphi _{1}f_{121}^{(+)}(\varphi )+\varphi
_{2}f_{122}^{(+)}(\varphi )+\varphi _{3}f_{123}^{(+)}(\varphi )  \nonumber \\
&&+\varphi _{4}f_{124}^{(+)}(\varphi )+\varphi _{5}f_{125}^{(+)}(\varphi
)+\varphi _{6}f_{126}^{(+)}(\varphi )+\varphi _{7}f_{127}^{(+)}(\varphi ), 
\nonumber \\
\varphi _{2} &=&\varphi _{2}f_{123}^{(+)}(\varphi )-\varphi
_{0}f_{121}^{(+)}(\varphi )-\varphi _{3}f_{122}^{(+)}(\varphi )  \nonumber \\
&&-\varphi _{5}f_{124}^{(+)}(\varphi )+\varphi _{4}f_{125}^{(+)}(\varphi
)+\varphi _{7}f_{126}^{(+)}(\varphi )-\varphi _{6}f_{127}^{(+)}(\varphi ), 
\nonumber \\
\varphi _{1} &=&\varphi _{0}f_{122}^{(+)}(\varphi )-\varphi
_{3}f_{121}^{(+)}(\varphi )+\varphi _{1}f_{123}^{(+)}(\varphi )  \nonumber \\
&&+\varphi _{6}f_{124}^{(+)}(\varphi )+\varphi _{7}f_{125}^{(+)}(\varphi
)-\varphi _{4}f_{126}^{(+)}(\varphi )-\varphi _{5}f_{127}^{(+)}(\varphi ), 
\nonumber \\
\varphi _{0} &=&\varphi _{2}f_{121}^{(+)}(\varphi )-\varphi
_{1}f_{122}^{(+)}(\varphi )+\varphi _{0}f_{123}^{(+)}(\varphi )  \nonumber \\
&&+\varphi _{7}f_{124}^{(+)}(\varphi )-\varphi _{6}f_{125}^{(+)}(\varphi
)+\varphi _{5}f_{126}^{(+)}(\varphi )-\varphi _{4}f_{127}^{(+)}(\varphi ), 
\nonumber \\
\varphi _{7} &=&\varphi _{0}f_{124}^{(+)}(\varphi )-\varphi
_{5}f_{121}^{(+)}(\varphi )-\varphi _{6}f_{122}^{(+)}(\varphi )  \nonumber \\
&&-\varphi _{7}f_{123}^{(+)}(\varphi )+\varphi _{1}f_{125}^{(+)}(\varphi
)+\varphi _{2}f_{126}^{(+)}(\varphi )+\varphi _{3}f_{127}^{(+)}(\varphi ), 
\nonumber \\
\varphi _{6} &=&\varphi _{7}f_{122}^{(+)}(\varphi )-\varphi
_{4}f_{121}^{(+)}(\varphi )-\varphi _{6}f_{123}^{(+)}(\varphi )  \nonumber \\
&&+\varphi _{1}f_{124}^{(+)}(\varphi )-\varphi _{0}f_{125}^{(+)}(\varphi
)+\varphi _{3}f_{126}^{(+)}(\varphi )-\varphi _{2}f_{127}^{(+)}(\varphi ), 
\nonumber \\
\varphi _{5} &=&\varphi _{7}f_{121}^{(+)}(\varphi )+\varphi
_{4}f_{122}^{(+)}(\varphi )-\varphi _{5}f_{123}^{(+)}(\varphi )  \nonumber \\
&&-\varphi _{2}f_{124}^{(+)}(\varphi )+\varphi _{3}f_{125}^{(+)}(\varphi
)+\varphi _{0}f_{126}^{(+)}(\varphi )-\varphi _{1}f_{127}^{(+)}(\varphi ), 
\nonumber \\
\varphi _{4} &=&\varphi _{6}f_{121}^{(+)}(\varphi )-\varphi
_{5}f_{122}^{(+)}(\varphi )-\varphi _{4}f_{123}^{(+)}(\varphi )  \nonumber \\
&&+\varphi _{3}f_{124}^{(+)}(\varphi )+\varphi _{2}f_{125}^{(+)}(\varphi
)-\varphi _{1}f_{126}^{(+)}(\varphi )-\varphi _{0}f_{127}^{(+)}(\varphi ). 
\nonumber \\
&&  \label{eqns}
\end{eqnarray}
We now solve these equations for the seven unknown $f_{12i}^{(+)}\left(
\varphi \right) $. \ We find 
\begin{equation}
f_{121}^{(+)}\left( \varphi \right) =f_{122}^{(+)}\left( \varphi \right) =0,
\end{equation}
\[
f_{123}^{(+)}\left( \varphi \right) ={\frac{{\varphi _{0}}^{2}-{\varphi _{6}}%
^{2}-{\varphi _{5}}^{2}+{\varphi _{2}}^{2}-{\varphi _{4}}^{2}+{\varphi _{1}}%
^{2}+{\varphi _{3}}^{2}-{\varphi _{7}}^{2}}{r^{2}}} 
\]
and 
\begin{eqnarray}
f_{124}^{(+)}\left( \varphi \right) &=&+2{\frac{{\varphi _{0}}{\varphi _{7}}-%
{\varphi _{5}}{\varphi _{2}}+{\varphi _{6}}{\varphi _{1}}+{\varphi _{3}}{%
\varphi _{4}}}{r^{2}},} \\
f_{125}^{(+)}\left( \varphi \right) &=&-2{\frac{{\varphi _{0}}{\varphi _{6}}-%
{\varphi _{3}}{\varphi _{5}}-{\varphi _{1}}{\varphi _{7}}-{\varphi _{2}}{%
\varphi _{4}}}{r^{2}},} \\
f_{126}^{(+)}\left( \varphi \right) &=&+2{\frac{{\varphi _{0}}{\varphi _{5}}-%
{\varphi _{1}}{\varphi _{4}}+{\varphi _{7}}{\varphi _{2}}+{\varphi _{3}}{%
\varphi _{6}}}{r^{2}},} \\
f_{127}^{(+)}\left( \varphi \right) &=&-2{\frac{{\varphi _{0}}{\varphi _{4}}+%
{\varphi _{6}}{\varphi _{2}}+{\varphi _{1}}{\varphi _{5}}-{\varphi _{3}}{%
\varphi _{7}}}{r^{2}},}
\end{eqnarray}
where 
\begin{equation}
r^{2}=(\varphi _{0}^{2}{\ +}\varphi _{1}^{2}{\ +}\varphi _{2}^{2}{\ +}%
\varphi _{3}^{2}{\ +}\varphi _{4}^{2}{\ +}\varphi _{5}^{2}{\ +}\varphi
_{6}^{2}{\ +}\varphi _{7}^{2}).
\end{equation}
Along the same lines we can calculate all the structure functions, we give
all of them in Appendix C. What we have just calculated is commonly called
the (+) torsion\cite{rom}, we can find the (--) torsion by replacing the
left by right multiplication in (\ref{tor1}) 
\begin{equation}
\lbrack 1|\Bbb{E}_{i},1|\Bbb{E}_{j}]{\ \varphi }=2f_{ijk}^{(-)}(\varphi )1|%
\Bbb{E}_{k}{\ \varphi }.
\end{equation}
we find 
\begin{equation}
f_{121}^{(-)}\left( \varphi \right) =f_{122}^{(-)}\left( \varphi \right) =0,
\end{equation}
\[
f_{123}^{(-)}\left( \varphi \right) =-{\frac{{\varphi _{0}}^{2}-{\varphi _{6}%
}^{2}-{\varphi _{4}}^{2}+{\varphi _{2}}^{2}+{\varphi _{1}}^{2}-{\varphi _{7}}%
^{2}-{\varphi _{5}}^{2}+{\varphi _{3}}^{2}}{r^{2}},} 
\]
and 
\begin{eqnarray}
f_{124}^{(-)}\left( \varphi \right) &=&+2{\frac{{\varphi _{0}}{\varphi _{7}}+%
{\varphi _{5}}{\varphi _{2}}-{\varphi _{6}}{\varphi _{1}}-{\varphi _{3}}\ ,{%
\varphi _{4}}}{r^{2}},} \\
f_{125}^{(-)}\left( \varphi \right) &=&-2{\frac{{\varphi _{0}}{\varphi _{6}}+%
{\varphi _{3}}{\varphi _{5}}+{\varphi _{1}}{\varphi _{7}}+{\varphi _{2}}{%
\varphi _{4}}}{r^{2}},} \\
f_{126}^{(-)}\left( \varphi \right) &=&+2{\frac{{\varphi _{0}}{\varphi _{5}}+%
{\varphi _{1}}{\varphi _{4}}-{\varphi _{7}}{\varphi _{2}}-{\varphi _{3}}\ ,{%
\varphi _{6}}}{r^{2}},} \\
f_{127}^{(-)}\left( \varphi \right) &=&-2{\frac{{\varphi _{0}}{\varphi _{4}}-%
{\varphi _{6}}{\varphi _{2}}-{\varphi _{1}}{\varphi _{5}}+{\varphi _{3}}{%
\varphi _{7}}}{r^{2}},}
\end{eqnarray}
the remaining $f_{ijk}^{(-)}\left( \varphi \right) $ are listed in appendix
C.

Let's pause for a moment and note some of the evident features of these $%
f_{ijk}^{\left( \pm \right) }(\varphi )$,

\begin{itemize}
\item  One notices immediately that at $\varphi ^{t}$=(1,0,0,0,0,0,0,0) /
(-1,0,0,0,0,0,0,0), the north / south pole (NP/SP), we recover the
octonionic structure constants: $%
f_{ijk}^{(+)}(NP/SP)=-f_{ijk}^{(-)}(NP/SP)=f_{ijk}$ and any non-standard
cycles vanishes e.g. $f_{567}^{\left( \pm \right) }(NP/SP)=0$.

\item  Our construction started from a given multiplication table and as
there are different choices \cite{manog1}\cite{manog2}, we can have
different families.

\item  Restricting ourselves to $S^{3}$, we have the quaternionic structure
constants i.e. for $\varphi _{0}^{2}+\varphi _{1}^{2}+\varphi
_{2}^{2}+\varphi _{3}^{2}=1$, and $\varphi _{4}=\varphi _{5}=\varphi
_{6}=\varphi _{7}=0$ we have $f_{123}(\varphi )=\epsilon _{123}$ given in \
eq. (\ref{q2}) and all other $f$'s \ vanish.

\item  Our matrix representation $\left\{ \Bbb{E}_{0},\Bbb{E}_{i}\right\} $
is 8 dimensional and isomorphic to $\left\{ \delta _{0},\delta _{i}\right\} $%
. Of course constraining ourselves to a seven sphere of radius $r,$ \ ($%
\varphi ^{\mu }\varphi _{\mu }=r^{2})$, $\Bbb{E}_{i}$ or $1|\Bbb{E}_{i}$ $%
\left( i=1..7\right) $ are completely legitimate representation of the seven
imaginary octonionic units as our $\Bbb{E}_{i}$ or $1|\Bbb{E}_{i}$ $\left(
i=1,2,3\right) $ defined in the introduction are a representation of
imaginary quaternion units.

\item  Over S$^{7}$, $\partial _{\mbox{S}^{7}}f_{ijk}^{\left( \pm \right)
}(\varphi )\neq 0.$ This is a very important characteristic of the seven
sphere.
\end{itemize}

To manifest the $\varphi $ dependence, let's give some examples. To simplify
the notations, we use here $(i,j,k)^{\left( \pm \right) }$ for $%
f_{ijk}^{\left( \pm \right) }(\varphi )$

\begin{itemize}
\item  As we said before at ($\varphi _{0}=1,\varphi _{i}=0$), the
nonvanishing cocycles are 
\[
(1,2,3)^{(+)}=(1,4,5)^{(+)}=(1,7,6)^{(+)}=(2,4,6)^{(+)}=(2,5,7)^{(+)}=1 
\]
\begin{equation}
(3,4,7)^{(+)}=(3,6,5)^{(+)}=1,
\end{equation}
\[
(1,2,3)^{(-)}=(1,4,5)^{(-)}=(1,7,6)^{(-)}=(2,4,6)^{(-)}=(2,5,7)^{(-)}=-1 
\]
\begin{equation}
(3,4,7)^{(-)}=(3,6,5)^{(-)}=-1,
\end{equation}
and zero otherwise. Another non-trivial example is at $\left( \varphi _{\mu
}=\frac{\mu +1}{\sqrt{204}}\right) $, we find 
\begin{equation}
\begin{array}{c}
\begin{array}{lll}
(1,2,3)^{(+)}=-12/17, & (2,5,7)^{(+)}=4/51, & (1,5,6)^{(+)}=1/51, \\ 
(1,4,5)^{(+)}=-6/17, & (1,7,6)^{(+)}=8/51, & (3,6,5)^{(+)}=0, \\ 
(4,3,7)^{(+)}=-2/51, & (4,2,6)^{(+)}=3/17, & 
\end{array}
\\ 
\begin{array}{lll}
(1,2,4)^{(+)}=4/17, & (1,5,2)^{(+)}=-8/17, & (3,5,4)^{(+)}=-44/51, \\ 
(5,6,7)^{(+)}=-10/17, & (1,3,4)^{(+)}=-5/17, & (4,1,6)^{(+)}=-14/17, \\ 
(1,5,7)^{(+)}=-40/51, & (3,5,7)^{(+)}=-2/17, & (3,1,6)^{(+)}=-14/51, \\ 
(2,3,5)^{(+)}=-23/51, & (1,7,4)^{(+)}=-4/17, & (4,5,6)^{(+)}=-16/51, \\ 
(2,6,5)^{(+)}=-38/51, & (1,6,2)^{(+)}=-8/17, & (6,3,2)^{(+)}=-22/51, \\ 
(1,3,5)^{(+)}=-10/51, & (2,4,3)^{(+)}=-2/17, & (4,3,6)^{(+)}=-20/51, \\ 
(3,7,6)^{(+)}=-13/17, & (1,2,7)^{(+)}=-1/17, & (4,2,7)^{(+)}=-16/17, \\ 
(2,7,3)^{(+)}=-16/17, & (2,6,7)^{(+)}=-4/51, & (7,1,3)^{(+)}=-28/51, \\ 
(2,4,5)^{(+)}=2/17, & (4,7,5)^{(+)}=-7/51, & (4,6,7)^{(+)}=-10/51,
\end{array}
\end{array}
\label{xxx1}
\end{equation}
and 
\begin{equation}
\begin{array}{c}
\begin{array}{lll}
(1,2,3)^{(-)}=12/17, & (1,4,5)^{(-)}=6/17, & (1,7,6)^{(-)}=-8/51, \\ 
(2,4,6)^{(-)}=3/17, & (3,5,7)^{(-)}=8/51, & (3,4,7)^{(-)}=-2/51, \\ 
(4,6,7)^{(-)}=4/51, & (1,7,3)^{(-)}=-2/3, & (1,4,3)^{(-)}=-8/51,
\end{array}
\\ 
\begin{array}{lll}
(1,2,4)^{(-)}=-4/51, & (1,5,6)^{(-)}=-4/51, & (1,2,5)^{(-)}=-31/51, \\ 
(1,2,7)^{(-)}=-2/51, & (3,1,5)^{(-)}=-2/51, & (1,4,6)^{(-)}=-46/51, \\ 
(1,4,7)^{(-)}=-3/17, & (2,5,7)^{(-)}=-4/51, & (2,4,7)^{(-)}=-50/51, \\ 
(2,5,3)^{(-)}=-6/17, & (4,6,5)^{(-)}=-8/51, & (1,7,5)^{(-)}=-12/17, \\ 
(3,6,1)^{(-)}=-3/17, &  & (3,6,2)^{(-)}=-10/17, \\ 
(2,4,5)^{(-)}=-2/51, & (2,4,3)^{(-)}=0, & (3,4,5)^{(-)}=-47/51 \\ 
(3,4,6)^{(-)}=-6/17, & (3,6,5)^{(-)}=0, & (2,5,6)^{(-)}=-12/17, \\ 
(2,3,7)^{(-)}=-3/17, & (2,6,7)^{(-)}=0, & (3,6,7)^{(-)}=-12/17, \\ 
(1,2,6)^{(-)}=-6/17, & (4,5,7)^{(-)}=0, & (6,5,7)^{(-)}=-35/51.
\end{array}
\end{array}
\label{xxx2}
\end{equation}
We have some kind of dynamical Lie algebra of seven generators with
structure ``constants'' that change their values from one point to another.
Let us emphasis the difference between considering $\Bbb{E}_{i}$ as an open
algebra or as elements of a soft seven sphere, observe that 
\begin{equation}
\begin{array}{lll}
\left[ \Bbb{E}_{1},\Bbb{E}_{2}\right] & =2\Bbb{E}_{3}-2\left[ \Bbb{E}_{1},1|%
\Bbb{E}_{2}\right] , & 
\end{array}
\end{equation}
but 
\begin{equation}
\begin{array}{lll}
\left[ \Bbb{E}_{1},\Bbb{E}_{2}\right] \Phi & =2f_{123}^{\left( +\right)
}\left( \varphi \right) \Bbb{E}_{3}\Phi & +2f_{124}^{\left( +\right) }\left(
\varphi \right) \Bbb{E}_{4}\Phi \\ 
& +2f_{125}^{\left( +\right) }\left( \varphi \right) \Bbb{E}_{5}\Phi & 
+2f_{126}^{\left( +\right) }\left( \varphi \right) \Bbb{E}_{6}\Phi
+2f_{127}^{\left( +\right) }\left( \varphi \right) \Bbb{E}_{7}\Phi .
\end{array}
\label{soft}
\end{equation}
At the NP 
\begin{equation}
\Phi _{NP}^{t}=\left( 
\begin{array}{llllllll}
1 & 0 & 0 & 0 & 0 & 0 & 0 & 0
\end{array}
\right)
\end{equation}
we still have 
\begin{equation}
\left[ \Bbb{E}_{1},\Bbb{E}_{2}\right] =2\Bbb{E}_{3}-2\left[ \Bbb{E}_{1},1|%
\Bbb{E}_{2}\right]
\end{equation}
whereas 
\begin{eqnarray*}
\left[ \Bbb{E}_{1},\Bbb{E}_{2}\right] \Phi _{NP} &=&2f_{12k}^{\left(
+\right) }(\varphi _{NP})\Bbb{E}_{k}\Phi _{NP} \\
&=&2\Bbb{E}_{3}\Phi _{NP}
\end{eqnarray*}
\begin{eqnarray*}
&&\left( 
\begin{array}{llllllll}
0 & 0 & 0 & -2 & 0 & 0 & 0 & 0 \\ 
0 & 0 & -2 & 0 & 0 & 0 & 0 & 0 \\ 
0 & 2 & 0 & 0 & 0 & 0 & 0 & 0 \\ 
2 & 0 & 0 & 0 & 0 & 0 & 0 & 0 \\ 
0 & 0 & 0 & 0 & 0 & 0 & 0 & 2 \\ 
0 & 0 & 0 & 0 & 0 & 0 & -2 & 0 \\ 
0 & 0 & 0 & 0 & 0 & 2 & 0 & 0 \\ 
0 & 0 & 0 & 0 & -2 & 0 & 0 & 0
\end{array}
\right) \left( 
\begin{array}{l}
1 \\ 
0 \\ 
0 \\ 
0 \\ 
0 \\ 
0 \\ 
0 \\ 
0
\end{array}
\right) =\left( 
\begin{array}{l}
0 \\ 
0 \\ 
0 \\ 
2 \\ 
0 \\ 
0 \\ 
0 \\ 
0
\end{array}
\right) \\
&=&\left( 
\begin{array}{llllllll}
0 & 0 & 0 & -2 & 0 & 0 & 0 & 0 \\ 
0 & 0 & -2 & 0 & 0 & 0 & 0 & 0 \\ 
0 & 2 & 0 & 0 & 0 & 0 & 0 & 0 \\ 
2 & 0 & 0 & 0 & 0 & 0 & 0 & 0 \\ 
0 & 0 & 0 & 0 & 0 & 0 & 0 & -2 \\ 
0 & 0 & 0 & 0 & 0 & 0 & 2 & 0 \\ 
0 & 0 & 0 & 0 & 0 & -2 & 0 & 0 \\ 
0 & 0 & 0 & 0 & 2 & 0 & 0 & 0
\end{array}
\right) \left( 
\begin{array}{l}
1 \\ 
0 \\ 
0 \\ 
0 \\ 
0 \\ 
0 \\ 
0 \\ 
0
\end{array}
\right) .
\end{eqnarray*}
\end{itemize}

At $\Phi _{W}\equiv \left( \varphi _{\mu }=\frac{\mu +1}{\sqrt{204}}\right)
, $ we still have 
\begin{equation}
\left[ \Bbb{E}_{1},\Bbb{E}_{2}\right] =2\Bbb{E}_{3}-2\left[ \Bbb{E}_{1},1|%
\Bbb{E}_{2}\right]
\end{equation}
we find that 
\begin{eqnarray*}
\left[ \Bbb{E}_{1},\Bbb{E}_{2}\right] \Phi _{W} &=&2f_{12k}^{\left( +\right)
}(\varphi _{W})\Bbb{E}_{k}\Phi _{W} \\
&=&\left( -\frac{24}{17}\Bbb{E}_{3}+\frac{8}{17}\Bbb{E}_{4}+\frac{16}{17}%
\Bbb{E}_{5}+\frac{16}{17}\Bbb{E}_{6}-\frac{2}{17}\Bbb{E}_{7}\right) \Phi _{W}
\end{eqnarray*}
\begin{eqnarray*}
&&\left( 
\begin{array}{llllllll}
0 & 0 & 0 & -2 & 0 & 0 & 0 & 0 \\ 
0 & 0 & -2 & 0 & 0 & 0 & 0 & 0 \\ 
0 & 2 & 0 & 0 & 0 & 0 & 0 & 0 \\ 
2 & 0 & 0 & 0 & 0 & 0 & 0 & 0 \\ 
0 & 0 & 0 & 0 & 0 & 0 & 0 & 2 \\ 
0 & 0 & 0 & 0 & 0 & 0 & -2 & 0 \\ 
0 & 0 & 0 & 0 & 0 & 2 & 0 & 0 \\ 
0 & 0 & 0 & 0 & -2 & 0 & 0 & 0
\end{array}
\right) \left( 
\begin{array}{l}
\frac{1}{\sqrt{204}} \\ 
\frac{2}{\sqrt{204}} \\ 
\frac{3}{\sqrt{204}} \\ 
\frac{4}{\sqrt{204}} \\ 
\frac{5}{\sqrt{204}} \\ 
\frac{6}{\sqrt{204}} \\ 
\frac{7}{\sqrt{204}} \\ 
\frac{8}{\sqrt{204}}
\end{array}
\right) =\left( 
\begin{array}{l}
\frac{-4}{51}\sqrt{51} \\ 
\frac{-1}{17}\sqrt{51} \\ 
\frac{2}{51}\sqrt{51} \\ 
\frac{1}{51}\sqrt{51} \\ 
\frac{8}{51}\sqrt{51} \\ 
\frac{-7}{51}\sqrt{51} \\ 
\frac{2}{17}\sqrt{51} \\ 
\frac{-5}{51}\sqrt{51}
\end{array}
\right) \\
&=&\left( 
\begin{array}{llllllll}
0 & 0 & 0 & \frac{24}{17} & \frac{-8}{17} & \frac{-16}{17} & \frac{-16}{17}
& \frac{2}{17} \\ 
0 & 0 & \frac{24}{17} & 0 & \frac{-16}{17} & \frac{8}{17} & \frac{-2}{17} & 
\frac{-16}{17} \\ 
0 & \frac{-24}{17} & 0 & 0 & \frac{-16}{17} & \frac{2}{17} & \frac{8}{17} & 
\frac{16}{17} \\ 
\frac{-24}{17} & 0 & 0 & 0 & \frac{2}{17} & \frac{16}{17} & \frac{-16}{17} & 
\frac{8}{17} \\ 
\frac{8}{17} & \frac{16}{17} & \frac{16}{17} & \frac{-2}{17} & 0 & 0 & 0 & 
\frac{24}{17} \\ 
\frac{16}{17} & \frac{-8}{17} & \frac{-2}{17} & \frac{-16}{17} & 0 & 0 & 
\frac{-24}{17} & 0 \\ 
\frac{16}{17} & \frac{2}{17} & \frac{-8}{17} & \frac{16}{17} & 0 & \frac{24}{%
17} & 0 & 0 \\ 
\frac{-2}{17} & \frac{16}{17} & \frac{-16}{17} & \frac{-8}{17} & \frac{-24}{%
17} & 0 & 0 & 0
\end{array}
\right) \left( 
\begin{array}{l}
\frac{1}{\sqrt{204}} \\ 
\frac{2}{\sqrt{204}} \\ 
\frac{3}{\sqrt{204}} \\ 
\frac{4}{\sqrt{204}} \\ 
\frac{5}{\sqrt{204}} \\ 
\frac{6}{\sqrt{204}} \\ 
\frac{7}{\sqrt{204}} \\ 
\frac{8}{\sqrt{204}}
\end{array}
\right) .
\end{eqnarray*}

We are not making a projection but a reformulation of the algebra. This fact
should always be kept in mind. The same happens in a non-trivial way for the
Jacobi identity, i.e. 
\begin{equation}
\left( f_{ijm}(\varphi )f_{mkt}(\varphi )+f_{jkm}(\varphi )f_{mit}(\varphi
)+f_{kim}(\varphi )f_{mjt}(\varphi )\right) \Bbb{E}_{t}\varphi =0,
\end{equation}
but, in general, 
\begin{equation}
\left( f_{ijm}(\varphi )f_{mkt}(\varphi )+f_{jkm}(\varphi )f_{mit}(\varphi
)+f_{kim}(\varphi )f_{mjt}(\varphi )\right) \neq 0.
\end{equation}
Another important feature is 
\begin{equation}
\left[ \Bbb{E}_{i},1|\Bbb{E}_{j}\right] =-2f_{ijk}\left( \varphi \right) 
\Bbb{E}_{k}+\left[ \Bbb{E}_{i},\Bbb{E}_{j}\right] =-2f_{ijk}\left( \varphi
\right) 1|\Bbb{E}_{k}+\left[ 1|\Bbb{E}_{i},1|\Bbb{E}_{j}\right]
\end{equation}
which is equal to zero iff $i=j$ but for the soft seven sphere, $\left[ \Bbb{%
E}_{i},1|\Bbb{E}_{j}\right] \varphi =0$ not only for $i=j$ but also at the
NP/SP for any i,j.

Lastly over any group manifold the left torsion equals minus the right
torsion, but for $S^{7}$ this is not in general true.

\section{More General Solutions}

In the previous section , we have used brute force to calculate $%
f_{ijk}^{\left( +\right) }\left( \varphi \right) $. There is another way,
smarter and easier. We have the following situation 
\begin{equation}
\Bbb{E}_{i}\Bbb{E}_{j}\;\varphi =\left( -\delta _{ij}+f_{ijk}^{\left(
+\right) }\left( \varphi \right) \Bbb{E}_{k}\right) \varphi
\end{equation}
but our $\Bbb{E}_{i}$ defines what is called a pure spinor as we mentioned
at the end of chapter II, 
\begin{equation}
\varphi ^{t}\Bbb{E}_{i}\varphi =0
\end{equation}
thus 
\begin{equation}
\varphi ^{t}\left( \Bbb{E}_{i}\Bbb{E}_{j}\right) \varphi =\varphi ^{t}\left(
-\delta _{ij}\right) \varphi ,
\end{equation}
using 
\begin{equation}
\left( \Bbb{E}_{k}\right) ^{-1}=-\Bbb{E}_{k}
\end{equation}
we find 
\begin{equation}
\varphi ^{t}\left( -\Bbb{E}_{k}\Bbb{E}_{i}\Bbb{E}_{j}\right) \varphi
=\varphi ^{t}\left( f_{ijk}^{\left( +\right) }\left( \varphi \right) \right)
\varphi
\end{equation}
but 
\begin{equation}
\varphi ^{t}\varphi =r^{2}
\end{equation}
which gives us 
\begin{equation}
f_{ijk}^{\left( +\right) }\left( \varphi \right) =\frac{\varphi ^{t}\left( -%
\Bbb{E}_{k}\Bbb{E}_{i}\Bbb{E}_{j}\right) \varphi }{r^{2}}.
\end{equation}
and 
\begin{equation}
f_{ijk}^{\left( -\right) }\left( \varphi \right) =\frac{\varphi ^{t}\left(
-1|\Bbb{E}_{k}\;\;1|\Bbb{E}_{i}\;\;1|\Bbb{E}_{j}\right) \varphi }{r^{2}}.
\end{equation}
There is another interesting property to note 
\begin{equation}
\varphi ^{t}\left[ \Bbb{E}_{i},1|\Bbb{E}_{j}\right] \varphi =0
\end{equation}
which may be the generalization of the standard Lie algebra relation, left
and right action commute everywhere over the group manifold.

The left and right torsions that we have constructed are not the only
parallelizable torsions of S$^{7}$. Our $\Bbb{E}_{i}$ and $1|\Bbb{E}_{i}$
are given in terms of the octonionic structure constants (\ref{ee1}) i.e.
the torsion at NP/SP. Considering two new points, we may define new sets of $%
\Bbb{E}_{i}$ and $1|\Bbb{E}_{i}$. As S$^{7}$ contains an infinity of points,
practically, we have an infinity of parallelizable torsion. If our method is
self contained and sufficient, we should be able to construct these infinity
of pointwise structures. Indeed, $\Bbb{E}_{i}\left( \varphi \right) $ and $1|%
\Bbb{E}_{i}\left( \varphi \right) $ are in general 
\begin{equation}
\begin{array}{ccccc}
\delta _{i} & \Longleftrightarrow & (\Bbb{E}_{i}(\varphi ))_{\mu \nu } & = & 
\delta _{0\mu }\delta _{i\nu }-\delta _{0\nu }\delta _{i\mu }-f_{i\mu \nu
}^{\left( +\right) }(\varphi ), \\ 
1|\delta _{i} & \Longleftrightarrow & (1|\Bbb{E}_{i}(\varphi ))_{\mu \nu } & 
= & \delta _{0\mu }\delta _{i\nu }-\delta _{0\nu }\delta _{i\mu }+f_{i\mu
\nu }^{\left( -\right) }(\varphi ),
\end{array}
\label{octon}
\end{equation}
in complete analogy with (\ref{jk2},\ref{ll2},\ref{ll3},\ref{ll4}). Of
course the soft Algebra idea should hold here as well as for the special $%
\left( \Bbb{E}_{i},1|\Bbb{E}_{i}\right) $ constructed in terms of the north
pole torsion. Repeating the calculation in terms of $\left( \Bbb{E}%
_{i}(\varphi ),1|\Bbb{E}_{i}(\varphi )\right) $. Let us introduce a new
vector field $\lambda $, 
\begin{equation}
\lambda ^{t}=\left( 
\begin{array}{llllllll}
\lambda _{0} & \lambda _{1} & \lambda _{2} & \lambda _{3} & \lambda _{4} & 
\lambda _{5} & \lambda _{6} & \lambda _{7}
\end{array}
\right) .
\end{equation}
We define two new generalized structure functions 
\begin{eqnarray}
\left[ \Bbb{E}_{i}(\varphi ),\Bbb{E}_{j}(\varphi )\right] \lambda
&=&2f_{ijk}^{(++)}(\varphi ,\lambda )\Bbb{E}_{k}(\varphi )\lambda \\
\left[ 1|\Bbb{E}_{i}(\varphi ),1|\Bbb{E}_{j}(\varphi )\right] \lambda
&=&2f_{ijk}^{(-\;-)}(\varphi ,\lambda )1|\Bbb{E}_{k}(\varphi )\lambda
\end{eqnarray}
where $f_{ijk}^{\left( \pm \;\pm \right) }\left( \varphi ,\lambda \right) $
have a very complicated structure, 
\begin{eqnarray}
f_{ijk}^{\left( ++\right) }\left( \varphi ,\lambda \right) &=&\frac{\lambda
^{t}\left( -\Bbb{E}_{k}\left( \varphi \right) \Bbb{E}_{i}\left( \varphi
\right) \Bbb{E}_{j}\left( \varphi \right) \right) \lambda }{r^{2}}, \\
f_{ijk}^{\left( -\;-\right) }\left( \varphi ,\lambda \right) &=&\frac{%
\lambda ^{t}\left( -1|\Bbb{E}_{k}\left( \varphi \right) \;\;1|\Bbb{E}%
_{i}\left( \varphi \right) \;\;1|\Bbb{E}_{j}\left( \varphi \right) \right)
\lambda }{r^{2}}
\end{eqnarray}
as examples, we list four of them in Appendix D. We will use them later when
we study some applications.

\section{Some Group Theory}

An arbitrary octonion can be associated to $\Bbb{R}^{8}=\Bbb{R}\oplus \Bbb{R}%
^{7}$ \cite{luk} where $\Bbb{R}$ denotes the subspace spanned by the
identity $e_{0}=1$. Octonions with unit length define the octonionic unit
sphere $S^{7}$. The isometries of octonions is described by $O(8)$ which may
be decomposed as 
\begin{equation}
O(8):\quad \quad H\oplus K\oplus E
\end{equation}
where $H$ is the 14 parameters $G_{2}$ algebra of the automorphism group of $%
octonions,$ K is the torsionful seven sphere $SO(7)/G_{2}$ and our E is the
round seven sphere $SO(8)/SO(7)$. In fact the different three non-equivalent
representation of O(8) - the vectorial so(8) and the two different spinorial 
$spin^{L}(8)$ and $spin^{R}(8)$, which are related by triality, can be
realized by suitable left and right octonionic multiplication. The reduction
of O(8) to O(7) induces $so(8)\longrightarrow so(7)\oplus 1$, $%
spin^{R}(8)\longrightarrow spin(7)$ and $spin^{L}(8)\longrightarrow spin(7)$.

We would like to show how to generate these different Lie algebras entirely
from our canonical left/right octonionic structures. We start from the $%
8\times 8$ gamma matrices $\gamma _{\mu \nu }^{i}$ in seven dimensions,
using $\delta _{ij}$ as our flat Euclidean metric, 
\begin{equation}
\{\gamma ^{i},\gamma ^{j}\}=2\delta ^{ij}\mathbf{1}_{8}\qquad ,
\end{equation}
where $i,j,\ldots =1,2,\ldots 7$ and $\mu ,\nu ,\ldots =0,1,2,\ldots 7$. We
can use either of the following choices 
\begin{equation}
\gamma _{+}^{j}=i\Bbb{E}_{j}\quad or\quad \gamma _{-}^{j}=i1|\Bbb{E}%
_{j}\qquad ,  \label{cliffo}
\end{equation}
of course the $i$ in the right hand sides is the imaginary complex unit.
This relates our antisymmetric, Hermitian and \ hence purely imaginary gamma
matrices to the canonical octonionic left/right structures. The
antisymmetric product of two gamma matrices will be denoted by 
\begin{equation}
\gamma ^{ij}=\gamma ^{[i}\gamma ^{j]}\qquad ,
\end{equation}
and we have \footnote{%
It is interesting to note that this equation may be used as an alternative
definition for the octonionic multiplication table.} 
\begin{equation}
\gamma ^{i}\gamma ^{j}\gamma ^{k}=\frac{1}{4!}\epsilon ^{ijklmnp}\gamma
^{l}\gamma ^{m}\gamma ^{n}\gamma ^{p}\qquad .
\end{equation}

The matrices $\gamma ^{ij}$ span the 21 generators $J^{ij}$of spin(7) in its
eight-dimensional spinor representation. The spinorial representation of
spin(7) can be enlarged to the left/right handed spinor representation of
spin(8) by different ways. The easiest one is to include either of $\pm \Bbb{%
E}_{i}$ or $\pm 1|\Bbb{E}_{i}$ \cite{c10}\cite{gunn1}\cite{gunket} defining $%
J^{i}=J^{i0}$, so(8) can be written as 
\begin{eqnarray}
\lbrack J^{i},J^{i}] &=&2J^{ij} \\
\lbrack J^{i},J^{mn}] &=&2\delta ^{im}J^{n}-2\delta ^{in}J^{m} \\
\lbrack J^{ij},J^{kl}] &=&2\delta ^{jk}J^{il}+2\delta ^{il}J^{jk}-2\delta
^{ik}J^{jl}-2\delta ^{jl}J^{ik}.
\end{eqnarray}

The automorphism group of octonions is $G_{2}\subset SO(7)\subset SO(8)$. A
suitable basis for $G_{2}$ is \cite{c10}\cite{luk}\cite{gunn1}\cite{gunket} 
\begin{equation}
H_{ij}=f_{ijk}\left( \Bbb{E}_{k}-1|\Bbb{E}_{k}\right) -\frac{3}{2}\left[ 
\Bbb{E}_{i},1|\Bbb{E}_{j}\right] \qquad ,
\end{equation}
which implies the linear relations 
\begin{equation}
f_{ijk}H_{jk}=0\qquad ,
\end{equation}
These constraints enforce $H_{ij}$ to generate the 14 dimensional vector
space of $G_{2}$. There are different ways to represent the remaining seven
generators, denoted here by K, 
\begin{eqnarray}
\frac{so(7)}{G_{2}} &:&K_{v}^{\pm i}=\pm \frac{1}{2}\left( \Bbb{E}_{i}-1|%
\Bbb{E}_{i}\right) \qquad ,  \label{lo1} \\
\frac{spin(7)}{G_{2}} &:&K_{s}^{\pm i}=\pm \left( \frac{1}{2}\Bbb{E}_{i}+1|%
\Bbb{E}_{i}\right) \qquad ,  \label{lo2} \\
\frac{\overline{spin(7)}}{G_{2}} &:&{\overline{K}}_{s}^{\ \pm i}=\mp \left( 
\Bbb{E}_{i}+\frac{1}{2}1|\Bbb{E}_{i}\right) \qquad ,  \label{lo3}
\end{eqnarray}
Defining the conjugate representation\footnote{%
n.b. this definition is not matrix conjugation.} \ by 
\begin{equation}
{\overline{\Bbb{E}}}=-1|\Bbb{E}\qquad \mbox{and\qquad }{\overline{1|\Bbb{E}}}%
=-\Bbb{E}\qquad ,
\end{equation}
(\ref{lo1}) is self-conjugate while (\ref{lo2}) is octonionic-conjugate to (%
\ref{lo3}). The vector representation so(7) generated by $H_{ij}\oplus
K_{v}^{\pm i}$ is seven dimensional because $K_{v}^{\pm i}e_{0}=0$ whereas
the spin(7) representation generated by $H_{ij}\oplus K_{s}^{\pm i}$ is
eight dimensional.

To make apparent the role of the automorphism group $G_{2}$, the different
commutators of $\Bbb{E}$ and $1|\Bbb{E}$ may be written as 
\begin{eqnarray}
\left[ \Bbb{E}_{i},\Bbb{E}_{j}\right] &=&\frac{1}{3}\left( 4H_{ij}+2f_{ijk}%
\Bbb{E}_{k}+4f_{ijk}1|\Bbb{E}_{k}\right) \qquad , \\
\left[ 1|\Bbb{E}_{i},1|\Bbb{E}_{j}\right] &=&\frac{1}{3}\left(
4H_{ij}-4f_{ijk}\Bbb{E}_{k}-2f_{ijk}1|\Bbb{E}_{k}\right) \qquad , \\
\left[ \Bbb{E}_{i},1|\Bbb{E}_{j}\right] &=&\frac{1}{3}\left(
-2H_{ij}+2f_{ijk}\Bbb{E}_{k}-2f_{ijk}1|\Bbb{E}_{k}\right) \qquad .
\end{eqnarray}
or $G_{2}$ given by 
\begin{equation}
H_{ij}=\frac{1}{2}\left( [\Bbb{E}_{i},\Bbb{E}_{j}]+[1|\Bbb{E}_{i},1|\Bbb{E}%
_{j}]+[\Bbb{E}_{i},1|\Bbb{E}_{j}]\right) \qquad .
\end{equation}
Thus as we promised, the $\Bbb{E}$ and $1|\Bbb{E}$ are the necessary and the
sufficient building blocks for expressing the different Lie algebras and
coset representations related to the seven sphere. Note that all the
constructions given in this section start from the Clifford algebra relation
(\ref{cliffo}), and the formulation holds equally for $\Bbb{E}\left( \varphi
\right) $ and $1|\Bbb{E}\left( \varphi \right) $.

\chapter{Soft Seven Sphere Self Duality}

The word instanton has been coined for solutions of elliptic non linear
field equations in Euclidean space time, with boundary conditions at
infinity in such a way that stable topological properties emerge \cite{bpst}%
. The study of Euclidean Yang-Mills fields involves many mathematical items
falling under the headings : differential geometry (fiber bundles,
connections), differential topology (characteristic classes, index theory),
and algebraic geometry (twistors, holomorphic bundles) which makes it a rich
and unique subject. We review the standard $d=4$ dimensional instanton
solution in the first section.

In higher dimensions $d>4$, there is no unique way to define self-duality.
It is a matter of prejudice. One tries to conserve as much as possible of
the four dimensions duality characteristics. Particularly in 8 dimensions,
there have been a lot of proposals \cite{gks},\cite{wrd}---\cite{landi}.
There are some relationships between these apparently distinct
constructions. Each author of these proposals concentrated on a certain
aspect of the 4 dimensional self-duality which they considered a good
starting point for the generalization to 8 dimensions.

In the second section of this chapter, we list the main features of the GKS
instanton, then in the last section we present the soft seven sphere
instanton. We will show how the soft seven sphere can be used to reformulate
the GKS in a way very similar to the four dimensional case.

\section{The 4 Dimensional Instanton}

Consider an $SU\left( 2\right) $ classical gauge field over four dimensional
Euclidean space $\Bbb{R}^{4}$. Let the gauge potential be $A_{\mu }=A_{\mu
}^{a}t^{a}$, while the field strength is defined as the commutator of the
covariant derivative $\left( D_{\mu }\equiv \partial _{\mu }+A_{\mu }\right)
,$%
\begin{equation}
F_{\mu \nu }\equiv \left[ D_{\mu },D_{\nu }\right] =F_{\mu \nu
}^{a}t^{a}=\partial _{\mu }A_{\nu }-\partial _{\nu }A_{\mu }+\left[ A_{\mu
},A_{\nu }\right] ,
\end{equation}
where the generators of the Lie algebra are $t^{a}=-i\frac{\sigma ^{a}}{2},$
and $\left[ t^{a},t^{b}\right] =i\epsilon ^{abc}t^{c}$. We want to find the $%
A_{\mu }^{a}$ which minimizes the Euclidean action 
\begin{equation}
S_{E}=\int d^{4}x\frac{1}{4}F_{\mu \nu }^{a}F^{a\mu \nu }=-\frac{1}{2}\int
d^{4}x\;tr\left( F_{\mu \nu }F^{\mu \nu }\right) .
\end{equation}
We rewrite it as 
\begin{equation}
S_{E}=\frac{1}{8}\int d^{4}x\left\{ \left[ F_{\mu \nu }^{a}\pm {}^{*}F_{\mu
\nu }^{a}\right] ^{2}\mp 2F_{\mu \nu }^{a}{}^{*}F^{a\mu \nu }\right\}
\label{bogo4}
\end{equation}
where the dual field strength $^{*}F$ is defined by (the four dimensional
Levi--Cevita tensor $\epsilon _{0123}=1$) 
\begin{equation}
^{\ast }F_{\mu \nu }^{a}=\frac{1}{2}\epsilon _{\mu \nu \alpha \beta
}F^{a\alpha \beta }.
\end{equation}
Hence the lower bound for the action is 
\begin{equation}
S_{E}\geq \mp \frac{1}{4}\int d^{4}xF_{\mu \nu }^{a}{}^{*}F^{a\mu \nu }
\label{lowbou}
\end{equation}
The equality sign holds iff 
\begin{equation}
F_{\mu \nu }^{a}\pm {}^{*}F_{\mu \nu }^{a}=0  \label{seldua}
\end{equation}
which is the self/antiself duality condition (SD/ASD). The self/antiself
duality condition when combined with the Bianchi identities 
\begin{equation}
D_{\mu }F_{\nu \rho }+D_{\rho }F_{\mu \nu }+D_{\nu }F_{\rho \mu
}=0\Leftrightarrow \epsilon ^{\omega \mu \nu \rho }D_{\mu }F_{\nu \rho }=0,
\end{equation}
yields the equations of motion 
\begin{equation}
\epsilon ^{\omega \mu \nu \rho }D_{\mu }F_{\nu \rho }=D_{\mu }\;^{*}F^{\mu
\omega }=\pm D_{\mu }F^{\mu \omega }=0.
\end{equation}
Since the self/antiself duality equation is only first order in derivatives,
it is much easier to solve than the second order field equations.

In order for $S_{E}$ to remain finite, we require 
\begin{equation}
F_{\mu \nu }\stackrel{\;\;\left| x\right| {\ \rightarrow \infty }\quad }{%
\longrightarrow }0
\end{equation}
which is automatically valid if 
\begin{equation}
A_{\mu }\stackrel{\;\;\left| x\right| {\ \rightarrow \infty }\quad }{%
\longrightarrow }g^{-1}\left( x\right) \partial _{\mu }g\left( x\right)
\end{equation}
where $g\left( x\right) \in SU\left( 2\right) $ is a gauge transformation.
Thus we have the following situation: In general $F_{\mu \nu }\neq 0$ inside
a volume $\Bbb{R}^{4}$, but vanishes at the infinite boundary $\partial
E^{4}=S_{\infty }^{3}$, a three dimensional sphere, where the gauge
potential $A_{\mu }$ approaches a pure gauge. At the boundary $x\in
S_{\infty }^{3}$, our gauge transformation $g\left( x\right) \in SU\left(
2\right) ,\;$are mappings 
\begin{equation}
g:S_{\infty }^{3}\longrightarrow SU\left( 2\right) \simeq S^{3}.
\end{equation}
These mappings are classified according to the homotopy classes determined
by the topological winding number $w$ 
\begin{equation}
\pi _{3}\left( SU\left( 2\right) \right) \simeq \pi _{3}\left( S^{3}\right)
\simeq \left\{ w\right\} =\Bbb{Z}.
\end{equation}
The topological number $w$ is related to the Chern number 
\begin{equation}
Ch_{2}=\frac{1}{8\pi ^{2}}\int d^{4}x\;\epsilon ^{\mu \nu \alpha \beta
}F_{\mu \nu }^{a}F_{\alpha \beta }^{a}.
\end{equation}
Defining the topological charge 
\begin{equation}
N=\int d^{4}x\;n(x)\hspace{2cm}n(x)=\frac{e^{2}}{32\pi ^{2}}F_{\mu \nu
}^{a}{}^{*}F^{a\mu \nu }  \label{topchaeu}
\end{equation}
this is then an integer, since it is the topological number corresponding to
the mapping of the three-dimensional sphere into the gauge group $SU(2)$.
One can then get the following expression for the euclidean action if (\ref
{seldua}) is satisfied 
\begin{equation}
S_{E}=\mp \frac{8\pi ^{2}}{e^{2}}N.  \label{eucliac3}
\end{equation}

The simplest solution to the ASD condition ($F_{\mu \nu }=-\frac{1}{2}%
\epsilon _{\mu \nu \alpha \beta }F^{\alpha \beta }$), corresponding to a
value of the topological charge $N=1$, \cite{bpst}\cite{hooft1} is given by: 
\begin{equation}
A_{\mu }^{a}=\frac{x^{2}}{x^{2}+\lambda ^{2}}\left( g^{-1}\left( x\right)
\partial _{\mu }g\left( x\right) \right) ^{a}=-\frac{2\left( \Bbb{E}%
^{a}\right) _{\mu \nu }x^{\nu }}{x^{2}+\lambda ^{2}}
\end{equation}
where 
\begin{equation}
g\left( x\right) =\frac{x_{0}\mathbf{1}_{4}+ix.\sigma }{\sqrt{x^{2}}}
\end{equation}
and $\Bbb{E}^{a}$ are the canonical left quaternionic structures given in
eq.(\ref{cqs1}). Leading to 
\begin{equation}
F_{\mu \nu }^{a}=\frac{4\lambda ^{2}\left( \Bbb{E}^{a}\right) _{\mu \nu }}{%
\left( x^{2}+\lambda ^{2}\right) ^{2}}.
\end{equation}

Such a solution has a natural quaternionic formulation \cite{nash}\cite{adhm}%
. Consider 
\begin{equation}
g\left( x\right) =\frac{x^{\mu }e_{\mu }}{\left| x\right| }
\end{equation}
where $e_{\mu }$ are our four quaternionic units, introducing the self dual $%
SO\left( 4\right) $ basis 
\begin{equation}
\vartheta _{\mu \nu }=\frac{1}{2}\left( \bar{e}_{\mu }e_{\nu }-\bar{e}_{\nu
}e_{\mu }\right)
\end{equation}
such that $\left( \bar{e}_{0}=e_{0},\bar{e}_{i}=-e_{i}\right) $, we find 
\begin{equation}
\vartheta _{\mu \nu }=-\frac{1}{2}\epsilon _{\mu \nu \alpha \beta }\vartheta
^{\alpha \beta }
\end{equation}
and 
\begin{equation}
\left[ \vartheta _{\mu \nu },\vartheta _{\alpha \beta }\right] =2\left(
\delta _{\alpha \mu }\vartheta _{\beta \nu }-\delta _{\alpha \nu }\vartheta
_{\beta \mu }-\delta _{\beta \mu }\vartheta _{\alpha \nu }+\delta _{\beta
\nu }\vartheta _{\alpha \mu }\right) .
\end{equation}
For $A_{\mu }$ given by 
\begin{equation}
A_{\mu }=\frac{x^{2}}{x^{2}+\lambda ^{2}}\left( g^{-1}\left( x\right)
\partial _{\mu }g\left( x\right) \right) =-\frac{2\Bbb{\vartheta }_{\mu \nu
}x^{\nu }}{x^{2}+\lambda ^{2}}
\end{equation}
we have 
\begin{equation}
F_{\mu \nu }=\frac{4\lambda ^{2}\Bbb{\vartheta }_{\mu \nu }}{\left(
x^{2}+\lambda ^{2}\right) ^{2}}\Longrightarrow F_{\mu \nu }=-\frac{1}{2}%
\epsilon _{\mu \nu \alpha \beta }F^{\alpha \beta }.
\end{equation}
The self-dual solution $F_{\mu \nu }=\frac{1}{2}\epsilon _{\mu \nu \alpha
\beta }F^{\alpha \beta }$ can be found simply by replacing $e_{\mu
}\rightarrow 1|e_{\mu }$ leading to 
\begin{equation}
g\left( x\right) =\frac{x^{\mu }1|e_{\mu }}{\left| x\right| }.
\end{equation}
The corresponding $SO\left( 4\right) $ antiself--dual basis are 
\begin{equation}
\bar{\vartheta}_{\mu \nu }=\frac{1}{2}\left( 1|\bar{e}_{\mu }1|e_{\nu }-1|%
\bar{e}_{\nu }1|e_{\mu }\right) \Longrightarrow \bar{\vartheta}_{\mu \nu }=%
\frac{1}{2}\epsilon _{\mu \nu \alpha \beta }\bar{\vartheta}^{\alpha \beta }.
\end{equation}
Then with the choice 
\begin{equation}
A_{\mu }=\frac{x^{2}}{x^{2}+\lambda ^{2}}\left( g^{-1}\left( x\right)
\partial _{\mu }g\left( x\right) \right) =-\frac{2\bar{\Bbb{\vartheta }}%
_{\mu \nu }x^{\nu }}{x^{2}+\lambda ^{2}},
\end{equation}
we have 
\begin{equation}
F_{\mu \nu }=\frac{4\lambda ^{2}\bar{\Bbb{\vartheta }}_{\mu \nu }}{\left(
x^{2}+\lambda ^{2}\right) ^{2}}\Longrightarrow F_{\mu \nu }=\frac{1}{2}%
\epsilon _{\mu \nu \alpha \beta }F^{\alpha \beta }
\end{equation}

The higher $n>1$ instantons, Atiyah, Drinfield, Hitchin and Manin (ADHM)
solutions, are given naturally in terms of quaternions \cite{adhm}.

Lastly, we would like to show how to get static monopole solutions by field
redefinition of the instanton problem. We consider only static monopoles,
purely magnetic and which are solutions of an equation called the Bogomolny
condition\cite{nash}. The model is defined over $\Bbb{R}^{3}$. Such
monopoles are called BPS states, they correspond to an $su\left( 2\right) $
valued pair of a gauge field $A_{i}$ $\left( i=1,2,3\right) $ and a scalar
field $\phi $. The action of the monopole system is 
\begin{equation}
\mathcal{L}=\frac{1}{2}\int d^{3}x\left[ B^{i}B_{i}+\left( D_{i}\phi \right)
\left( D^{i}\phi \right) -\lambda \left( \phi ^{2}-a^{2}\right) ^{2}\right] ,
\end{equation}
$B_{i}$ and $D_{i}\phi $ are defined by 
\begin{equation}
B_{i}=\epsilon _{ijk}\left( \partial ^{j}A^{k}+A^{j}A^{k}\right)
,\;\;\;\;\;\;\;\;\;\;\;\;\;\;\;D_{i}\phi =\partial _{i}\phi +\left[
A_{i},\phi \right] .
\end{equation}
The field equations for this static system are 
\begin{eqnarray}
\epsilon _{ijk}D^{j}B^{k} &=&\left[ \phi ,D_{i}\phi \right] ,  \label{klop1}
\\
D^{i}D_{i}\phi &=&2\lambda \phi ^{3}-2\lambda a^{2}\phi  \label{klop2}
\end{eqnarray}
whereas the Bianchi identiy is 
\begin{equation}
D_{i}B_{i}=0,  \label{klop3}
\end{equation}
Being a second order system of partial differential equations (\ref{klop1}--%
\ref{klop3}), they are hard to solve. However, if we take the limit $\phi
^{2}=a^{2}$, we impose a first order equation 
\begin{equation}
B_{i}=D_{i}\phi  \label{bg}
\end{equation}
which is the Bogomolny equation. We can relate the self duality condition (%
\ref{seldua}) to the Bogomolny condition if we redefine $\phi $ as a fourth
component $A_{0}$ of the gauge field $A_{i}$ i.e. $A_{\mu }\equiv \left(
A_{0}=\phi ,A_{i}\right) $. We then have 
\begin{equation}
D_{i}\phi =F_{i0},\;\;\;\;\;\;\;\;\;\;\;\;\;\;\;B_{i}=\frac{1}{2}\epsilon
_{ijk}F^{jk}
\end{equation}
by substitution in (\ref{bg}), we recover the self duality condition.

\section{The Grossman--Kephart--Stasheff Instanton}

The four dimensional self--duality notion proved to be a very powerful tool
both of physics and mathematics, so it is natural to investigate the
occurrence of a similar condition in higher dimensions. As we have already
said, there is no standard way to express the self duality equation in $d>4$
dimensions. In a generic $d$ dimensions, a $p$ form $v$ is 
\begin{equation}
v=\frac{1}{p!}v_{\alpha _{1}\cdots \alpha _{p}}dx^{\alpha _{1}}\wedge \cdots
\wedge dx^{\alpha _{p}},\;\;\;\;\;\;\;\;\;\;\;\;\;\;\;\;\;\;\;\;\;\;{\
\alpha }_{1}{\ ,\ldots ,\alpha }_{p}{\ =1\ldots d}
\end{equation}
and the dual form is 
\begin{equation}
^{\ast }v=\frac{1}{d!}\epsilon ^{\alpha _{1}\cdots \alpha _{d}}v_{\alpha
_{1}\cdots \alpha _{p}}dx^{\alpha _{p+1}}\wedge \cdots \wedge dx^{\alpha
_{d}}.
\end{equation}
which means, in $d$ dimensions the dual of a $p$ form is a $d-p$ form.
Knowing that the Yang-Mills field strength can be written as a two form 
\begin{equation}
F=\frac{1}{2}F_{\alpha _{1}\alpha _{2}}dx^{\alpha _{1}}\wedge dx^{\alpha
_{2}},
\end{equation}
then the dual of a 2 form is another 2 form iff $d=4$. Constraining
ourselves to Yang-Mills models, we have at disposal just the one form gauge
field $A=A_{i}dx^{i}$ or the 2 form field strength $F.$

In eight dimensions, it is not obvious how we should proceed. There have
been different suggestions. For example:

\begin{itemize}
\item  The Fubini--Nicolai \cite{fub} or the Corrigan-Devchand-Fairlie-Nuyts 
\cite{cdfn} instanton: the authors insist on the existence of ``squaring''
i.e. the action can be written as the square of self-dual fields.

\item  There exist also some promising generic higher dimensional
self-duality conditions that are not just restricted to 8 dimensions but
also go beyond that limit. For example, the Ivanon-Popov \cite{IP} proposal
where a Clifford-algebraic structure is used. Another example is the
geometric Bais-Batenburg \ \cite{bb} self-duality which is based upon
hypercomplex structures over appropriate manifolds.
\end{itemize}

Here, we would like to concentrate on another proposal.
Grossman--Kephart-Stasheff \cite{gks} suggested a condition, for eight
dimensions, that has deep topological roots: The last Hopf map $S^{15}{%
\stackrel{\quad S^{7}\quad }{\longrightarrow }}S^{8}$, conformal invariance
and spin structure over $S^{8}$\cite{landi}.

Working over the 8 dimensional euclidean space $\Bbb{R}^{8}$. The (GKS) self
duality condition\ is (the eight dimensional Levi--Cevita tensor $\epsilon
_{01234567}=1$) 
\begin{equation}
F_{\alpha _{1}\alpha _{2}}^{a}F_{\alpha _{3}\alpha _{4}}^{a}=\frac{1}{4!}%
\epsilon _{\alpha _{1}\ldots \alpha _{8}}F^{a\alpha _{5}\alpha
_{6}}F^{a\alpha _{7}\alpha _{8}}.  \label{ax}
\end{equation}
where there is summation over the Lie algebra indices $a$. Grossman, Kephart
and Stasheff insisted upon the conformal invariance of the Yang-Mills action
in 4 dimensions. In eight dimensions the Yang-Mills action is not
conformally invariant hence they considered the functional 
\begin{equation}
\mathcal{A}=\int d^{8}x\;\left( F_{\mu _{1}\mu _{2}}^{a}F_{\mu _{3}\mu
_{4}}^{a}F^{b\mu _{1}\mu _{2}}F^{b\mu _{3}\mu _{4}}\right)
\end{equation}
which upon the use of the self duality condition (\ref{ax}), takes a form
similar to the \ fourth Chern class 
\begin{equation}
\int_{S^{8}}\epsilon _{\alpha _{1}\ldots \alpha _{8}}F^{a\alpha _{1}\alpha
_{2}}F^{a\alpha _{3}\alpha _{4}}F^{b\alpha _{5}\alpha _{6}}F^{b\alpha
_{7}\alpha _{8}}.
\end{equation}
In the search for solutions of the GKS duality condition and requiring that $%
F_{\mu _{1}\mu _{2}}\stackrel{\;\;S_{\infty }^{7}\sim \left\{ \left|
x\right| \rightarrow \infty \right\} \quad }{\longrightarrow }0$, so that $%
A_{\mu _{1}}$ must be a pure gauge at infinity 
\begin{equation}
A_{\mu }\stackrel{\;\;S_{\infty }^{7}\quad }{\longrightarrow }g^{-1}\left(
x\right) \partial _{\mu }g\left( x\right) .
\end{equation}
where $g\left( x\right) $ is a gauge transformation. GKS assumed the
following form for $g\left( x\right) $%
\begin{equation}
g\left( x\right) =\frac{x^{\mu }\Bbb{E}_{\mu }}{\sqrt{x^{2}}}
\end{equation}
where $\Bbb{E}_{\mu }$ are given by (\ref{ee1}), whence, 
\begin{equation}
g^{\dagger }\left( x\right) g\left( x\right)
=1\;\;\;\;\;\;\;\;\;and\;\;\;\;\;\;\;\;\;Det\left( g\left( x\right) \right)
=1.
\end{equation}
Note that, for $\hat{x}^{\mu }=\frac{x^{\mu }}{\sqrt{x^{2}}},$ we have $\hat{%
x}^{\mu }\hat{x}_{\mu }=1$ i.e. $g\left( x\right) $ parameterize a unit $%
S^{7}.$ For the boundary condition given above, we have the following
situation: 
\begin{equation}
g\left( x\right) :S_{\infty }^{7}\longrightarrow S^{7}
\end{equation}
Such a class of maps are classified according to the seventh homotopy group
of the seven sphere 
\begin{equation}
\pi _{7}\left( S^{7}\right) =n\in \Bbb{Z}
\end{equation}
i.e. there is no map between solutions of different n. They lie in different
classes. In particular, there can be no map between the trivial $n=0$ ($%
A_{\mu }=0$) field configuration and $n>0$ ($A_{\mu }\neq 0$). Any $n\neq 0$
is stable and will never decay. GKS proposed the following ansatz for $%
A_{\mu }$ ($n=1$) solution 
\begin{equation}
A_{\mu }=\frac{x^{2}}{x^{2}+\lambda ^{2}}\left( g^{-1}\left( x\right)
\partial _{\mu }g\left( x\right) \right) .
\end{equation}
which solves (\ref{ax}). Now, let's mention the difference between this GKS
duality and the standard four dimensional duality considered in the previous
section.

\begin{itemize}
\item  It is not derived from an action. \ $\mathcal{A}$\ \ \ \ has no
quadratic term in the derivatives i.e. there is no kinetic term.

\item  The GKS duality is valid only for a specific representation $8$ of a
specific group $SO\left( 8\right) $ in contrast to the self--duality
condition which is well defined for any representation of any simple Lie
group.

\item  The GKS solution is not a solution of the $\Bbb{R}^{8}$ Yang--Mills
field equations.
\end{itemize}

\section{Eight Dimensional Soft Self Duality}

Now we would like to reformulate in a quadratic form the GKS self duality
condition. We work over $\Bbb{R}^{8}$. Contrary to the standard Yang-Mills
gauge field, the soft gauge field strength carries a dependence upon the
internal manifold. It is important to know where we stand over the $%
S_{gauge}^{7}$, as the structure functions vary from one point to another.
For a soft gauge field $A_{\mu }(x)$ $\equiv A_{\mu }^{i}\Bbb{E}_{i}$, the
insertion of the $\varphi $ is essential for the closure of the commutator 
\begin{eqnarray}
\left[ A_{\mu }(x),A_{\nu }(x)\right] \varphi &\equiv &A_{\mu }^{i}(x)A_{\nu
}^{j}(x)\left[ \Bbb{E}_{i},\Bbb{E}_{j}\right] \varphi \\
&=&2f_{ijk}^{\left( +\right) }\left( \varphi \right) A_{\mu }^{i}(x)A_{\nu
}^{j}(x)\Bbb{E}_{k}\varphi .
\end{eqnarray}
thus \ the field strength is given by 
\begin{eqnarray}
F_{\mu \nu }(x,{\varphi }) &=&F_{\mu \nu }(x){\ \varphi }  \nonumber \\
&=&F_{\mu \nu }^{i}\left( x\right) \Bbb{E}_{i}\varphi  \nonumber \\
{} &=&{}\left( \frac{\partial A_{\nu }(x)}{\partial x^{\mu }}-\frac{\partial
A_{\mu }(x)}{\partial x^{\nu }}+\left[ A_{\mu }(x),A_{\nu }(x)\right]
\right) {\varphi }.  \label{fmn}
\end{eqnarray}

The critical point for the self-duality condition is the existence of a
fourth rank tensor. Adding a zero index to extend $f_{ijk}^{\left( \pm
\right) }({\varphi })$ from $\Bbb{R}^{7}$ to $\Bbb{R}^{8}$, we define a
fourth rank tensor $\eta _{\alpha \beta \mu \nu }({\varphi })$ which is
equal to 
\begin{equation}
\eta _{0ijk}^{\left( \pm \right) }({\varphi })=f_{ijk}^{\left( \pm \right) }(%
{\varphi })\qquad ,  \label{sel}
\end{equation}
and zero elsewhere. The proposed generalization of the four dimensional self
duality is the following \emph{soft self duality condition} 
\begin{equation}
F{\ (x,{\varphi })}=\star F{\ (x,{\varphi })},
\end{equation}
or in terms of components 
\begin{equation}
F_{0i}{\ (x,{\varphi })}=\frac{1}{2}\eta _{0ijk}^{\left( \pm \right) }({%
\varphi })F^{jk}(x,{\varphi })\qquad ,  \label{a3}
\end{equation}
note that $\eta _{\alpha \beta \mu \nu }({\varphi })$ varies over the seven
sphere. To proceed, we require the vanishing of $F_{\mu \nu }$ at the
infinite $S_{\infty }^{7}$, thus $A_{\mu }$ (at $S_{\infty }^{7}$) must be a
pure gauge $A_{\mu }=g^{-1}\left( x\right) \partial _{\mu }g\left( x\right) $%
, where our gauge transformation $g\left( x\right) $ is a map from the
spatial $S_{\infty }^{7}$ to the gauge space $S_{gauge}^{7}$ \footnote{%
Soft seven sphere gauge transformations reduce to the standard Yang-Mills
theory at any single point over the seven sphere. The soft seven sphere
gauge field $A_{\mu }$ transforms as $A\mu \left( x\right) \varphi
\rightarrow g^{-1}\left( A_{\mu }\left( x\right) +\partial _{\mu }\right)
g\varphi $ and the Field strength $F_{\mu \nu }\left( x\right) $ transforms
as $F_{\mu \nu }\left( x,\varphi \right) \rightarrow g^{-1}\left( F_{\mu \nu
}\left( x\right) \right) g\varphi .$ The presence of the $\varphi $ is
essential.}. Consider an $S^{7}$ element 
\begin{equation}
g(x)=\frac{\Bbb{E}_{\mu }x^{\mu }}{\sqrt{x^{2}}},  \label{gg}
\end{equation}
the self-dual gauge solution of the self dual condition is exactly the GKS
ansatz 
\begin{equation}
A_{\mu }^{(+)}(x)=\frac{x^{2}}{x^{2}+\lambda ^{2}}g^{-1}(x)\partial _{\mu
}g(x)=-\frac{\Xi _{\mu \nu }^{\left( +\right) }x^{\nu }}{x^{2}+\lambda ^{2}}
\label{sol1}
\end{equation}
where the $\Xi _{\mu \nu }^{\left( +\right) }$ is given by 
\begin{equation}
\Xi _{\mu \nu }^{\left( +\right) }=\frac{1}{2}\left( \Bbb{E}_{\mu }^{t}\Bbb{E%
}_{\nu }-\Bbb{E}_{\nu }^{t}\Bbb{E}_{\mu }\right) .
\end{equation}
\ We call $\Xi _{\mu \nu }^{\left( +\right) }$ the self dual tensor, because 
\begin{equation}
\Xi _{oi}^{\left( +\right) }\varphi =\frac{1}{2}\eta _{0ijk}^{\left(
+\right) }({\varphi })\Xi ^{\left( +\right) jk}\varphi .
\end{equation}
After substituting $A_{\mu }^{(+)}(x)$ into (\ref{fmn}), we find 
\begin{equation}
F_{\mu \nu }^{\left( +\right) }(x,{\varphi })=2\frac{(\Xi _{\mu \nu
}^{\left( +\right) })\lambda ^{2}}{(\lambda ^{2}+x^{2})^{2}}\varphi
\end{equation}
which is obviously soft self dual (\ref{a3}).

Another problem of the GKS instanton that can be overcame in the soft seven
sphere framework is the compatibility of the equation of motion and the self
duality condition. We find by explicit calculation that our solution (\ref
{sol1}) satisfies the Yang-Mills equation of motion for a soft gauge field 
\begin{equation}
D_{\mu }F_{\mu \nu }\left( x\right) {\varphi }=\partial _{\mu }F_{\mu \nu
}\left( x\right) {\varphi }+[A_{\mu }\left( x\right) ,F_{\mu \nu }\left(
x\right) ]{\varphi }=0.
\end{equation}
Of course the four dimensional case is more powerful because the self
duality is related directly to the Bianchi identity which does not hold in
higher dimensions. However, in our case the soft self duality is compatible
with the equation of motion whereas for Grossman-Kephart-Stasheff instanton,
one must work over curved space--time with certain condition for the metric
in order to satisfy both the self duality and the equation of motion. To
construct a static seven dimensional monopole, we proceed by static
dimensional reduction from $\Bbb{R}^{8}$ to $\Bbb{R}^{7}$. Identifying 
\begin{eqnarray}
A_{0} &=&\phi , \\
F_{ij} &=&f_{ijk}\left( \varphi \right) B^{k}\;\;\;\;\;\;\;\;\;\;\;\;\;{%
i,j,k=1..7}
\end{eqnarray}
then the self duality will be reduced to 
\begin{equation}
\left( B_{i}\right) \varphi =\left( D_{i}\phi \right) \varphi .
\end{equation}

Using the soft seven sphere, we can easily generate new solutions of GKS
dualities. Working with $\Bbb{E}\left( \varphi \right) $, replacing and $%
g\left( x\right) $ by 
\begin{equation}
g\left( x,\varphi \right) =\frac{\Bbb{E}_{\mu }\left( \varphi \right) x^{\mu
}}{\sqrt{x^{2}}}
\end{equation}
the resulting gauge field given in terms 
\begin{equation}
\Xi _{\mu \nu }^{\left( ++\right) }\left( \varphi \right) =\frac{1}{2}\left( 
\Bbb{E}_{\mu }^{t}\left( \varphi \right) \Bbb{E}_{\nu }\left( \varphi
\right) -\Bbb{E}_{\nu }^{t}\left( \varphi \right) \Bbb{E}_{\mu }\left(
\varphi \right) \right)
\end{equation}
leads to 
\begin{eqnarray}
A_{\mu }^{(++)}(x) &=&\frac{x^{2}}{x^{2}+\lambda ^{2}}g^{-1}(x,\varphi
)\partial _{\mu }g(x,\varphi )=-\frac{\Xi _{\mu \nu }^{\left( ++\right)
}\left( \varphi \right) x^{\nu }}{x^{2}+\lambda ^{2}} \\
F_{\mu \nu }^{\left( ++\right) }(x,{\varphi }) &=&2\frac{(\Xi _{\mu \nu
}^{\left( ++\right) }\left( \varphi \right) )_{\mu \nu ~}\lambda ^{2}}{%
(\lambda ^{2}+x^{2})^{2}}  \label{fff}
\end{eqnarray}
which also satisfy (\ref{ax}) and (\ref{a3}).

\chapter{Hypercomplex SSYM Models}

Day after day, supersymmetry consolidates its position in theoretical
physics. Even if it was introduced more than 25 years ago, there are still
problems with the geometric basis of extended $\left( N>1\right) $
supersymmetry. The situation of the extended superspace is far less
satisfactory than the original N=1 superspace. At the level of the algebra
the on-shell formalism closes up to modulo of the classical equations of
motion. This fact seems odd at the quantum level since the equations of
motion receive loop corrections\footnote{%
Also, the supersymmetry transformations receive corrections and one should
test the closure of \ the algebra order by order in perturbation theory.}.
The superspace introduces an elegant supermanifold with different enlarged
superconnections, where some are truly integrable in the sense of having
zero supercurvature. In principle, the extended superspace should be a very
powerful tool for quantum calculations.

Before starting, we feel obliged to mention something about the history of
the following conjecture: Ring Division Algebras $\Bbb{K}\equiv $\{ real $%
\Bbb{R},\;$complex $\Bbb{C},\;$quaternions $\Bbb{H},\;$octonions $\Bbb{O}$
\} are relevant to simple supersymmetric Yang-Mills. The first hint, as
mentioned by Schwarz \cite{r1} comes from the number of propagating Bose and
Fermi degrees of freedom which is one for $d=3$, two for $d=4$, four for $%
d=6 $ and eight for $d=10$ suggesting a correspondence with real $\Bbb{R}$,
complex $\Bbb{C}$, quaternions $\Bbb{H}$ and octonions $\Bbb{O}$. Kugo and
Townsend \cite{r2} investigated in detail the relationship between $\Bbb{K}$
and the irreducible spinorial representation of the Lorentz group in $%
d=3,4,6,10$, building upon the following chain of isomorphisms 
\begin{eqnarray*}
so\left( 2,1\right) &\Longleftrightarrow &sl(2,\Bbb{R}) \\
so(3,1) &\Longleftrightarrow &sl(2,\Bbb{C}) \\
so(5,1) &\Longleftrightarrow &sl(2,\Bbb{H}).
\end{eqnarray*}
They conjectured that $so(9,1)\Longleftrightarrow sl(2,\Bbb{O})$, the
correct relation turned out to be 
\[
so(9,1)\Longleftrightarrow sl(2,\Bbb{O})\oplus G_{2} 
\]
as has been shown by Chung and Sudbery \cite{r3}, i.e. the dimension of $%
Sl\left( 2,\Bbb{O}\right) $ is 31. Also in \cite{r2}, a quaternionic
treatment of the $d=6$ case is presented. Later, Evans made a systematic
investigation of the relationship between SSYM and ring division algebra in
a couple of papers. In the first \cite{r4}, he simplified the construction
of SSYM by proving a very important identity between gamma matrices by using
the intrinsic triality of ring division algebra instead of the ``tour de
force'' used originally by Brink, Scherck and Schwarz \cite{r0} via Fierz
identities generalized to $d>4$ dimensions. Then, in the second paper \cite
{r6}, Evans made the connection even clearer by showing how the auxiliary
fields are really related to ring division algebras. For $d=3,4,6,10$ we
need $k=0,1,3,7$ auxiliary fields respectively. An alternative approach for
the octonionic case was introduced by Berkovits \cite{r7} who invented a
larger supersymmetric transformation called generalized supersymmetry in 
\cite{r8}. There has also been a twistor attempt by Bengtsson and Cederwall 
\cite{r9}. For more references about the octonionic case and ten dimensional
physics one may consult references in \cite{r10} and its extension to
p-branes by Belecowe and Duff \cite{r11}. The early work of Nilsson may be
relevant \cite{r12}\cite{r13} too.

As a first step towards an extended superspace, we address the point of the
algebraic auxiliary fields for simple N=1 supersymmetric Yang-Mills (SSYM)
definable only in $d=3,4,6$ and $10$ dimensions \cite{r0}. The important
point is: \emph{While the physical fields couple to ring division left
action the auxiliary ones couple to right action (or vice versa)}. To admit
a closed off-shell supersymmetric algebra, left and right action must
commute i.e. we should have a parallelizable associative algebra. For d = 6,
quaternions work fine but for d = 10, the only associative seven dimensional
algebra that is known is the soft seven sphere. We shall show below how this
works. In this chapter, we use the same symbols $\left( \mbox{left action}%
\equiv \Bbb{E}_{j}\mbox{,
right action}\equiv 1|\Bbb{E}_{j}\right) $ for either complex, quaternionic
or octonionic numbers and each case should be distinguished by the range of
the indices $j$ which run from $1$ to $\left( 1,3,7\right) $ for complex,
quaternions and octonions respectively.

\section{On--Shell SSYM in $d=3,4,6,10$}

In this section we follow a notation which is a mixture between that of
Evans \cite{r4}\cite{r6} and Supersolutions \cite{supersol}. The Minkowskian
metric has signature $\eta _{MN}\equiv \left( -,+,\ldots ,+\right) $ and
spinorial indices have range $n$. Our generalized gamma matrices $\Gamma $
and $\tilde{\Gamma}$ are \emph{real symmetric} and we will never raise or
lower the spinors indices. We define $\left( \Gamma _{M}\right) _{ab}$ of
lower spinorial indices, whereas the upper ones are defined by $\left( 
\tilde{\Gamma}_{0}\right) ^{ab}=-\left( \Gamma _{0}\right) _{ab}$ and $%
\left( \tilde{\Gamma}\right) ^{ab}=\left( \Gamma \right) _{ab},$ whence 
\[
\Gamma ^{M}\tilde{\Gamma}^{N}+\Gamma ^{N}\tilde{\Gamma}^{M}=2\eta ^{MN} 
\]
or in terms of components 
\begin{equation}
\left( \Gamma ^{M}\right) _{ab}\left( \tilde{\Gamma}^{N}\right) ^{bc}+\left(
\Gamma ^{N}\right) _{ab}\left( \tilde{\Gamma}^{M}\right) ^{bc}=2\eta
^{MN}\delta _{a}^{c}.
\end{equation}
with $a,b$ of range $n$.

Simple supersymmetric Yang-Mills models are composed of gauge fields $A_{M}$%
, spinor fields $\Psi ^{a}$ and the Lagrangian density is 
\begin{equation}
\mathcal{L}=-\frac{1}{4}F_{MN}F^{MN}+\frac{1}{2}\Psi ^{t}\Gamma ^{M}\nabla
_{M}\Psi ,  \label{ssym1}
\end{equation}
where $\nabla _{M}\equiv \partial _{M}+A_{M};$ $F_{MN}\equiv \left[ \nabla
_{M},\nabla _{N}\right] $ and we have suppressed here the Lie algebra
indices. We first ask in which dimensions $d$ and for what type of spinorial
field $\Psi $ the action (\ref{ssym1}) is supersymmetric? Assume the
spinorial field has $\ n$ components. Upon quantization, the gauge field has 
$d-2$ \emph{physical degrees of freedom} while the spinor field has $n/2$
physical degrees of freedom. A supersymmetric model must have equal number
of bosonic and fermionic degrees of freedom, hence $d=2+\frac{n}{2}$ and
since for spinors $n=2,4,8,16$ \ this lead to $d=3,4,6,10.$ We work with 
\emph{real bases for spinors and vectors} \cite{r6}. Introducing an n
components Grassmann variable $\xi $, we postulate the supersymmetry
transformation $\delta _{\xi }$ to be 
\begin{eqnarray}
\delta _{\xi }A_{M} &=&\xi ^{a}\left( \Gamma _{M}\right) _{ab}\Psi ^{b}, 
\nonumber \\
\delta _{\xi }\Psi ^{a} &=&\frac{1}{2}\xi ^{b}\left( \tilde{\Gamma}%
^{M}\right) ^{ac}\left( \Gamma ^{N}\right) _{cb}F_{MN}.  \nonumber \\
&&  \label{trs}
\end{eqnarray}

We have to check the invariance of the Lagrangian. The variation of $F_{MN}$
is 
\begin{equation}
\delta _{\xi }F_{MN}=\xi ^{a}\left( \Gamma _{N}\right) _{ab}\nabla _{M}\Psi
^{b}-\xi ^{a}\left( \Gamma _{M}\right) _{ab}\nabla _{N}\Psi ^{b},
\end{equation}
the variation of the first term of our $\mathcal{L}$ $\ $gives 
\begin{equation}
\delta _{\xi }\left( -\frac{1}{4}F_{MN}F^{MN}\right) =-\xi ^{a}\left( \Gamma
_{M}\right) _{ab}F^{MN}\nabla _{N}\Psi ^{b}.
\end{equation}
Now, taking into account $\nabla _{M}=\partial _{M}+A_{M}$, making the
variation of the second term of the Lagrangian (\ref{ssym1}) and using as a
basis of our Lie algebra $\Psi ^{z}t^{z}$ with $\left[ t^{x},t^{y}\right]
=c^{xyz}t^{z},$%
\begin{eqnarray}
\delta _{\xi }\left( \frac{1}{2}\left( \Gamma ^{M}\right) _{ab}\Psi
^{a}\nabla _{M}\Psi ^{b}\right) &=&\xi ^{a}\left( \Gamma ^{N}\right)
_{ab}F_{MN}\nabla ^{M}\Psi ^{b}  \nonumber \\
&&-\frac{1}{2}\xi ^{c}\left( \Gamma _{M}\right) _{ab}\left( \Gamma
^{M}\right) _{cd}\left( \Psi ^{az}c_{xyz}\Psi ^{dx}\Psi ^{by}\right) 
\nonumber \\
&&+total\;derivative.
\end{eqnarray}
We have used the Bianchi identity in the above derivation. The SSYM action (%
\ref{ssym1}) is invariant under (\ref{trs}) if 
\begin{equation}
\xi ^{c}\left( \Gamma _{M}\right) _{ab}\left( \Gamma ^{M}\right) _{cd}\left(
\Psi ^{az}c_{xyz}\Psi ^{dx}\Psi ^{by}\right) =0  \label{abb}
\end{equation}
\begin{equation}
\Rightarrow Q_{abcd}=\left( \Gamma _{M}\right) _{ab}\left( \Gamma
^{M}\right) _{cd}+\left( \Gamma _{M}\right) _{bd}\left( \Gamma ^{M}\right)
_{ca}+\left( \Gamma _{M}\right) _{da}\left( \Gamma ^{M}\right) _{cb}=0.
\label{ident}
\end{equation}
We conclude from this calculation that $\mathcal{L}$\ \ \ is invariant\
under (\ref{trs}) in any dimension for abelian algebras with any spin
representation because (\ref{abb}) is trivially zero for $c_{xyz}=0$. Now,
we want to find the solution of (\ref{ident}). As our complex numbers,
quaternions, octonions form a Clifford algebra of signature $%
Cliff(0,1),Cliff\left( 0,3\right) ,Cliff\left( 0,7\right) $ respectively, we
have 
\begin{eqnarray}
(\Bbb{E}_{k})_{\mu \nu }(\Bbb{E}_{j})_{\lambda \nu }+(\Bbb{E}_{j})_{\mu \nu
}(\Bbb{E}_{k})_{\lambda \nu } &=&2\delta _{kj}\delta _{\mu \lambda }, 
\nonumber \\
(\Bbb{E}_{k})_{\mu \nu }(\Bbb{E}_{j})_{\mu \lambda }+(\Bbb{E}_{j})_{\mu \nu
}(\Bbb{E}_{k})_{\mu \lambda } &=&2\delta _{kj}\delta _{\nu \lambda }, 
\nonumber \\
(\Bbb{E}_{k})_{\mu \nu }(\Bbb{E}_{k})_{\lambda \zeta }+(\Bbb{E}%
_{k})_{\lambda \nu }(\Bbb{E}_{k})_{\mu \zeta } &=&2\delta _{\mu \lambda
}\delta _{\nu \zeta },  \label{trl}
\end{eqnarray}
and the same holds equally well for $(1|\Bbb{E}_{j})$. As had been noticed
by Evans, these are direct consequences of the ring division triality\cite
{r4}. We can construct immediately two sets of gamma matrices as follows 
\[
\left( \Gamma _{0}\right) =\left( 
\begin{array}{cc}
-\mathbf{1}_{\frac{n}{2}} & 0 \\ 
0 & -\mathbf{1}_{\frac{n}{2}}
\end{array}
\right) ; 
\]
\[
\left( \Gamma _{j}\right) =\left( 
\begin{array}{cc}
0 & \Bbb{E}_{j} \\ 
-\Bbb{E}_{j} & 0
\end{array}
\right) ;\;\;\;\left( 1|\Gamma _{j}\right) =\left( 
\begin{array}{cc}
0 & 1|\Bbb{E}_{j} \\ 
-1|\Bbb{E}_{j} & 0
\end{array}
\right) , 
\]
\begin{eqnarray}
\left( \Gamma _{d-2}\right) &=&\left( 
\begin{array}{cc}
0 & \mathbf{1}_{\frac{n}{2}} \\ 
\mathbf{1}_{\frac{n}{2}} & 0
\end{array}
\right) ;\;\;\;\left( \Gamma _{d-1}\right) =\left( 
\begin{array}{cc}
\mathbf{1}_{\frac{n}{2}} & 0 \\ 
0 & -\mathbf{1}_{\frac{n}{2}}
\end{array}
\right) ,  \nonumber \\
&&  \label{gmm}
\end{eqnarray}
where $j=1..d-3,$ thus 
\[
\Gamma ^{M}\tilde{\Gamma}^{N}+\Gamma ^{N}\tilde{\Gamma}^{M}=1|\Gamma ^{M}1|%
\tilde{\Gamma}^{N}+1|\Gamma ^{N}1|\tilde{\Gamma}^{M}=2\eta ^{MN} 
\]
or in terms of components 
\begin{eqnarray}
\left( \Gamma ^{M}\right) _{ab}\left( \tilde{\Gamma}^{N}\right) ^{bc}+\left(
\Gamma ^{N}\right) _{ab}\left( \tilde{\Gamma}^{M}\right) ^{bc} &=&\left(
1|\Gamma ^{M}\right) _{ab}\left( 1|\tilde{\Gamma}^{N}\right) ^{bc}+\left(
1|\Gamma ^{N}\right) _{ab}\left( 1|\tilde{\Gamma}^{M}\right) ^{bc}  \nonumber
\\
&=&2\eta ^{MN}\delta _{a}^{c}.
\end{eqnarray}
automatically, our $\Gamma $'s satisfy the identity (\ref{ident}) for both
left and right actions. Indeed by direct calculation using (\ref{trl}), we
see that 
\begin{equation}
\Gamma _{Ma(b}\Gamma _{cd)}^{M}=1|\Gamma _{Ma(b}1|\Gamma _{cd)}^{M}=0.
\end{equation}
Consequently, the spin representation decomposes into 
\begin{eqnarray}
SPIN\left( 1,2\right) &\Longleftrightarrow &SL\left( 2,R\right) \\
SPIN\left( 1,3\right) &\Longleftrightarrow &SL\left( 2,C\right) \\
SPIN\left( 1,5\right) &\Longleftrightarrow &SL\left( 2,H\right)
\end{eqnarray}
and using the soft sphere, it seems that $soft\;SPIN\left( 1,9\right)
\Longleftrightarrow SL\left( 2,soft\;S^{7}\right) $.

We still have to show that the commutators of (\ref{trs}) close to a
supersymmetric algebra. For any arbitrary field $V$ in our Lagrangian, we
need to check that 
\begin{equation}
\fbox{$\left[ \delta _{\xi },\delta _{\chi }\right] V=2\xi ^{a}\chi
^{b}\left( \Gamma ^{M}\right) _{ab}\partial _{M}V$\ .}  \label{salg}
\end{equation}
As we are working on--shell, the algebra should closes modulo the fermionic
equation of motion 
\begin{equation}
\left( \Gamma ^{M}\right) _{ab}\nabla _{M}\Psi ^{a}=0,\;\;\;\;\;\forall b
\end{equation}
and gauge transformation. Using (\ref{trs}), we can easily check the closure
for the gauge field $A_{M}$%
\begin{eqnarray}
\left[ \delta _{\xi },\delta _{\chi }\right] A_{M} &=&-\frac{1}{2}\xi
^{a}\chi ^{b}\left( \left( \tilde{\Gamma}^{P}\right) ^{cd}\left( \Gamma
_{M}\right) _{bc}\left( \Gamma ^{N}\right) _{ad}+\left( \tilde{\Gamma}%
^{P}\right) ^{cd}\left( \Gamma _{M}\right) _{ac}\left( \Gamma ^{N}\right)
_{bd}\right) F_{PN}  \nonumber \\
&=&2\xi ^{a}\chi ^{b}\left( \Gamma ^{N}\right) _{ab}F_{NM}  \nonumber \\
&&
\end{eqnarray}
where we have used $F_{PN}=-F_{NP}$. To check the close for the fermionic
field is a little bit lengthy, but straightforward 
\begin{eqnarray}
\left[ \delta _{\xi },\delta _{\chi }\right] \Psi ^{c} &=&\underline{\xi
^{a}\chi ^{b}\left( \tilde{\Gamma}^{M}\right) ^{cd}\left( \Gamma ^{N}\right)
_{de}\left( \Gamma _{N}\right) _{ab}\nabla _{M}\Psi ^{e}}-\xi ^{a}\chi
^{b}\left( \tilde{\Gamma}^{M}\right) ^{cd}Q_{abde}\nabla _{M}\Psi ^{e} 
\nonumber \\
&=&2\xi ^{a}\chi ^{b}\left( \Gamma ^{M}\right) _{ab}\nabla _{M}\Psi ^{c}-\xi
^{a}\chi ^{b}\left( \tilde{\Gamma}^{M}\right) ^{cd}Q_{abde}\nabla _{M}\Psi
^{e},  \nonumber \\
&&
\end{eqnarray}
where we have used the fermionic equation of motion to simplify the
underlined term. Thus the supersymmetry closes iff $Q_{abde}=0.$ This is
true both for the abelian and nonabelian cases. In summary to \emph{close
the algebra} \emph{and} \emph{to have an invariant Lagrangian} $Q_{abde}$ 
\emph{must vanishes }for both the abelian and the non-abelian case.

\section{Representation of the Supersymmetry Algebra}

For a theory to be supersymmetric, it is necessary that its particle content
form a representation of the supersymmetry algebra. Using the gamma matrices
representation given in the previous section, we show how to describe the
representations of our supersymmetry algebra in $d=3,4,6,10$. From (\ref
{salg}), we deduce our supersymmetry algebra as 
\begin{equation}
\fbox{$\{Q_{a},Q_{b}\}=2\left( \Gamma \right) _{ab}\partial _{\mu }\,\equiv
-2\left( \Gamma \right) _{ab}P_{\mu }$}  \label{susy-noC}
\end{equation}
where $Q_{a}$ are the supersymmetry generators and transform as spin-half
operators under the angular momentum algebra. Moreover, the supersymmetry
generators commute with the momentum operator $P_{\mu }$ and hence, with $%
P^{2}$. Therefore, all states in a given representation of the algebra have
the same mass. For our case we will be concerned with the massless
representation only. For massless states, we can always go to a frame where $%
P^{\mu }=M(1,..,1)$. Then the supersymmetry algebra becomes 
\[
\fbox{$\{Q_{a},Q_{b}\}=\left( 
\begin{array}{lr}
0 & 0 \\ 
0 & 4M
\end{array}
\right) =-2M\left( \Gamma _{+}\right) _{ab}$}\,. 
\]
where 
\begin{equation}
\left( \Gamma _{+}\right) =\left( \Gamma _{0}\right) +\left( \Gamma
_{d-1}\right) =-2\left( 
\begin{array}{ll}
0 & 0 \\ 
0 & \mathbf{1}_{\frac{n}{2}}
\end{array}
\right) .
\end{equation}
It is convenient to rescale our generators as 
\[
a_{\mu }=\frac{1}{\sqrt{2M}}Q_{\mu }\,, 
\]
for $\mu =0..(\frac{n}{2}-1)$, where $(\frac{n}{2}-1)=0,1,3,7$ for $%
d=3,4,6,10$ respectively. Then, the supersymmetry algebra takes the form 
\[
\{a_{\mu },a_{\nu }\}=-\delta _{\mu \nu }\,, 
\]
This is a Clifford algebra with $\frac{n}{2}$ generators. We can now proceed
in two different ways:

\begin{itemize}
\item[1-]  To retrieve the standard complex representation of our
supersymmetry algebra, we have to pair our generators 
\begin{equation}
\begin{array}{llll}
d=3 & a_{0} &  & 
\begin{array}{l}
\end{array}
\\ 
d=4 & b=a_{0}+ia_{1} & b^{*}=a_{0}-ia_{1} & 
\begin{array}{l}
\end{array}
\\ 
d=6 & \left\{ 
\begin{array}{l}
b_{1}=a_{0}+ia_{1} \\ 
b_{1}^{*}=a_{0}-ia_{1}
\end{array}
\right. & 
\begin{array}{l}
b_{2}=a_{3}+ia_{4} \\ 
b_{2}^{*}=a_{3}-ia_{4}
\end{array}
& 
\begin{array}{l}
\end{array}
\\ 
d=10 & \left\{ 
\begin{array}{c}
\begin{array}{l}
b_{1}=a_{0}+ia_{1} \\ 
b_{1}^{*}=a_{0}-ia_{1}
\end{array}
\\ 
\begin{array}{l}
b_{2}=a_{3}+ia_{4} \\ 
b_{2}^{*}=a_{3}-ia_{4}
\end{array}
\end{array}
\right. & 
\begin{array}{l}
\begin{array}{l}
b_{3}=a_{5}+ia_{6} \\ 
b_{3}^{*}=a_{5}-ia_{6}
\end{array}
\\ 
\begin{array}{l}
b_{4}=a_{7}+ia_{8} \\ 
b_{4}^{*}=a_{7}-ia_{8}
\end{array}
\end{array}
& 
\begin{array}{l}
\end{array}
\end{array}
\end{equation}
leading to $case\;I$%
\begin{equation}
\begin{array}{llllllll}
d=3 & a_{0} &  &  &  &  &  & 
\begin{array}{l}
\end{array}
\\ 
d=4 & \left\{ b,b\right\} & = & \left\{ b^{*},b^{*}\right\} & =0, & \left\{
b,b^{*}\right\} & =-2 & 
\begin{array}{l}
\end{array}
\\ 
d=6 & \left\{ b_{i},b_{j}\right\} & = & \left\{ b_{i}^{*},b_{j}^{*}\right\}
& =0, & \left\{ b_{i},b_{j}^{*}\right\} & =-2\delta _{ij} & 
\begin{array}{l}
\\ 
{\ i,j=1,2}
\end{array}
\\ 
d=10 & \left\{ b_{i},b_{j}\right\} & = & \left\{ b_{i}^{*},b_{j}^{*}\right\}
& =0, & \left\{ b_{i},b_{j}^{*}\right\} & =-2\delta _{ij} & 
\begin{array}{l}
\\ 
{\ i,j=1..4}
\end{array}
\end{array}
\end{equation}

\item[2-]  We can work with hypercomplex numbers, then we have $case\;II$%
\begin{equation}
\begin{array}{llll}
d=3 & a_{0} & 
\begin{array}{l}
\end{array}
&  \\ 
d=4 & \left\{ 
\begin{array}{l}
b=a_{0}+e_{1}a_{1} \\ 
b^{*}=a_{0}-e_{1}a_{1}
\end{array}
\right. & 
\begin{array}{l}
\\ 
e_{1}\;is\;the\;imaginary\;complex\;unit
\end{array}
&  \\ 
d=6 & \left\{ 
\begin{array}{l}
b_{1}=a_{0}+e_{i}a_{i} \\ 
b_{1}^{*}=a_{0}-e_{i}a_{i}
\end{array}
\right. & 
\begin{array}{l}
\\ 
e_{i}\;are\;imaginary\;quaternion\;units \\ 
{\ i=1\cdots 3}
\end{array}
&  \\ 
d=10 & \left\{ 
\begin{array}{l}
b_{1}=a_{0}+e_{i}a_{i} \\ 
b_{1}^{*}=a_{0}-e_{i}a_{i}
\end{array}
\right. & 
\begin{array}{l}
\\ 
e_{i}\;are\;imaginary\;octonionic\;units \\ 
{\ i=1\cdots 7}
\end{array}
& 
\end{array}
\;
\end{equation}
\end{itemize}

\section{The SSYM Auxiliary Fields Problem}

One may ask oneself: Why are auxiliary fields important? There are many
convincing reasons. Let us mention jut five of them.

1- Only in the presence of auxiliary fields is the supersymmetry manifest.
Indeed, when we use superspace, we \ can write our supersymmetric models in
a form clearly invariant under Lorentz transformation as well as
supersymmetry. The clearest example is the superspace formulation of $N=1$
supersymmetric theories.

2- As we saw in the first section of this chapter, the closure of the
supersymmetric algebra is achieved only by using the field equations of
motion. At the quantum level, the equations of motion get corrected and
consequently the supersymmetric \ algebra will be realized in a highly
non-trivial fashion, Look to \cite{superspace}.

3- The use of the Lagrangian formulation of field theory is usually
advocated on the basis of symmetries arguments. Hence making the symmetry
manifest is a priority.

4- Feynman diagrams with superfields explains naturally many of the
``miracle'' cancellations in supersymmetric models.

5- Supersymmetry is only constructed systematically when we use superspace.
In principle, the superspace formulation should provide us with all the
details, the supersymmetry transformations, the full interaction Lagrangian,
even the constraints must be derived in agreement with the super Bianchi
identities.

Unfortunately, we don't have a complete superspace treatment in $d>4$. The
number of auxiliary fields can be counted easily.

For SSYM, we have the following off--shell degrees of freedom (ofdf) 
\begin{equation}
\begin{array}{lllllll}
d\backslash ofdf &  & \Psi &  & A_{\mu } &  &  \\ 
3 &  & 2 & - & 2 & = & 0 \\ 
4 &  & 4 & - & 3 & = & 1 \\ 
6 &  & 8 & - & 5 & = & 3 \\ 
10 &  & 16 & - & 9 & = & 7
\end{array}
\end{equation}
The $d=3$ case is trivial in the sense that it contains no auxiliary fields.
For $d=4$, a superspace formalism based on $SL\left( 2,\Bbb{C}\right) $ is
needed to formulate our supersymmetric YM model in a manifestly invariant
way. Such a superspace treatment provides us automatically with the needed
single auxiliary field. In $d=6,10$, it is conjectured that an $SL\left( 2,%
\Bbb{H}\right) $ and $SL\left( 2,\Bbb{O}\right) $ are needed. Here, we try
to support this conjecture by a different argument and we hope that the
tools presented may help in the future to find the full superspace
formulation.

Using Evans ansatz \cite{r4}, SSYM are composed of: Gauge fields $A_{M}$,
spinors $\Psi ^{a}$, $j\left( =1..d-3\right) $ algebraic auxiliary fields $%
K^{j}$. The gauge group indices will be suppressed in the following. The
Lagrangian density is 
\begin{equation}
\mathcal{L}=-\frac{1}{4}F_{MN}F^{MN}+\frac{i}{2}\Psi ^{t}\Gamma ^{M}\nabla
_{M}\Psi +\frac{1}{2}K^{2},  \label{ddd}
\end{equation}
where $\nabla _{M}\equiv \partial _{M}+A_{M};$ $F_{MN}\equiv \left[ \nabla
_{M},\nabla _{N}\right] $ and the $\Gamma $ are given in (\ref{gmm}). The
Lagrangian is invariant up to a total derivative iff (\ref{ident}) holds.
Our supersymmetry transformations are\footnote{%
Contrary to \cite{r6}, we set $\Lambda _{P}=\tilde{\Lambda}^{P}$ from the
strart.} 
\begin{eqnarray}
\delta _{\eta }A_{M} &=&i\eta \Gamma _{M}\Psi ,  \nonumber \\
\delta _{\eta }\Psi ^{\alpha } &=&\frac{1}{2}F_{MN}\left( \Gamma _{MN}~\eta
\right) ^{\alpha }+K^{j}\left( \Lambda _{j}\right) _{\beta }^{\alpha }\eta
^{\beta },  \nonumber \\
\delta _{\eta }K_{j} &=&i\left( \Gamma ^{M}\nabla _{M}\Psi \right) _{\alpha
}\left( \Lambda _{j}\right) _{\beta }^{\alpha }\eta ^{\beta },  \nonumber \\
&&  \label{dddd}
\end{eqnarray}
where $\Lambda _{j}$ are some real antisymmetric matrices $\left( \Lambda
_{j}\right) ^{t}=-\left( \Lambda _{j}\right) \footnote{%
We will mention shortly how to relax this condition.}$ and Lorentz
transformations are generated by $\Gamma _{MN}\equiv \tilde{\Gamma}%
_{[M}\Gamma _{N]}$. Imposing the closure of the supersymmetry infinitesimal
transformations 
\begin{equation}
\left[ \delta _{\epsilon },\delta _{\eta }\right] =2i\epsilon ^{t}\Gamma
^{M}\eta \partial _{M}\;.  \label{alg}
\end{equation}
The closure on $A_{M}$ yields 
\begin{equation}
\Gamma _{M}\Lambda _{j}-\Lambda _{j}\Gamma _{M}=0.  \label{c1}
\end{equation}
In addition to this condition the closure on $K^{j}$ also requires 
\begin{equation}
\Lambda _{j}\Lambda _{h}+\Lambda _{h}\Lambda _{j}=-2\delta _{jh}.  \label{c2}
\end{equation}
While closure on the fermionic field $\Psi ^{\alpha }$ holds iff 
\[
\left( \Gamma ^{M}\right) _{\alpha \beta }\left( \tilde{\Gamma}_{M}\right)
^{\gamma \delta }=2\delta _{(\alpha }^{\gamma }\delta _{\beta )}^{\delta
}+2\left( \Lambda _{j}\right) _{(\alpha }^{\gamma }\left( \Lambda
_{j}\right) _{\beta )}^{\delta }. 
\]
Now, we continue in a different way to Evans. To construct $\Lambda _{j}$,
we first notice from (\ref{c2}) that the $\Lambda _{j}$ form a real Clifford
algebra, and from (\ref{c1}) that they commute with our space-time $\Gamma
_{M}$ Clifford algebra. The solution of the auxiliary field problem for $%
d=3,4,6$ dimensions is then simply 
\begin{equation}
\Lambda _{j}=\left( 
\begin{array}{cc}
1|\Bbb{E}_{j} & 0 \\ 
0 & 1|\Bbb{E}_{j}
\end{array}
\right) ,  \label{lam}
\end{equation}
because 
\begin{equation}
\left\{ 1|\Bbb{E}_{j},1|\Bbb{E}_{h}\right\} =-2\delta _{jh},
\end{equation}
and 
\begin{equation}
\left[ \Bbb{E}_{j},1|\Bbb{E}_{h}\right] =0.
\end{equation}
Of course this solution is not unique. For example, if someone had started
with $1|\Gamma _{M}$, he would have found $\Lambda _{j}=\left( 
\begin{array}{ll}
\Bbb{E}_{j} & 0 \\ 
0 & \Bbb{E}_{j}
\end{array}
\right) $. In general, we can relax the conditions \ of antisymmetricity of $%
\Lambda $ and the symmetricity of $\Gamma .$ One writes any $\Gamma $ and
expand it in terms left/right action $\left( \Bbb{E}_{i,}1|\Bbb{E}_{j},\Bbb{E%
}_{m}|\Bbb{E}_{n}\right) $ then the $\Lambda $ will be given in terms of $%
\left( 1|\Bbb{E}_{i,}\Bbb{E}_{j},\Bbb{E}_{n}|\Bbb{E}_{m}\right) $.

For $d=10$, working with octonions the situation is different. From chapter
3, we know that octonionic left and right action commutes only when applied
to $\varphi $, 
\begin{equation}
\varphi ^{t}\left[ \Bbb{E}_{j},1|\Bbb{E}_{h}\right] \varphi =0,
\end{equation}
and $\varphi $ is just an 8 dimensional column matrix. Up to now, we have
not restricted $\varphi $ by any other conditions. With two different $%
\varphi $, $\left( \varphi ^{\left( 1\right) },\varphi ^{\left( 2\right)
}\right) $, we impose now the conditions that $\varphi ^{\left( i\right) }$
be fermionic fields. We express our 16 dimensional Grassmanian variables $%
\epsilon ,\eta $ of eqn.(\ref{alg}) in terms of $\varphi $, 
\begin{eqnarray}
\epsilon &=&\eta ^{t}  \nonumber \\
&\Downarrow &  \nonumber \\
\epsilon &=&\left( 
\begin{array}{ll}
\varphi ^{\left( 1\right) } & \varphi ^{\left( 2\right) }
\end{array}
\right) ;\quad \quad \eta =\left( 
\begin{array}{l}
\varphi ^{\left( 1\right) } \\ 
\varphi ^{\left( 2\right) }
\end{array}
\right)  \label{cond1}
\end{eqnarray}
We now rederive (\ref{alg}) for the octonions. The closure conditions of our
algebra, \emph{without omitting the Grassmanian variables} are 
\begin{eqnarray}
\eta ^{t}\left( \Gamma _{M}\Lambda _{j}-\Lambda _{j}\Gamma _{M}\right) \eta
&=&0,  \nonumber \\
\eta ^{t}\left( \Lambda _{j}\Lambda _{h}+\Lambda _{h}\Lambda _{j}\right)
\eta &=&\eta ^{t}\left( -2\delta _{jh}\right) \eta ,  \nonumber \\
\eta ^{t}\left( \left( \Gamma ^{M}\right) _{\alpha \beta }\left( \tilde{%
\Gamma}_{M}\right) ^{\gamma \delta }\right) \eta &=&\eta ^{t}\left( 2\delta
_{(\alpha }^{\gamma }\delta _{\beta )}^{\delta }+2\left( \Lambda _{j}\right)
_{(\alpha }^{\gamma }\left( \Lambda _{j}\right) _{\beta )}^{\delta }\right)
\eta ,  \label{algoct}
\end{eqnarray}
which are satisfied for the octonionic representation 
\begin{equation}
\left( \Gamma _{j}\right) _{ab}=\left( 
\begin{array}{cc}
0 & \Bbb{E}_{j} \\ 
-\Bbb{E}_{j} & 0
\end{array}
\right) ,\quad \quad \Lambda _{j}=\left( 
\begin{array}{cc}
1|\Bbb{E}_{j} & 0 \\ 
0 & 1|\Bbb{E}_{j}
\end{array}
\right) .
\end{equation}
By interchanging left/right action, we have different solutions as in the
quaternionic case. In summary, while the fermionic fields couple to
left/right action through the gamma matrices, the auxiliary fields couple to
right/left action through the $\Lambda $. For the octonionic case \emph{the
presence of the Grassmanian variables is essential.} Contrary to the
standard supersymmetry transformation, our Grassman variables are the same,
which is identical to the result obtained by Berkovits in \cite{r7}. \
According to Evans \cite{r8}, the attractive feature of this scheme is that
the Lagrangian (\ref{ddd}) and the transformation (\ref{dddd}) are
manifestly invariant under the generalized Lorentz group $SO\left(
1,9\right) $. In our formulation, we can show some additional
characteristic. In some cases, the \ (\ref{cond1}) condition may be relaxed,
for equal $j$ or $h$ (no summation) 
\begin{equation}
\left. 
\begin{array}{c}
\varphi ^{t}\;\Bbb{E}_{j}\;\left[ \Bbb{E}_{j},1|\Bbb{E}_{h}\right] \varphi
\\ 
\varphi ^{t}\;\;1|\Bbb{E}_{i}\;\;\left[ \Bbb{E}_{j},1|\Bbb{E}_{h}\right]
\varphi \\ 
\varphi ^{t}\;E_{h}\;\left[ \Bbb{E}_{j},1|\Bbb{E}_{h}\right] \varphi \\ 
\varphi ^{t}\;\;1|E_{h}\;\;\left[ \Bbb{E}_{j},1|\Bbb{E}_{h}\right] \varphi
\end{array}
\right\} =0.
\end{equation}
i.e. relating $\epsilon $ and $\eta $ by an $S^{7}$ is also allowed.

Now, Let us show what will happen to $spin\left( 1,9\right) $ when we
transform it to $soft\;spin\left( 1,9\right) $ 
\begin{eqnarray}
soft\;spin\left( 1,9\right) &\sim &\left[ \Gamma _{i},\Gamma _{j}\right] \eta
\nonumber \\
&=&\left[ \left( 
\begin{array}{cc}
0 & \Bbb{E}_{i} \\ 
-\Bbb{E}_{i} & 0
\end{array}
\right) ,\left( 
\begin{array}{cc}
0 & \Bbb{E}_{j} \\ 
-\Bbb{E}_{j} & 0
\end{array}
\right) \right] \left( 
\begin{array}{l}
\varphi ^{\left( 1\right) } \\ 
\varphi ^{\left( 2\right) }
\end{array}
\right)  \nonumber \\
&=&-\left( 
\begin{array}{cc}
0 & \left[ \Bbb{E}_{i},\Bbb{E}_{j}\right] \\ 
\left[ \Bbb{E}_{i},\Bbb{E}_{j}\right] & 0
\end{array}
\right) \left( 
\begin{array}{l}
\varphi ^{\left( 1\right) } \\ 
\varphi ^{\left( 2\right) }
\end{array}
\right)  \nonumber \\
&=&-\left( 
\begin{array}{ll}
0 & f_{ijk}^{\left( +\right) }\left( \varphi ^{\left( 2\right) }\right) \Bbb{%
E}_{k} \\ 
f_{ijk}^{\left( +\right) }\left( \varphi ^{\left( 1\right) }\right) \Bbb{E}%
_{k} & 0
\end{array}
\right) \left( 
\begin{array}{l}
\varphi ^{\left( 1\right) } \\ 
\varphi ^{\left( 2\right) }
\end{array}
\right) .  \nonumber \\
&&
\end{eqnarray}

Lastly, let us make some comments about a possible superspace. It seems that
the best way to find the $d=6,10$ superspace for SSYM is by defining some
quaternionic and octonionic Grassmann variables that decompose the
corresponding spinors into an $SL\left( 2,H\right) $ and an $SL\left(
2,soft\;S^{7}\right) $ respectively 
\begin{equation}
\left\{ \theta _{\alpha },\theta _{\beta }\right\} =\left\{ \bar{\theta}_{%
\dot{\alpha}},\bar{\theta}_{\dot{\beta}}\right\} =\left\{ \theta _{\alpha },%
\bar{\theta}_{\dot{\beta}}\right\} =0,
\end{equation}
where $\alpha =1,2$ over quaternions or octonions. We know that the
supersymmetry generators $Q_{\alpha }$ are derived from right multiplication 
\begin{eqnarray}
{Q}_{\alpha } &=&\left( \partial _{\alpha }-1|\Gamma _{\alpha \dot{\beta}%
}^{\mu }\bar{\theta}^{\dot{\beta}}P_{\mu }\right)  \\
{Q}^{\alpha } &=&\left( -\partial ^{{\alpha }}+\bar{\theta}_{\dot{\beta}}1|%
\tilde{\Gamma}^{\mu \dot{\beta}\alpha }P_{\mu }\right) 
\end{eqnarray}
also 
\begin{equation}
{\bar{Q}}^{\dot{\alpha}}=\left( \partial ^{\dot{\alpha}}-1|\tilde{\Gamma}%
^{\mu \dot{\alpha}\alpha }\theta _{\alpha }P_{\mu }\right) 
\end{equation}
\begin{equation}
{\bar{Q}}_{\dot{\alpha}}=\left( -\partial _{\dot{\alpha}}+\theta ^{\alpha
}1|\Gamma _{\alpha \dot{\alpha}}P_{\mu }\right) 
\end{equation}
whereas the covariant derivative $D_{\alpha }$ are obtained by left action 
\begin{eqnarray}
{D}_{\alpha } &=&\left( \partial _{\alpha }+\Gamma _{\alpha \dot{\beta}%
}^{\mu }\bar{\theta}^{\dot{\beta}}P_{\mu }\right)  \\
{D}^{\alpha } &=&\left( -\partial ^{{\alpha }}-\bar{\theta}_{\dot{\beta}}%
\tilde{\Gamma}^{\mu \dot{\beta}\alpha }P_{\mu }\right) 
\end{eqnarray}
also 
\begin{equation}
{\bar{D}}^{\dot{\alpha}}=\left( \partial ^{\dot{\alpha}}+\tilde{\Gamma}^{\mu 
\dot{\alpha}\alpha }\theta _{\alpha }P_{\mu }\right) 
\end{equation}
\begin{equation}
{\bar{D}}_{\dot{\alpha}}=\left( -\partial _{\dot{\alpha}}-\theta ^{\alpha
}\Gamma _{\alpha \dot{\alpha}}P_{\mu }\right) 
\end{equation}
Leading to a result acceptable but different from the standard $N=1$, $d=4$
superspace, 
\begin{eqnarray}
\{Q_{\alpha },\bar{Q}_{\dot{\alpha}}\} &=&-2\left( 1|\Gamma _{\alpha {\dot{%
\alpha}}}^{\mu }\right) P_{\mu }\;,  \nonumber \\
\{Q_{\alpha },Q_{\beta }\} &=&\{\bar{Q}_{\dot{\alpha}},\bar{Q}_{\dot{\beta}%
}\}\;=\;0\;,  \nonumber
\end{eqnarray}
\begin{eqnarray}
\{D_{\alpha },\bar{D}_{\dot{\alpha}}\} &=&2\Gamma _{\alpha {\dot{\alpha}}%
}^{\mu }P_{\mu }\;,  \nonumber \\
\{D_{\alpha },D_{\beta }\} &=&\{\bar{D}_{\dot{\alpha}},\bar{D}_{\dot{\beta}%
}\}\;=\;0\;,  \nonumber
\end{eqnarray}
and iff left and right action commute, we restore 
\begin{eqnarray}
\{Q_{\alpha },\bar{D}_{\dot{\alpha}}\} &=&\{D_{\alpha },\bar{Q}_{\dot{\alpha}%
}\}=0\;,  \nonumber \\
\{Q_{\alpha },D_{\beta }\} &=&\{\bar{D}_{\dot{\alpha}},\bar{Q}_{\dot{\beta}%
}\}\;=\;0\;.  \nonumber
\end{eqnarray}
On the other hand for octonions we would have the weaker conditions, 
\begin{eqnarray}
\left( 
\begin{array}{ll}
\varphi ^{\left( 1\right) } & \varphi ^{\left( 2\right) }
\end{array}
\right) \{Q_{\alpha },\bar{D}_{\dot{\alpha}}\}\left( 
\begin{array}{l}
\varphi ^{\left( 1\right) } \\ 
\varphi ^{\left( 2\right) }
\end{array}
\right)  &=&\left( 
\begin{array}{ll}
\varphi ^{\left( 1\right) } & \varphi ^{\left( 2\right) }
\end{array}
\right) \{D_{\alpha },\bar{Q}_{\dot{\alpha}}\}\left( 
\begin{array}{l}
\varphi ^{\left( 1\right) } \\ 
\varphi ^{\left( 2\right) }
\end{array}
\right) =0\;,  \nonumber \\
\left( 
\begin{array}{ll}
\varphi ^{\left( 1\right) } & \varphi ^{\left( 2\right) }
\end{array}
\right) \{Q_{\alpha },D_{\beta }\}\left( 
\begin{array}{l}
\varphi ^{\left( 1\right) } \\ 
\varphi ^{\left( 2\right) }
\end{array}
\right)  &=&\left( 
\begin{array}{ll}
\varphi ^{\left( 1\right) } & \varphi ^{\left( 2\right) }
\end{array}
\right) \{\bar{D}_{\dot{\alpha}},\bar{Q}_{\dot{\beta}}\}\left( 
\begin{array}{l}
\varphi ^{\left( 1\right) } \\ 
\varphi ^{\left( 2\right) }
\end{array}
\right) =\;0\;.  \nonumber
\end{eqnarray}
The commutation of left and right actions is not just needed for
associativity but for the invariance under supersymmetry transformation 
\begin{equation}
\delta _{\xi }\equiv \xi Q+\bar{\xi}\bar{Q}
\end{equation}
because only the associativity ensures 
\begin{equation}
\left( 
\begin{array}{ll}
\varphi ^{\left( 1\right) } & \varphi ^{\left( 2\right) }
\end{array}
\right) \left[ \delta _{\xi },D_{\alpha }\right] \left( 
\begin{array}{l}
\varphi _{1} \\ 
\varphi _{2}
\end{array}
\right) =\left( 
\begin{array}{ll}
\varphi ^{\left( 1\right) } & \varphi ^{\left( 2\right) }
\end{array}
\right) \left[ \delta _{\xi },\bar{D}_{\dot{\alpha}}\right] \left( 
\begin{array}{l}
\varphi _{1} \\ 
\varphi _{2}
\end{array}
\right) =0,
\end{equation}
since $\delta _{\xi }$ is left action and $D_{\alpha }$ is right action
which is a very important relation in the standard $N=1$ superspace for the
invariance of the Lagrangian under supersymmetry transformation. We hope to
return to this point in a future work.

\chapter{Conclusions}


In this thesis, we have presented a systematic study of the hidden faithful
Clifford algebraic structure in the different types of ring division
algebras. This relationship had been elaborated by going beyond octonions to
hexagonions. We have then dedicated a complete chapter to octonions. They
are not as useless as often believed.. They may be safely employed once the
non-associativity has been bypassed. The necessary ingredients are:

\begin{itemize}
\item  \emph{Fixing the direction of action by introducing the }$\delta $%
\emph{\ operator.}

\item  \emph{Closing the }$\delta $\emph{\ algebra by using structure
functions }$f_{ijk}^{\left( +\right) }\left( \varphi \right) $\emph{.}

\item  \emph{Matrix representation of the }$\delta $\emph{\ algebra. The }$%
\Bbb{E}$\emph{\ or }$\Bbb{E}\left( \varphi \right) $\emph{\ can be found and
their structure functions can be computed easily.}
\end{itemize}

During this analysis, we have introduced and discussed the soft seven
sphere. There maybe different applications of the soft seven sphere in
physics \cite{berk1} \cite{berk2} \cite{brink} \cite{spt}. We have given two
such cases where the ring division algebras occupies a special position.
Self-duality and SSYM are two promising places where the soft seven sphere
proves to be useful and indeed essential. In our formulation, \emph{we find
a new eight dimensional feature, that had never appeared before, the
existence of an infinite family of dualities. By moving over the gauged
seven sphere, we define \ new conditions and we have new solutions. We have
parameterized all these conditions and solutions in terms of the coordinate
system over the gauged seven sphere.}

By defining the soft--duality condition, we have tried to retain as much as
possible of the four dimensional case. We then showed how new solutions of
the GKS condition can be easily found .

\emph{For SSYM, the new and old off--shell formulations can be rederived in
a systematic and uniform fashion.} We believe that the interplay between
left and right ring division operators is not a coincidence but an intrinsic
property of supersymmetry that needs further study. By interchanging left
and right action, we have many different solutions. Again the \ octonionic
ten dimensional case is very special. It will be interesting to apply the
ideas presented here into the GS context.

We hope that this work constitute a first step in the correct direction for
further application of the soft seven sphere algebra.

\appendix


\chapter{The First Appendix}


In this appendix we give the translation rules between octonionic left-right
barred operators and $8\times 8$ real matrices. In order to simplify our
presentation we introduce the following notation:

\begin{eqnarray}
\{~a,\;b,\;c,\;d~\}_{(1)} &\equiv &\left( 
\begin{array}{cccc}
{a} & {0} & {0} & {0} \\ 
0 & b & 0 & 0 \\ 
0 & 0 & c & 0 \\ 
0 & 0 & 0 & d
\end{array}
\right) ,\;\;\{~a,\;b,\;c,\;d~\}_{(2)}\equiv \left( 
\begin{array}{cccc}
{0} & {a} & {0} & {0} \\ 
b & 0 & 0 & 0 \\ 
0 & 0 & 0 & c \\ 
0 & 0 & d & 0
\end{array}
\right) ,  \nonumber \\
&& \\
\{~a,\;b,\;c,\;d~\}_{(3)} &\equiv &\left( 
\begin{array}{cccc}
{0} & {0} & {a} & {0} \\ 
0 & 0 & 0 & b \\ 
c & 0 & 0 & 0 \\ 
0 & d & 0 & 0
\end{array}
\right) ,\;\;\{~a,\;b,\;c,\;d~\}_{(4)}\equiv \left( 
\begin{array}{cccc}
{0} & {0} & {0} & {a} \\ 
0 & 0 & b & 0 \\ 
0 & c & 0 & 0 \\ 
d & 0 & 0 & 0
\end{array}
\right) ,  \nonumber \\
&&
\end{eqnarray}

where $a,\;b,\;c,\;d$ and $0$ represent $2\times 2$ real matrices. As
elsewhere by $\sigma _{1}$, $\sigma _{2}$, $\sigma _{3}$ we mean the
standard Pauli matrices: 
\begin{equation}
\sigma _{1}=\left( 
\begin{array}{cc}
{0} & {1} \\ 
{1} & {0}
\end{array}
\right) \quad ,\quad \sigma _{2}=\left( 
\begin{array}{cc}
{0} & {-i} \\ 
{i} & {0}
\end{array}
\right) \quad ,\quad \sigma _{3}=\left( 
\begin{array}{cc}
{1} & {0} \\ 
{0} & {-1}
\end{array}
\right) \quad .
\end{equation}

The only necessary translation rules that we need to know explicitly are the
following

\begin{center}
\begin{tabular}{llllll}
&  &  &  &  &  \\ 
${e_{1}}$ & $~\leftrightarrow ~\{$ & $-i\sigma _{2}$, & ~$-i\sigma _{2}$, & ~%
$-i\sigma _{2}$, & ~$i\sigma _{2}~\}_{(1)}$ \\ 
${e_{2}}$ & $~\leftrightarrow ~\{$ & $-\sigma _{3}$, & ~$\sigma _{3}$, & ~$-%
\mathbf{1}$, & ~$\mathbf{1}~\}_{(2)}$ \\ 
${e_{3}}$ & $~\leftrightarrow ~\{$ & $-\sigma _{1}$, & ~$\sigma _{1}$, & ~$%
-i\sigma _{2}$, & ~$-i\sigma _{2}~\}_{(2)}$ \\ 
${e_{4}}$ & $~\leftrightarrow ~\{$ & $-\sigma _{3}$, & ~$\mathbf{1}$, & ~$%
\sigma _{3}$, & ~$-\mathbf{1}~\}_{(3)}$ \\ 
${e_{5}}$ & $~\leftrightarrow ~\{$ & $-\sigma _{1}$, & ~$i\sigma _{2}$, & ~$%
\sigma _{1}$, & ~$i\sigma _{2}~\}_{(3)}$ \\ 
${e_{6}}$ & $~\leftrightarrow ~\{$ & $-\mathbf{1}$, & ~$-\sigma _{3}$, & ~$%
\sigma _{3}$, & ~$\mathbf{1}~\}_{(4)}$ \\ 
${e_{7}}$ & $~\leftrightarrow ~\{$ & $-i\sigma _{2}$, & ~$-\sigma _{1}$, & ~$%
\sigma _{1}$, & ~$-i\sigma _{2}~\}_{(4)}$ \\ 
${1\mid e_{1}}$ & $~\leftrightarrow ~\{$ & $-i\sigma _{2}$, & ~$i\sigma _{2}$%
, & ~$i\sigma _{2}$, & ~$-i\sigma _{2}~\}_{(1)}$ \\ 
${1\mid e_{2}}$ & $~\leftrightarrow ~\{$ & $-\mathbf{1}$, & ~$\mathbf{1}$, & 
~$\mathbf{1}$, & ~$-\mathbf{1}~\}_{(2)}$ \\ 
${1\mid e_{3}}$ & $~\leftrightarrow ~\{$ & $-i\sigma _{2}$, & ~$-i\sigma
_{2} $, & ~$i\sigma _{2}$, & ~$i\sigma _{2}~\}_{(2)}$ \\ 
${1\mid e_{4}}$ & $~\leftrightarrow ~\{$ & $-\mathbf{1}$, & ~$-\mathbf{1}$,
& ~$\mathbf{1}$, & ~$\mathbf{1}~\}_{(3)}$ \\ 
${1\mid e_{5}}$ & $~\leftrightarrow ~\{$ & $-i\sigma _{2}$, & ~$-i\sigma
_{2} $, & ~$-i\sigma _{2}$, & ~$-i\sigma _{2}~\}_{(3)}$ \\ 
${1\mid e_{6}}$ & $~\leftrightarrow ~\{$ & $-\sigma _{3}$, & ~$\sigma _{3}$,
& ~$-\sigma _{3}$, & ~$\sigma _{3}~\}_{(4)}$ \\ 
${1\mid e_{7}}$ & $~\leftrightarrow ~\{$ & $-\sigma _{1}$, & ~$\sigma _{1}$,
& ~$-\sigma _{1}$, & ~$\sigma _{1}~\}_{(4)}$%
\end{tabular}
\end{center}

The remaining rules can be easily constructed remembering that 
\begin{eqnarray*}
{e_{m}\mid e_{m}} &&~~\leftrightarrow ~1|E_{m}\;\;E_{m}\quad , \\
&&~\leftrightarrow ~~E_{m}\;\;1|E_{m}\quad , \\
{e_{m}~\mathbf{)}~e_{n}} &&~~\leftrightarrow ~1|E_{n}\;\;E_{m}\quad , \\
{e_{m}~\mathbf{(}~e_{n}} &&~~\leftrightarrow ~E_{m}\;\;1|E_{n}\quad .
\end{eqnarray*}
For example, 
\[
\begin{array}{llll}
{e_{1}\mid e_{1}} & \leftrightarrow & ~\left( 
\begin{array}{cccc}
-i\sigma _{2} & 0 & 0 & 0 \\ 
0 & -i\sigma _{2} & 0 & 0 \\ 
0 & 0 & -i\sigma _{2} & 0 \\ 
0 & 0 & 0 & i\sigma _{2}
\end{array}
\right) & \left( 
\begin{array}{cccc}
-i\sigma _{2} & 0 & 0 & 0 \\ 
0 & i\sigma _{2} & 0 & 0 \\ 
0 & 0 & i\sigma _{2} & 0 \\ 
0 & 0 & 0 & -i\sigma _{2}
\end{array}
\right) \\ 
&  & =\{\;-\mathbf{1},\;\mathbf{1},\;\mathbf{1},\;\mathbf{1}\;\}_{(1)}\quad ,
& 
\end{array}
~ 
\]
\[
\begin{array}{llll}
{e_{3}~\mathbf{)}~e_{1}} & \leftrightarrow & \left( 
\begin{array}{cccc}
-i\sigma _{2} & 0 & 0 & 0 \\ 
0 & i\sigma _{2} & 0 & 0 \\ 
0 & 0 & i\sigma _{2} & 0 \\ 
0 & 0 & 0 & -i\sigma _{2}
\end{array}
\right) & \left( 
\begin{array}{cccc}
0 & -\sigma _{1} & 0 & 0 \\ 
\sigma _{1} & 0 & 0 & 0 \\ 
0 & 0 & 0 & -i\sigma _{2} \\ 
0 & 0 & -i\sigma _{2} & 0
\end{array}
\right) \\ 
&  & =~\{\;\sigma _{3},\;\sigma _{3},\;\mathbf{1},\;-\mathbf{1}%
\;\}_{(2)}\quad , & 
\end{array}
\]
and 
\[
\begin{array}{llll}
{e_{3}~(~e_{1}}~ & \leftrightarrow & ~\left( 
\begin{array}{cccc}
0 & -\sigma _{1} & 0 & 0 \\ 
\sigma _{1} & 0 & 0 & 0 \\ 
0 & 0 & 0 & -i\sigma _{2} \\ 
0 & 0 & -i\sigma _{2} & 0
\end{array}
\right) & \left( 
\begin{array}{cccc}
-i\sigma _{2} & 0^{0} & 0 &  \\ 
0 & i\sigma _{2} & 0 & 0 \\ 
0 & 0 & i\sigma _{2} & 0 \\ 
0 & 0 & 0 & -i\sigma _{2}
\end{array}
\right) \\ 
&  & =~\{\;\sigma _{3},\;\sigma _{3},\;-\mathbf{1},\;\mathbf{1}\}_{(2)}\quad
. & 
\end{array}
\]
Following this procedure any matrix representation of right/left barred
operators can be obtained. Using Mathematica~\cite{math}, we have proved the
linear independence of the 64 elements which represent the most general
octonionic operator 
\[
\Bbb{O}_{0}+\sum_{m=1}^{7}\Bbb{O}_{m}~)~e_{m}\quad . 
\]
So our barred operators form a complete basis for any $8\times 8$ real
matrix and this establishes the isomorphism between $GL(8,\Bbb{R})$ and
barred octonions.

\newpage

\chapter{The Second Appendix}


We have given the action of barred operators on the octonionic functions
(states) 
\[
\psi =\psi _{1}+e_{2}\psi _{2}+e_{4}\psi _{3}+e_{6}\psi _{4}\quad \quad
[~\psi _{1,...,4}\in C(1,\;e_{1})~]\quad . 
\]
In the following we will use the notation 
\[
e_{2}~\rightarrow ~\{-\psi _{2},\;\psi _{1},\;-\psi _{4}^{*},\;\psi
_{3}^{*}\}\quad , 
\]
to indicate 
\[
e_{2}\psi ~=~-\psi _{2}+e_{2}\psi _{1}-e_{4}\psi _{4}^{*}+e_{6}\psi
_{3}^{*}\quad . 
\]

As occurred in the previous appendix we need to know only the action of the
barred operators ${e_{m}}$ and ${1\mid e_{m}}$

\begin{center}
\begin{tabular}{llllll}
${e_{1}}$ & $~\rightarrow ~\{$ & $e_{1}\psi _{1}$, & ~$-e_{1}\psi _{2}$, & ~$%
-e_{1}\psi _{3}$, & ~$-e_{1}\psi _{4}~\}$ \\ 
${e_{2}}$ & $~\rightarrow ~\{$ & $-\psi _{2}$, & ~$\psi _{1}$, & ~$-\psi
_{4}^{*}$, & ~$\psi _{3}^{*}~\}$ \\ 
${e_{3}}$ & $~\rightarrow ~\{$ & $-e_{1}\psi _{2}$, & ~$-e_{1}\psi _{1}$, & ~%
$-e_{1}\psi _{4}^{*}$, & ~$e_{1}\psi _{3}^{*}~\}$ \\ 
${e_{4}}$ & $~\rightarrow ~\{$ & $-\psi _{3}$, & ~$\psi _{4}^{*}$, & ~$\psi
_{1}$, & ~$-\psi _{2}^{*}~\}$ \\ 
${e_{5}}$ & $~\rightarrow ~\{$ & $-e_{1}\psi _{3}$, & ~$e_{1}\psi _{4}^{*}$,
& ~$-e_{1}\psi _{1}$, & ~$-e_{1}\psi _{2}^{*}~\}$ \\ 
${e_{6}}$ & $~\rightarrow ~\{$ & $-\psi _{4}$, & ~$-\psi _{3}^{*}$, & ~$\psi
_{2}^{*}$, & ~$\psi _{1}~\}$ \\ 
${e_{7}}$ & $~\rightarrow ~\{$ & $e_{1}\psi _{4}$, & ~$e_{1}\psi _{3}^{*}$,
& ~$-e_{1}\psi _{2}^{*}$, & ~$e_{1}\psi _{1}~\}$ \\ 
${1\mid e_{1}}$ & $~\rightarrow ~\{$ & $e_{1}\psi _{1}$, & ~$e_{1}\psi _{2}$,
& ~$e_{1}\psi _{3}$, & ~$e_{1}\psi _{4}~\}$ \\ 
${1\mid e_{2}}$ & $~\rightarrow ~\{$ & $-\psi _{2}^{*}$, & ~$\psi _{1}^{*}$,
& ~$\psi _{4}^{*}$, & ~$-\psi _{3}^{*}~\}$ \\ 
${1\mid e_{3}}$ & $~\rightarrow ~\{$ & $e_{1}\psi _{2}^{*}$, & ~$-e_{1}\psi
_{1}^{*}$, & ~$e_{1}\psi _{4}^{*}$, & ~$-e_{1}\psi _{3}^{*}~\}$ \\ 
${1\mid e_{4}}$ & $~\rightarrow ~\{$ & $-\psi _{3}^{*}$, & ~$-\psi _{4}^{*}$,
& ~$\psi _{1}^{*}$, & ~$\psi _{2}^{*}~\}$ \\ 
${1\mid e_{5}}$ & $~\rightarrow ~\{$ & $e_{1}\psi _{3}^{*}$, & ~$-e_{1}\psi
_{4}^{*}$, & ~$-e_{1}\psi _{1}^{*}$, & ~$e_{1}\psi _{2}^{*}~\}$ \\ 
${1\mid e_{6}}$ & $~\rightarrow ~\{$ & $-\psi _{4}^{*}$, & ~$\psi _{3}^{*}$,
& ~$-\psi _{2}^{*}$, & ~$\psi _{1}^{*}~\}$ \\ 
${1\mid e_{7}}$ & $~\rightarrow ~\{$ & $-e_{1}\psi _{4}^{*}$, & ~$-e_{1}\psi
_{3}^{*}$, & ~$e_{1}\psi _{2}^{*}$, & ~$e_{1}\psi _{1}^{*}~\}$%
\end{tabular}
\end{center}

From the previous correspondence rules we immediately obtain the others
barred operators. We give, as an example, the construction of the operator ${%
e_{4}~)~e_{7}}$. We know that 
\[
{e_{4}}~\rightarrow ~\{-\psi _{3},\;\psi _{4}^{*},\;\psi _{1},\;-\psi
_{2}^{*}\} 
\]
and 
\begin{equation}
{1\mid e_{7}}~\rightarrow ~\{-e_{1}\psi _{4}^{*},\;-e_{1}\psi
_{3}^{*},\;e_{1}\psi _{2}^{*},\;e_{1}\psi _{1}^{*}\}\quad .
\end{equation}
Combining these operators we find 
\[
\{-e_{1}(-\psi _{2}^{*})^{*},\;-e_{1}\psi _{1}^{*},\;e_{1}(\psi
_{4}^{*})^{*},\;e_{1}(-\psi _{3})^{*}\}\quad , 
\]
and so 
\[
{e_{4}~)~e_{7}}~\rightarrow ~\{e_{1}\psi _{2},\;-e_{1}\psi
_{1}^{*},\;e_{1}\psi _{4},\;-e_{1}\psi _{3}^{*}\}\quad . 
\]

As remarked at the end of subsection IV-b, we can extract the 32 basis
elements of $GL(4,\Bbb{C})$ directly by suitable combinations of the 64
basis elements of $GL(8,\Bbb{R})$. We must choose the combination which have
only $\mathbf{1}_{2\times 2}$ and $-i\sigma _{2}$ as matrix elements.
Nevertheless we must take care in manipulating our octonionic barred
operators. If we wish to extract from $GL(8,\Bbb{R})$ the 32 elements which
characterize $GL(4,\Bbb{C})$ we need to change the octonionic basis of $GL(8,%
\Bbb{R})$. In fact, the natural choice for the symplectic octonionic
representation 
\[
\psi ~=~(\varphi _{0}+e_{1}\varphi _{1})+e_{2}(\varphi _{2}+e_{1}\varphi
_{3})+e_{4}(\varphi _{4}+e_{1}\varphi _{5})+e_{6}(\varphi _{6}+e_{1}\varphi
_{7})\quad , 
\]
requires the following real counterpart 
\[
\tilde{\varphi}~=~\varphi _{0}+e_{1}\varphi _{1}+e_{2}\varphi
_{2}-e_{3}\varphi _{3}+e_{4}\varphi _{4}-e_{5}\varphi _{5}+e_{6}\varphi
_{6}+e_{7}\varphi _{7}\quad . 
\]
whereas we used in subsection IV-a the following basis 
\[
\varphi ~=~\varphi _{0}+e_{1}\varphi _{1}+e_{2}\varphi _{2}+e_{3}\varphi
_{3}+e_{4}\varphi _{4}+e_{5}\varphi _{5}+e_{6}\varphi _{6}+e_{7}\varphi
_{7}\quad . 
\]

The changes in the signs of $e_{3}\varphi _{3}$ and $e_{5}\varphi _{5}$
implies a modification in the generators of $GL(8,\Bbb{R})$. For example, ${%
e_{2}}$ and ${e_{3}~)~e_{1}}$ now read 
\[
{e_{2}}~\equiv ~\{-\mathbf{1},\;\mathbf{1},\;-\sigma _{3},\;\sigma
_{3}\}_{(2)}\quad \quad \mbox{and}\quad \quad {e_{3}~)~e_{1}}~\equiv ~\{%
\mathbf{1},\;\mathbf{1},\;\sigma _{3},\;-\sigma _{3}\}_{(2)}\quad . 
\]
i.e. the change of basis induces the following modifications 
\[
\mathbf{1}\rightleftharpoons \sigma _{3}\quad . 
\]
Their appropriate combination gives 
\[
\frac{{e_{2}}+{e_{3}~)~e_{1}}}{2}~\equiv ~\{0,\;\mathbf{1},\;0,\;0\}_{(2)}~~%
\stackrel{complexifing}{\longrightarrow }~~\left( 
\begin{array}{cccc}
0 & 0 & 0 & 0 \\ 
1 & 0 & 0 & 0 \\ 
0 & 0 & 0 & 0 \\ 
0 & 0 & 0 & 0
\end{array}
\right) \quad , 
\]
as required by eq.~(\ref{ct}).\newpage

\chapter{The Third Appendix}


The seven standard cycles are given by

\begin{eqnarray}
{{f}_{123}^{(+)}{(\varphi )}} &=&{\frac{{\varphi _{0}}^{2}-{\varphi _{6}}%
^{2}-{\varphi _{5}}^{2}+{\varphi _{2}}^{2}-{\varphi _{4}}^{2}+{\varphi _{1}}%
^{2}+{\varphi _{3}}^{2}-{\varphi _{7}}^{2}}{r^{2}},}  \nonumber
\label{ptor1} \\
{{f}_{145}^{(+)}{(\varphi )}} &=&{\frac{{\varphi _{0}}^{2}-{\varphi _{6}}%
^{2}+{\varphi _{4}}^{2}-{\varphi _{2}}^{2}+{\varphi _{1}}^{2}-{\varphi _{7}}%
^{2}+{\varphi _{5}}^{2}-{\varphi _{3}}^{2}}{r^{2}},}  \nonumber \\
{{f}_{176}^{(+)}{(\varphi )}} &=&{\frac{{\varphi _{0}}^{2}+{\varphi _{6}}%
^{2}-{\varphi _{4}}^{2}-{\varphi _{2}}^{2}+{\varphi _{1}}^{2}+{\varphi _{7}}%
^{2}-{\varphi _{5}}^{2}-{\varphi _{3}}^{2}}{r^{2}},}  \nonumber \\
{{f}_{246}^{(+)}{(\varphi )}} &=&{\frac{{\varphi _{0}}^{2}+{\varphi _{6}}%
^{2}+{\varphi _{4}}^{2}+{\varphi _{2}}^{2}-{\varphi _{1}}^{2}-{\varphi _{7}}%
^{2}-{\varphi _{5}}^{2}-{\varphi _{3}}^{2}}{{r^{2}}},}  \nonumber \\
{{f}_{257}^{(+)}{(\varphi )}} &=&{\frac{{\varphi _{0}}^{2}-{\varphi _{6}}%
^{2}-{\varphi _{4}}^{2}+{\varphi _{2}}^{2}-{\varphi _{1}}^{2}+{\varphi _{7}}%
^{2}+{\varphi _{5}}^{2}-{\varphi _{3}}^{2}}{{r^{2}}},}  \nonumber \\
{{f}_{347}^{(+)}{(\varphi )}} &=&{\frac{{\varphi _{0}}^{2}-{\varphi _{6}}%
^{2}+{\varphi _{4}}^{2}-{\varphi _{2}}^{2}-{\varphi _{1}}^{2}+{\varphi _{7}}%
^{2}-{\varphi _{5}}^{2}+{\varphi _{3}}^{2}}{r^{2}},}  \nonumber \\
{{f}_{365}^{(+)}{(\varphi )}} &=&{\frac{{\varphi _{0}}^{2}+{\varphi _{6}}%
^{2}-{\varphi _{4}}^{2}-{\varphi _{2}}^{2}-{\varphi _{1}}^{2}-{\varphi _{7}}%
^{2}+{\varphi _{5}}^{2}+{\varphi _{3}}^{2}}{r^{2}},}  \nonumber \\
&&
\end{eqnarray}
where 
\begin{equation}
r^{2}=(\varphi _{0}^{2}{\ +}\varphi _{1}^{2}{\ +}\varphi _{2}^{2}{\ +}%
\varphi _{3}^{2}{\ +}\varphi _{4}^{2}{\ +}\varphi _{5}^{2}{\ +}\varphi
_{6}^{2}{\ +}\varphi _{7}^{2})
\end{equation}
and the non-standard subset 
\begin{eqnarray}
{{f}_{124}^{(+)}{(\varphi )}} &=&+2{\frac{{\varphi _{0}}{\varphi _{7}}-{%
\varphi _{5}}{\varphi _{2}}+{\varphi _{6}}{\varphi _{1}}+{\varphi _{3}}{%
\varphi _{4}}}{r^{2}},}  \nonumber \\
{\quad {f}_{125}^{(+)}{(\varphi )}} &=&-2{\frac{{\varphi _{0}}{\varphi _{6}}-%
{\varphi _{3}}{\varphi _{5}}-{\varphi _{1}}{\varphi _{7}}-{\varphi _{2}}{%
\varphi _{4}}}{r^{2}},}  \nonumber \\
{{f}_{126}^{(+)}{(\varphi )}} &=&+2{\frac{{\varphi _{0}}{\varphi _{5}}-{%
\varphi _{1}}{\varphi _{4}}+{\varphi _{7}}{\varphi _{2}}+{\varphi _{3}}{%
\varphi _{6}}}{r^{2}},}  \nonumber \\
{\quad {f}_{127}^{(+)}{(\varphi )}} &=&-2{\frac{{\varphi _{0}}{\varphi _{4}}+%
{\varphi _{6}}{\varphi _{2}}+{\varphi _{1}}{\varphi _{5}}-{\varphi _{3}}{%
\varphi _{7}}}{r^{2}},}  \nonumber \\
{{f}_{143}^{(+)}{(\varphi )}} &=&+2{\frac{{\varphi _{0}}{\varphi _{6}}+{%
\varphi _{3}}{\varphi _{5}}+{\varphi _{2}}{\varphi _{4}}-{\varphi _{1}}{%
\varphi _{7}}}{r^{2}},}  \nonumber \\
{\quad {f}_{146}^{(+)}{(\varphi )}} &=&-2{\frac{{\varphi _{3}}{\varphi _{0}}-%
{\varphi _{4}}{\varphi _{7}}-{\varphi _{1}}{\varphi _{2}}-{\varphi _{5}}{%
\varphi _{6}}}{r^{2}},}  \nonumber \\
{{f}_{175}^{(+)}{(\varphi )}} &=&+2{\frac{{\varphi _{3}}{\varphi _{0}}-{%
\varphi _{1}}{\varphi _{2}}+{\varphi _{5}}{\varphi _{6}}+{\varphi _{4}}{%
\varphi _{7}}}{r^{2}},}  \nonumber \\
{\quad {f}_{247}^{(+)}{(\varphi )}} &=&-2{\frac{{\varphi _{0}}{\varphi _{1}}-%
{\varphi _{7}}{\varphi _{6}}-{\varphi _{4}}{\varphi _{5}}-{\varphi _{3}}{%
\varphi _{2}}}{r^{2}},}  \nonumber \\
{{f}_{147}^{(+)}{(\varphi )}} &=&+2{\frac{{\varphi _{2}}{\varphi _{0}}-{%
\varphi _{4}}{\varphi _{6}}+{\varphi _{5}}{\varphi _{7}}+{\varphi _{1}}{%
\varphi _{3}}}{r^{2}},}  \nonumber \\
{\quad {f}_{243}^{(+)}{(\varphi )}} &=&-2{\frac{{\varphi _{0}}{\varphi _{5}}+%
{\varphi _{1}}{\varphi _{4}}+{\varphi _{7}}{\varphi _{2}}-{\varphi _{3}}{%
\varphi _{6}}}{r^{2}},}  \nonumber \\
{{f}_{253}^{(+)}{(\varphi )}} &=&+2{\frac{{\varphi _{0}}{\varphi _{4}}-{%
\varphi _{1}}{\varphi _{5}}+{\varphi _{6}}{\varphi _{2}}+{\varphi _{3}}{%
\varphi _{7}}}{r^{2}},}  \nonumber \\
{\quad {f}_{173}^{(+)}{(\varphi )}} &=&-2{\frac{{\varphi _{0}}{\varphi _{5}}-%
{\varphi _{7}}{\varphi _{2}}-{\varphi _{3}}{\varphi _{6}}-{\varphi _{1}}{%
\varphi _{4}}}{r^{2}},}  \nonumber \\
{{f}_{245}^{(+)}{(\varphi )}} &=&+2{\frac{{\varphi _{3}}{\varphi _{0}}+{%
\varphi _{5}}{\varphi _{6}}-{\varphi _{4}}{\varphi _{7}}+{\varphi _{1}}{%
\varphi _{2}}}{r^{2}},}  \nonumber \\
{\quad {f}_{256}^{(+)}{(\varphi )}} &=&+2{\frac{{\varphi _{0}}{\varphi _{1}}-%
{\varphi _{3}}{\varphi _{2}}+{\varphi _{7}}{\varphi _{6}}+{\varphi _{4}}{%
\varphi _{5}}}{r^{2}},}  \nonumber \\
{{f}_{361}^{(+)}{(\varphi )}} &=&+2{\frac{{\varphi _{0}}{\varphi _{4}}+{%
\varphi _{3}}{\varphi _{7}}+{\varphi _{1}}{\varphi _{5}}-{\varphi _{6}}{%
\varphi _{2}}}{r^{2}},}  \nonumber \\
{\quad {f}_{362}^{(+)}{(\varphi )}} &=&-2{\frac{{\varphi _{0}}{\varphi _{7}}-%
{\varphi _{3}}{\varphi _{4}}-{\varphi _{5}}{\varphi _{2}}-{\varphi _{6}}{%
\varphi _{1}}}{r^{2}},}  \nonumber \\
{{f}_{345}^{(+)}{(\varphi )}} &=&-2{\frac{{\varphi _{2}}{\varphi _{0}}-{%
\varphi _{5}}{\varphi _{7}}-{\varphi _{1}}{\varphi _{3}}-{\varphi _{4}}{%
\varphi _{6}}}{r^{2}},}  \nonumber \\
{\quad {f}_{346}^{(+)}{(\varphi )}} &=&+2{\frac{{\varphi _{0}}{\varphi _{1}}+%
{\varphi _{3}}{\varphi _{2}}+{\varphi _{7}}{\varphi _{6}}-{\varphi _{4}}{%
\varphi _{5}}}{r^{2}},}  \nonumber \\
{{f}_{367}^{(+)}{(\varphi )}} &=&+2{\frac{{\varphi _{2}}{\varphi _{0}}+{%
\varphi _{4}}{\varphi _{6}}+{\varphi _{5}}{\varphi _{7}}-{\varphi _{1}}{%
\varphi _{3}}}{{r^{2}}},}  \nonumber \\
{\quad {f}_{135}^{(+)}{(\varphi )}} &=&-2{\frac{{\varphi _{0}}{\varphi _{7}}-%
{\varphi _{3}}{\varphi _{4}}+{\varphi _{6}}{\varphi _{1}}+{\varphi _{5}}{%
\varphi _{2}}}{r^{2}},}  \nonumber \\
{{f}_{156}^{(+)}{(\varphi )}} &=&-2{\frac{{\varphi _{2}}{\varphi _{0}}+{%
\varphi _{1}}{\varphi _{3}}+{\varphi _{4}}{\varphi _{6}}-{\varphi _{5}}{%
\varphi _{7}}}{r^{2}},}  \nonumber \\
{\quad {f}_{237}^{(+)}{(\varphi )}} &=&+2{\frac{{\varphi _{0}}{\varphi _{6}}-%
{\varphi _{2}}{\varphi _{4}}+{\varphi _{3}}{\varphi _{5}}+{\varphi _{1}}{%
\varphi _{7}}}{r^{2}},}  \nonumber \\
{{f}_{267}^{(+)}{(\varphi )}} &=&-2{\frac{{\varphi _{3}}{\varphi _{0}}-{%
\varphi _{5}}{\varphi _{6}}+{\varphi _{4}}{\varphi _{7}}+{\varphi _{1}}{%
\varphi _{2}}}{r^{2}},}  \nonumber \\
{\quad {f}_{357}^{(+)}{(\varphi )}} &=&+2{\frac{{\varphi _{0}}{\varphi _{1}}+%
{\varphi _{4}}{\varphi _{5}}-{\varphi _{7}}{\varphi _{6}}+{\varphi _{3}}{%
\varphi _{2}}}{r^{2}},}  \nonumber \\
{{f}_{456}^{(+)}{(\varphi )}} &=&-2{\frac{{\varphi _{0}}{\varphi _{7}}+{%
\varphi _{5}}{\varphi _{2}}+{\varphi _{3}}{\varphi _{4}}-{\varphi _{6}}{%
\varphi _{1}}}{r^{2}},}  \nonumber \\
{\quad {f}_{457}^{(+)}{(\varphi )}} &=&+2{\frac{{\varphi _{0}}{\varphi _{6}}+%
{\varphi _{2}}{\varphi _{4}}+{\varphi _{1}}{\varphi _{7}}-{\varphi _{3}}{%
\varphi _{5}}}{r^{2}},}  \nonumber \\
{{f}_{467}^{(+)}{(\varphi )}} &=&-2{\frac{{\varphi _{0}}{\varphi _{5}}+{%
\varphi _{1}}{\varphi _{4}}+{\varphi _{3}}{\varphi _{6}}-{\varphi _{7}}{%
\varphi _{2}}}{r^{2}},}  \nonumber \\
{\quad {f}_{567}^{(+)}{(\varphi )}} &=&+2{\frac{{\varphi _{0}}{\varphi _{4}}-%
{\varphi _{6}}{\varphi _{2}}-{\varphi _{1}}{\varphi _{5}}-{\varphi _{3}}{%
\varphi _{7}}}{r^{2}}.}  \nonumber \\
&&
\end{eqnarray}
For right actions, the standard cocycles are 
\begin{eqnarray}
{{f}_{123}^{(-)}{(\varphi )}} &=&-{\frac{{\varphi _{0}}^{2}-{\varphi _{6}}%
^{2}-{\varphi _{4}}^{2}+{\varphi _{2}}^{2}+{\varphi _{1}}^{2}-{\varphi _{7}}%
^{2}-{\varphi _{5}}^{2}+{\varphi _{3}}^{2}}{r^{2}},}  \nonumber \\
{{f}_{145}^{(-)}{(\varphi )}} &=&-{\frac{{\varphi _{0}}^{2}-{\varphi _{6}}%
^{2}+{\varphi _{4}}^{2}-{\varphi _{2}}^{2}+{\varphi _{1}}^{2}-{\varphi _{7}}%
^{2}+{\varphi _{5}}^{2}-{\varphi _{3}}^{2}}{r^{2}},}  \nonumber \\
{{f}_{176}^{(-)}{(\varphi )}} &=&-{\frac{{\varphi _{0}}^{2}+{\varphi _{6}}%
^{2}-{\varphi _{4}}^{2}-{\varphi _{2}}^{2}+{\varphi _{1}}^{2}+{\varphi _{7}}%
^{2}-{\varphi _{5}}^{2}-{\varphi _{3}}^{2}}{r^{2}},}  \nonumber \\
{{f}_{246}^{(-)}{(\varphi )}} &=&-{\frac{{\varphi _{0}}^{2}+{\varphi _{6}}%
^{2}+{\varphi _{4}}^{2}+{\varphi _{2}}^{2}-{\varphi _{1}}^{2}-{\varphi _{7}}%
^{2}-{\varphi _{5}}^{2}-{\varphi _{3}}^{2}}{{r^{2}}},}  \nonumber \\
{{f}_{257}^{(-)}{(\varphi )}} &=&-{\frac{{\varphi _{0}}^{2}-{\varphi _{6}}%
^{2}-{\varphi _{4}}^{2}+{\varphi _{2}}^{2}-{\varphi _{1}}^{2}+{\varphi _{7}}%
^{2}+{\varphi _{5}}^{2}-{\varphi _{3}}^{2}}{{r^{2}}},}  \nonumber \\
{{f}_{347}^{(-)}{(\varphi )}} &=&-{\frac{{\varphi _{0}}^{2}-{\varphi _{6}}%
^{2}+{\varphi _{4}}^{2}-{\varphi _{2}}^{2}-{\varphi _{1}}^{2}+{\varphi _{7}}%
^{2}-{\varphi _{5}}^{2}+{\varphi _{3}}^{2}}{r^{2}},}  \nonumber \\
{{f}_{365}^{(-)}{(\varphi )}} &=&-{\frac{{\varphi _{0}}^{2}+{\varphi _{6}}%
^{2}-{\varphi _{4}}^{2}-{\varphi _{2}}^{2}-{\varphi _{1}}^{2}-{\varphi _{7}}%
^{2}+{\varphi _{5}}^{2}+{\varphi _{3}}^{2}}{r^{2}},}  \nonumber \\
&&
\end{eqnarray}
while the non-standard cocycles are 
\begin{eqnarray}
{{f}_{124}^{(-)}{(\varphi )}} &=&+2{\frac{{\varphi _{0}}{\varphi _{7}}+{%
\varphi _{5}}{\varphi _{2}}-{\varphi _{6}}{\varphi _{1}}-{\varphi
_{3}\varphi _{4}}}{r^{2}},}  \nonumber \\
{\quad {f}_{125}^{(-)}{(\varphi )}} &=&-2{\frac{{\varphi _{0}}{\varphi _{6}}+%
{\varphi _{3}}{\varphi _{5}}+{\varphi _{1}}{\varphi _{7}}+{\varphi _{2}}{%
\varphi _{4}}}{r^{2}},}  \nonumber \\
{{f}_{126}^{(-)}{(\varphi )}} &=&+2{\frac{{\varphi _{0}}{\varphi _{5}}+{%
\varphi _{1}}{\varphi _{4}}-{\varphi _{7}}{\varphi _{2}}-{\varphi
_{3}\varphi _{6}}}{r^{2}},}  \nonumber \\
{\quad {f}_{127}^{(-)}{(\varphi )}} &=&-2{\frac{{\varphi _{0}}{\varphi _{4}}-%
{\varphi _{6}}{\varphi _{2}}-{\varphi _{1}}{\varphi _{5}}+{\varphi _{3}}{%
\varphi _{7}}}{r^{2}},}  \nonumber \\
{{f}_{143}^{(-)}{(\varphi )}} &=&+2{\frac{{\varphi _{0}}{\varphi _{6}}-{%
\varphi _{3}}{\varphi _{5}}-{\varphi _{2}}{\varphi _{4}}+{\varphi
_{1}\varphi _{7}}}{r^{2}},}  \nonumber \\
{\quad {f}_{146}^{(-)}{(\varphi )}} &=&-2{\frac{{\varphi _{3}}{\varphi _{0}}+%
{\varphi _{4}}{\varphi _{7}}+{\varphi _{1}}{\varphi _{2}}+{\varphi _{5}}{%
\varphi _{6}}}{r^{2}},}  \nonumber \\
{{f}_{175}^{(-)}{(\varphi )}} &=&+2{\frac{{\varphi _{3}}{\varphi _{0}}+{%
\varphi _{1}}{\varphi _{2}}-{\varphi _{5}}{\varphi _{6}}-{\varphi
_{4}\varphi _{7}}}{r^{2}},}  \nonumber \\
{\quad {f}_{247}^{(-)}{(\varphi )}} &=&-2{\frac{{\varphi _{0}}{\varphi _{1}}+%
{\varphi _{7}}{\varphi _{6}}+{\varphi _{4}}{\varphi _{5}}+{\varphi _{3}}{%
\varphi _{2}}}{r^{2}},}  \nonumber \\
{{f}_{147}^{(-)}{(\varphi )}} &=&+2{\frac{{\varphi _{2}}{\varphi _{0}}+{%
\varphi _{4}}{\varphi _{6}}-{\varphi _{5}}{\varphi _{7}}-{\varphi
_{1}\varphi _{3}}}{r^{2}},}  \nonumber \\
{\quad {f}_{243}^{(-)}{(\varphi )}} &=&-2{\frac{{\varphi _{0}}{\varphi _{5}}-%
{\varphi _{1}}{\varphi _{4}}-{\varphi _{7}}{\varphi _{2}}+{\varphi _{3}}{%
\varphi _{6}}}{r^{2}},}  \nonumber \\
{{f}_{253}^{(-)}{(\varphi )}} &=&+2{\frac{{\varphi _{0}}{\varphi _{4}}+{%
\varphi _{1}}{\varphi _{5}}-{\varphi _{6}}{\varphi _{2}}-{\varphi
_{3}\varphi _{7}}}{r^{2}},}  \nonumber \\
{\quad {f}_{173}^{(-)}{(\varphi )}} &=&-2{\frac{{\varphi _{0}}{\varphi _{5}}+%
{\varphi _{7}}{\varphi _{2}}+{\varphi _{3}}{\varphi _{6}}+{\varphi _{1}}{%
\varphi _{4}}}{r^{2}},}  \nonumber \\
{{f}_{245}^{(-)}{(\varphi )}} &=&+2{\frac{{\varphi _{3}}{\varphi _{0}}-{%
\varphi _{5}}{\varphi _{6}}+{\varphi _{4}}{\varphi _{7}}-{\varphi
_{1}\varphi _{2}}}{r^{2}},}  \nonumber \\
{\quad {f}_{256}^{(-)}{(\varphi )}} &=&+2{\frac{{\varphi _{0}}{\varphi _{1}}+%
{\varphi _{3}}{\varphi _{2}}-{\varphi _{7}}{\varphi _{6}}-{\varphi _{4}}{%
\varphi _{5}}}{r^{2}},}  \nonumber \\
{{f}_{361}^{(-)}{(\varphi )}} &=&+2{\frac{{\varphi _{0}}{\varphi _{4}}-{%
\varphi _{3}}{\varphi _{7}}-{\varphi _{1}}{\varphi _{5}}+{\varphi
_{6}\varphi _{2}}}{r^{2}},}  \nonumber \\
{\quad {f}_{362}^{(-)}{(\varphi )}} &=&-2{\frac{{\varphi _{0}}{\varphi _{7}}+%
{\varphi _{3}}{\varphi _{4}}+{\varphi _{5}}{\varphi _{2}}+{\varphi _{6}}{%
\varphi _{1}}}{r^{2}},}  \nonumber \\
{{f}_{345}^{(-)}{(\varphi )}} &=&-2{\frac{{\varphi _{2}}{\varphi _{0}}+{%
\varphi _{5}}{\varphi _{7}}+{\varphi _{1}}{\varphi _{3}}+{\varphi
_{4}\varphi _{6}}}{r^{2}},}  \nonumber \\
{\quad {f}_{346}^{(-)}{(\varphi )}} &=&+2{\frac{{\varphi _{0}}{\varphi _{1}}-%
{\varphi _{3}}{\varphi _{2}}-{\varphi _{7}}{\varphi _{6}}+{\varphi _{4}}{%
\varphi _{5}}}{r^{2}},}  \nonumber \\
{{f}_{367}^{(-)}{(\varphi )}} &=&+2{\frac{{\varphi _{2}}{\varphi _{0}}-{%
\varphi _{4}}{\varphi _{6}}-{\varphi _{5}}{\varphi _{7}}+{\varphi
_{1}\varphi _{3}}}{{r^{2}}},}  \nonumber \\
{\quad {f}_{135}^{(-)}{(\varphi )}} &=&-2{\frac{{\varphi _{0}}{\varphi _{7}}+%
{\varphi _{3}}{\varphi _{4}}-{\varphi _{6}}{\varphi _{1}}-{\varphi _{5}}{%
\varphi _{2}}}{r^{2}},}  \nonumber \\
{{f}_{156}^{(-)}{(\varphi )}} &=&-2{\frac{{\varphi _{2}}{\varphi _{0}}-{%
\varphi _{1}}{\varphi _{3}}-{\varphi _{4}}{\varphi _{6}}+{\varphi
_{5}\varphi _{7}}}{r^{2}},}  \nonumber \\
{\quad {f}_{237}^{(-)}{(\varphi )}} &=&+2{\frac{{\varphi _{0}}{\varphi _{6}}+%
{\varphi _{2}}{\varphi _{4}}-{\varphi _{3}}{\varphi _{5}}-{\varphi _{1}}{%
\varphi _{7}}}{r^{2}},}  \nonumber \\
{{f}_{267}^{(-)}{(\varphi )}} &=&-2{\frac{{\varphi _{3}}{\varphi _{0}}+{%
\varphi _{5}}{\varphi _{6}}-{\varphi _{4}}{\varphi _{7}}-{\varphi
_{1}\varphi _{2}}}{r^{2}},}  \nonumber \\
{\quad {f}_{357}^{(-)}{(\varphi )}} &=&+2{\frac{{\varphi _{0}}{\varphi _{1}}-%
{\varphi _{4}}{\varphi _{5}}+{\varphi _{7}}{\varphi _{6}}-{\varphi _{3}}{%
\varphi _{2}}}{r^{2}},}  \nonumber \\
{{f}_{456}^{(-)}{(\varphi )}} &=&-2{\frac{{\varphi _{0}}{\varphi _{7}}-{%
\varphi _{5}}{\varphi _{2}}-{\varphi _{3}}{\varphi _{4}}+{\varphi
_{6}\varphi _{1}}}{r^{2}},}  \nonumber \\
{\quad {f}_{457}^{(-)}{(\varphi )}} &=&+2{\frac{{\varphi _{0}}{\varphi _{6}}-%
{\varphi _{2}}{\varphi _{4}}-{\varphi _{1}}{\varphi _{7}}+{\varphi _{3}}{%
\varphi _{5}}}{r^{2}},}  \nonumber \\
{{f}_{467}^{(-)}{(\varphi )}} &=&-2{\frac{{\varphi _{0}}{\varphi _{5}}-{%
\varphi _{1}}{\varphi _{4}}-{\varphi _{3}}{\varphi _{6}}+{\varphi
_{7}\varphi _{2}}}{r^{2}},}  \nonumber \\
{\quad {f}_{567}^{(-)}{(\varphi )}} &=&+2{\frac{{\varphi _{0}}{\varphi _{4}}+%
{\varphi _{6}}{\varphi _{2}}+{\varphi _{1}}{\varphi _{5}}+{\varphi _{3}}{%
\varphi _{7}}}{r^{2}}.}  \nonumber \\
&&
\end{eqnarray}

\newpage

\chapter{The Fourth Appendix}


We give below some examples of the torsionful structure functions, 
\begin{eqnarray*}
\lefteqn{f_{123}^{\left( ++\right) }\left( \varphi ,\lambda \right)
=\!\!(y_{0}{}^{2}\,x_{3}{}^{2}+y_{0}{}^{2}\,x_{2}{}^{2}-y_{0}{}^{2}%
\,x_{5}{}^{2}-y_{0}{}^{2}\,x_{4}{}^{2}+y_{0}{}^{2}\,x_{6}{}^{2}-y_{0}{}^{2}%
\,x_{7}{}^{2}-} \\
&&\mbox{}+{y_{0}}^{2}\,{x_{1}}^{2}+{y_{7}}^{2}\,{x_{4}}^{2}+{y_{7}}^{2}\,{%
x_{6}}^{2}-4\,{y_{0}}\,{y_{6}}\,{x_{1}}\,{x_{7}} \\
&&\mbox{}-{y_{3}}^{2}\,{x_{4}}^{2}-y_{0}{}^{2}\,x_{4}{}^{2}-y_{0}{}^{2}%
\,x_{6}{}^{2}-y_{0}{}^{2}\,x_{7}{}^{2} \\
&&\mbox{}-{y_{3}}^{2}\,{x_{6}}^{2}+4\,{y_{0}}\,{y_{4}}\,{x_{3}}\,{x_{7}}-4\,{%
y_{0}}\,{y_{7}}\,{x_{5}}\,{x_{2}}-4\,{y_{0}}\,{y_{7}}\,{x_{3}}\,{x_{4}} \\
&&\mbox{}+4\,{y_{0}}\,{y_{7}}\,{x_{6}}\,{x_{1}}+{y_{1}}^{2}\,{x_{1}}^{2}-{%
y_{1}}^{2}\,{x_{7}}^{2}-{y_{1}}^{2}\,{x_{5}}^{2}+{y_{1}}^{2}\,{x_{2}}^{2} \\
&&\mbox{}-{y_{1}}^{2}\,{x_{4}}^{2}-{y_{1}}^{2}\,{x_{6}}^{2}+{y_{1}}^{2}\,{%
x_{3}}^{2}-4\,{y_{0}}\,{y_{5}}\,{x_{1}}\,{x_{4}} \\
&&\mbox{}-4\,{y_{0}}\,{y_{5}}\,{x_{3}}\,{x_{6}}+4\,{y_{0}}\,{y_{4}}\,{x_{2}}%
\,{x_{6}}+4\,{y_{0}}\,{y_{6}}\,{x_{3}}\,{x_{5}} \\
&&\mbox{}-4\,{y_{0}}\,{y_{6}}\,{x_{2}}\,{x_{4}}+4\,{y_{1}}\,{y_{4}}\,{x_{7}}%
\,{x_{2}}+4\,{y_{0}}\,{y_{4}}\,{x_{1}}\,{x_{5}} \\
&&\mbox{}+4\,{y_{1}}\,{y_{5}}\,{x_{1}}\,{x_{5}}+4\,{y_{1}}\,{y_{4}}\,{x_{1}}%
\,{x_{4}}-{y_{2}}^{2}\,{x_{4}}^{2}-4\,{y_{1}}\,{y_{5}}\,{x_{3}}\,{x_{7}} \\
&&\mbox{}+{y_{2}}^{2}\,{x_{2}}^{2}+4\,{y_{1}}\,{y_{6}}\,{x_{3}}\,{x_{4}}+4\,{%
y_{1}}\,{y_{6}}\,{x_{5}}\,{x_{2}}+{y_{2}}^{2}\,{x_{1}}^{2} \\
&&\mbox{}+4\,{y_{1}}\,{y_{6}}\,{x_{6}}\,{x_{1}}-{y_{2}}^{2}\,{x_{6}}^{2}+4\,{%
y_{3}}\,{y_{5}}\,{x_{2}}\,{x_{4}}-{y_{7}}^{2}\,{x_{1}}^{2} \\
&&\mbox{}-{y_{2}}^{2}\,{x_{5}}^{2}+4\,{y_{3}}\,{y_{4}}\,{x_{3}}\,{x_{4}}-4\,{%
y_{3}}\,{y_{7}}\,{x_{6}}\,{x_{2}}+4\,{y_{3}}\,{y_{7}}\,{x_{3}}\,{x_{7}} \\
&&\mbox{}+4\,{y_{1}}\,{y_{7}}\,{x_{1}}\,{x_{7}}-4\,{y_{3}}\,{y_{4}}\,{x_{5}}%
\,{x_{2}}-{y_{6}}^{2}\,{x_{1}}^{2}+{y_{6}}^{2}\,{x_{5}}^{2} \\
&&\mbox{}-{y_{4}}^{2}\,{x_{3}}^{2}+4\,{y_{3}}\,{y_{4}}\,{x_{6}}\,{x_{1}}-{%
y_{6}}^{2}\,{x_{2}}^{2}+4\,{y_{3}}\,{y_{5}}\,{x_{3}}\,{x_{5}} \\
&&\mbox{}+4\,{y_{3}}\,{y_{6}}\,{x_{7}}\,{x_{2}}+{y_{6}}^{2}\,{x_{4}}^{2}+4\,{%
y_{2}}\,{y_{7}}\,{x_{1}}\,{x_{4}}+{y_{6}}^{2}\,{x_{6}}^{2} \\
&&\mbox{}+{y_{6}}^{2}\,{x_{7}}^{2}-4\,{y_{3}}\,{y_{6}}\,{x_{1}}\,{x_{4}}+4\,{%
y_{3}}\,{y_{6}}\,{x_{3}}\,{x_{6}}-{y_{6}}^{2}\,{x_{3}}^{2} \\
&&\mbox{}+{y_{5}}^{2}\,{x_{4}}^{2}+{y_{5}}^{2}\,{x_{6}}^{2}+{y_{5}}^{2}\,{%
x_{5}}^{2}+4\,{y_{3}}\,{y_{5}}\,{x_{1}}\,{x_{7}}+{y_{5}}^{2}\,{x_{7}}^{2} \\
&&\mbox{}-4\,{y_{1}}\,{y_{4}}\,{x_{3}}\,{x_{6}}+{y_{2}}^{2}\,{x_{3}}^{2}-{%
y_{3}}^{2}\,{x_{7}}^{2}-4\,{y_{1}}\,{y_{7}}\,{x_{2}}\,{x_{4}} \\
&&\mbox{}+4\,{y_{1}}\,{y_{7}}\,{x_{3}}\,{x_{5}}-{y_{2}}^{2}\,{x_{7}}^{2}-4\,{%
y_{1}}\,{y_{5}}\,{x_{6}}\,{x_{2}}-4\,{y_{3}}\,{y_{7}}\,{x_{1}}\,{x_{5}} \\
&&\mbox{}-{y_{5}}^{2}\,{x_{2}}^{2}-{y_{5}}^{2}\,{x_{3}}^{2}+4\,{y_{0}}\,{%
y_{5}}\,{x_{2}}\,{x_{7}}-4\,{y_{2}}\,{y_{6}}\,{x_{5}}\,{x_{1}} \\
&&\mbox{}+4\,{y_{2}}\,{y_{6}}\,{x_{6}}\,{x_{2}}+{y_{3}}^{2}\,{x_{1}}^{2}-{%
y_{3}}^{2}\,{x_{5}}^{2}+4\,{y_{2}}\,{y_{5}}\,{x_{5}}\,{x_{2}} \\
&&\mbox{}+{y_{7}}^{2}\,{x_{7}}^{2}-{y_{7}}^{2}\,{x_{2}}^{2}+{y_{4}}^{2}\,{%
x_{5}}^{2}-{y_{5}}^{2}\,{x_{1}}^{2}+{y_{7}}^{2}\,{x_{5}}^{2} \\
&&\mbox{}+4\,{y_{2}}\,{y_{7}}\,{x_{3}}\,{x_{6}}-{y_{7}}^{2}\,{x_{3}}^{2}+4\,{%
y_{2}}\,{y_{7}}\,{x_{7}}\,{x_{2}}+4\,{y_{2}}\,{y_{5}}\,{x_{6}}\,{x_{1}} \\
&&\mbox{}-4\,{y_{2}}\,{y_{5}}\,{x_{3}}\,{x_{4}}-4\,{y_{2}}\,{y_{6}}\,{x_{3}}%
\,{x_{7}}-4\,{y_{2}}\,{y_{4}}\,{x_{7}}\,{x_{1}}+{y_{4}}^{2}\,{x_{7}}^{2} \\
&&\mbox{}-4\,{x_{0}}\,{y_{1}}\,{y_{6}}\,{x_{7}}-4\,{x_{0}}\,{y_{3}}\,{y_{7}}%
\,{x_{4}}+4\,{x_{0}}\,{y_{2}}\,{y_{5}}\,{x_{7}} \\
&&\mbox{}-4\,{x_{0}}\,{y_{3}}\,{y_{5}}\,{x_{6}}+{x_{0}}^{2}\,{y_{0}}^{2}+{%
x_{0}}^{2}\,{y_{3}}^{2}+{x_{0}}^{2}\,{y_{1}}^{2}-{x_{0}}^{2}\,{y_{6}}^{2} \\
&&\mbox{}+{x_{0}}^{2}\,{y_{2}}^{2}-{x_{0}}^{2}\,{y_{4}}^{2}-{x_{0}}^{2}\,{%
y_{7}}^{2}-{x_{0}}^{2}\,{y_{5}}^{2}+{y_{4}}^{2}\,{x_{6}}^{2} \\
&&\mbox{}+{y_{4}}^{2}\,{x_{4}}^{2}-{y_{4}}^{2}\,{x_{2}}^{2}+4\,{y_{2}}\,{%
y_{4}}\,{x_{3}}\,{x_{5}}+4\,{y_{2}}\,{y_{4}}\,{x_{2}}\,{x_{4}} \\
&&\mbox{}-{y_{4}}^{2}\,{x_{1}}^{2}+{y_{3}}^{2}\,{x_{2}}^{2}+{y_{3}}^{2}\,{%
x_{3}}^{2}-4\,{x_{0}}\,{y_{2}}\,{y_{7}}\,{x_{5}} \\
&&\mbox{}+4\,{x_{0}}\,{y_{2}}\,{y_{4}}\,{x_{6}}-4\,{x_{0}}\,{y_{2}}\,{y_{6}}%
\,{x_{4}}+4\,{x_{0}}\,{y_{3}}\,{y_{6}}\,{x_{5}} \\
&&\mbox{}-4\,{x_{0}}\,{y_{0}}\,{y_{7}}\,{x_{7}}-4\,{x_{0}}\,{y_{1}}\,{y_{5}}%
\,{x_{4}}-4\,{x_{0}}\,{y_{0}}\,{y_{5}}\,{x_{5}} \\
&&\mbox{}+4\,{x_{0}}\,{y_{3}}\,{y_{4}}\,{x_{7}}-4\,{x_{0}}\,{y_{0}}\,{y_{4}}%
\,{x_{4}}+4\,{x_{0}}\,{y_{1}}\,{y_{7}}\,{x_{6}} \\
&&\mbox{}-4\,{x_{0}}\,{y_{0}}\,{y_{6}}\,{x_{6}}+4\,{x_{0}}\,{y_{1}}\,{y_{4}}%
\,{x_{5}})\!\;\;/\;\;\;\! \\
&&\mbox{}(({y_{0}}^{2}+{y_{1}}^{2}+{y_{2}}^{2}+{y_{3}}^{2}+{y_{4}}^{2}+{y_{5}%
}^{2}+{y_{6}}^{2}+{y_{7}}^{2}) \\
&&\mbox{}({x_{0}}^{2}+{x_{5}}^{2}+{x_{3}}^{2}+{x_{1}}^{2}+{x_{2}}^{2}+{x_{7}}%
^{2}+{x_{6}}^{2}+{x_{4}}^{2}))\!\!
\end{eqnarray*}
\begin{eqnarray*}
\lefteqn{f_{127}^{\left( ++\right) }\left( \varphi ,\lambda \right)
=\!\!-2(y_{0}\,y_{4}\,x_{5}{}^{2}-2y_{0}\,y_{6}x_{1}x_{3}-2y_{0}%
\,y_{5}x_{4}x_{5}} \\
&&\mbox{}+2\,{y_{0}}\,{y_{5}}\,{x_{6}}\,{x_{7}}+2\,{y_{0}}\,{y_{5}}\,{x_{2}}%
\,{x_{3}}+{y_{1}}^{2}\,{x_{6}}\,{x_{2}}+{y_{1}}^{2}\,{x_{5}}\,{x_{1}}%
_{1}\,x_{3}-2y_{0}\,y_{5}\,x_{4}\,x_{5} \\
&&\mbox{}-2\,{y_{1}}\,{y_{6}}\,{x_{1}}\,{x_{2}}-{y_{1}}\,{y_{5}}\,{x_{3}}%
^{2}-{y_{0}}\,{y_{4}}\,{x_{1}}^{2}+{y_{0}}\,{y_{4}}\,{x_{3}}^{2} \\
&&\mbox{}-2\,{y_{1}}\,{y_{6}}\,{x_{4}}\,{x_{7}}+2\,{y_{1}}\,{y_{6}}\,{x_{5}}%
\,{x_{6}}+2\,{y_{1}}\,{y_{4}}\,{x_{4}}\,{x_{5}} \\
&&\mbox{}+2\,{y_{1}}\,{y_{3}}\,{x_{1}}\,{x_{7}}+2\,{y_{1}}\,{y_{4}}\,{x_{6}}%
\,{x_{7}}+2\,{y_{1}}\,{y_{4}}\,{x_{2}}\,{x_{3}} \\
&&\mbox{}-{y_{1}}^{2}\,{x_{3}}\,{x_{7}}+2\,{y_{2}}\,{y_{5}}\,{x_{5}}\,{x_{6}}%
-{y_{3}}^{2}\,{x_{6}}\,{x_{2}}-{y_{3}}^{2}\,{x_{1}}\,{x_{5}} \\
&&\mbox{}+2\,{y_{1}}\,{y_{3}}\,{x_{3}}\,{x_{5}}-2\,{y_{1}}\,{y_{3}}\,{x_{2}}%
\,{x_{4}}-{y_{1}}\,{y_{5}}\,{x_{1}}^{2}+{y_{1}}\,{y_{5}}\,{x_{5}}^{2} \\
&&\mbox{}+{y_{1}}\,{y_{5}}\,{x_{7}}^{2}+{y_{2}}\,{y_{6}}\,{x_{6}}^{2}-{y_{2}}%
\,{y_{6}}\,{x_{4}}^{2}-{y_{1}}\,{y_{5}}\,{x_{6}}^{2} \\
&&\mbox{}-{y_{1}}\,{y_{5}}\,{x_{4}}^{2}+{y_{1}}\,{y_{5}}\,{x_{2}}^{2}-{y_{0}}%
^{2}\,{x_{3}}\,{x_{7}}+{y_{0}}^{2}\,{x_{5}}\,{x_{1}} \\
&&\mbox{}+{y_{0}}^{2}\,{x_{6}}\,{x_{2}}-2\,{y_{0}}\,{y_{6}}\,{x_{4}}\,{x_{6}}%
-2\,{y_{0}}\,{y_{6}}\,{x_{5}}\,{x_{7}}-{y_{0}}\,{y_{4}}\,{x_{2}}^{2} \\
&&\mbox{}+{y_{2}}^{2}\,{x_{6}}\,{x_{2}}+{y_{2}}\,{y_{6}}\,{x_{7}}^{2}-{y_{2}}%
\,{y_{6}}\,{x_{2}}^{2}-{y_{2}}\,{y_{6}}\,{x_{5}}^{2} \\
&&\mbox{}-{y_{2}}\,{y_{6}}\,{x_{3}}^{2}-{y_{0}}\,{y_{4}}\,{x_{7}}^{2}+{y_{0}}%
\,{y_{4}}\,{x_{6}}^{2}-{y_{0}}\,{y_{4}}\,{x_{4}}^{2} \\
&&\mbox{}+{y_{2}}^{2}\,{x_{5}}\,{x_{1}}-{y_{2}}^{2}\,{x_{3}}\,{x_{7}}+2\,{%
y_{0}}\,{y_{3}}\,{x_{6}}\,{x_{1}}+2\,{y_{2}}\,{y_{3}}\,{x_{1}}\,{x_{4}} \\
&&\mbox{}-2\,{y_{0}}\,{y_{3}}\,{x_{3}}\,{x_{4}}-2\,{y_{0}}\,{y_{3}}\,{x_{5}}%
\,{x_{2}}-2\,{y_{2}}\,{y_{4}}\,{x_{5}}\,{x_{7}} \\
&&\mbox{}+2\,{y_{2}}\,{y_{4}}\,{x_{4}}\,{x_{6}}-2\,{y_{2}}\,{y_{4}}\,{x_{1}}%
\,{x_{3}}-2\,{y_{4}}\,{y_{7}}\,{x_{6}}\,{x_{1}} \\
&&\mbox{}+2\,{y_{4}}\,{y_{7}}\,{x_{2}}\,{x_{5}}-2\,{y_{5}}\,{y_{7}}\,{x_{7}}%
\,{x_{1}}-2\,{y_{5}}\,{y_{7}}\,{x_{3}}\,{x_{5}} \\
&&\mbox{}-2\,{y_{5}}\,{y_{7}}\,{x_{2}}\,{x_{4}}-2\,{y_{6}}\,{y_{7}}\,{x_{7}}%
\,{x_{2}}+2\,{y_{6}}\,{y_{7}}\,{x_{1}}\,{x_{4}} \\
&&\mbox{}+{x_{0}}^{2}\,{y_{0}}\,{y_{4}}+2\,{y_{2}}\,{y_{3}}\,{x_{7}}\,{x_{2}}%
+2\,{y_{2}}\,{y_{3}}\,{x_{3}}\,{x_{6}}+{x_{0}}\,{y_{7}}^{2}\,{x_{4}} \\
&&\mbox{}-{x_{0}}\,{y_{3}}^{2}\,{x_{4}}-{x_{0}}\,{y_{4}}^{2}\,{x_{4}}+{x_{0}}%
\,{y_{2}}^{2}\,{x_{4}}+{x_{0}}\,{y_{0}}^{2}\,{x_{4}} \\
&&\mbox{}-{x_{0}}\,{y_{5}}^{2}\,{x_{4}}-2\,{y_{2}}\,{y_{5}}\,{x_{1}}\,{x_{2}}%
+2\,{y_{2}}\,{y_{5}}\,{x_{4}}\,{x_{7}} \\
&&\mbox{}-2\,{y_{4}}\,{y_{7}}\,{x_{3}}\,{x_{4}}+{y_{2}}\,{y_{6}}\,{x_{1}}%
^{2}+{y_{3}}^{2}\,{x_{3}}\,{x_{7}}+{y_{3}}\,{y_{7}}\,{x_{4}}^{2} \\
&&\mbox{}+{y_{3}}\,{y_{7}}\,{x_{6}}^{2}-{y_{3}}\,{y_{7}}\,{x_{2}}^{2}+{y_{3}}%
\,{y_{7}}\,{x_{5}}^{2}-{y_{3}}\,{y_{7}}\,{x_{3}}^{2} \\
&&\mbox{}-{y_{3}}\,{y_{7}}\,{x_{1}}^{2}+{y_{3}}\,{y_{7}}\,{x_{7}}^{2}-{y_{4}}%
^{2}\,{x_{6}}\,{x_{2}}+{y_{4}}^{2}\,{x_{7}}\,{x_{3}} \\
&&\mbox{}-{y_{4}}^{2}\,{x_{1}}\,{x_{5}}-{y_{5}}^{2}\,{x_{6}}\,{x_{2}}-{y_{5}}%
^{2}\,{x_{5}}\,{x_{1}}+{y_{5}}^{2}\,{x_{3}}\,{x_{7}} \\
&&\mbox{}+{y_{7}}^{2}\,{x_{6}}\,{x_{2}}+{y_{7}}^{2}\,{x_{5}}\,{x_{1}}-{y_{6}}%
^{2}\,{x_{5}}\,{x_{1}}-{y_{6}}^{2}\,{x_{2}}\,{x_{6}} \\
&&\mbox{}+{y_{6}}^{2}\,{x_{3}}\,{x_{7}}-{y_{7}}^{2}\,{x_{3}}\,{x_{7}}+{x_{0}}%
^{2}\,{y_{2}}\,{y_{6}}-2\,{y_{6}}\,{y_{7}}\,{x_{3}}\,{x_{6}} \\
&&\mbox{}-{x_{0}}^{2}\,{y_{3}}\,{y_{7}}+{x_{0}}^{2}\,{y_{1}}\,{y_{5}}-{x_{0}}%
\,{y_{6}}^{2}\,{x_{4}}+{x_{0}}\,{y_{1}}^{2}\,{x_{4}} \\
&&\mbox{}-2\,{x_{0}}\,{y_{0}}\,{y_{3}}\,{x_{7}}-2\,{x_{0}}\,{y_{4}}\,{y_{7}}%
\,{x_{7}}-2\,{x_{0}}\,{y_{1}}\,{y_{6}}\,{x_{3}} \\
&&\mbox{}-2\,{x_{0}}\,{y_{2}}\,{y_{3}}\,{x_{5}}+2\,{x_{0}}\,{y_{5}}\,{y_{7}}%
\,{x_{6}}+2\,{x_{0}}\,{y_{0}}\,{y_{5}}\,{x_{1}} \\
&&\mbox{}+2\,{x_{0}}\,{y_{0}}\,{y_{6}}\,{x_{2}}-2\,{x_{0}}\,{y_{2}}\,{y_{4}}%
\,{x_{2}}+2\,{x_{0}}\,{y_{1}}\,{y_{3}}\,{x_{6}} \\
&&\mbox{}+2\,{x_{0}}\,{y_{2}}\,{y_{5}}\,{x_{3}}-2\,{x_{0}}\,{y_{6}}\,{y_{7}}%
\,{x_{5}}-2\,{x_{0}}\,{y_{1}}\,{y_{4}}\,{x_{1}})\!\;\;/\;\; \\
&&\mbox{}(({y_{0}}^{2}+{y_{1}}^{2}+{y_{2}}^{2}+{y_{3}}^{2}+{y_{4}}^{2}+{y_{5}%
}^{2}+{y_{6}}^{2}+{y_{7}}^{2}) \\
&&\mbox{}({x_{0}}^{2}+{x_{5}}^{2}+{x_{3}}^{2}+{x_{1}}^{2}+{x_{2}}^{2}+{x_{7}}%
^{2}+{x_{6}}^{2}+{x_{4}}^{2}))\!\!
\end{eqnarray*}

\newpage


\end{document}